\documentclass[a4paper,fleqn,usenatbib,useAMS]{mnras}

\usepackage{amsmath}
\usepackage{graphicx} 
\usepackage{times}
\usepackage{enumerate}
\usepackage{enumitem}

\newcommand{\gsim}{\raisebox{-0.13cm}{~\shortstack{$>$ \\[-0.07cm]
      $\sim$}}~}
\usepackage[T1]{fontenc}
\usepackage{ae,aecompl}

\title[Spark Motion in IAR to study Subpulse Drifting]{A Mechanism of Spark Motion in Inner Acceleration Region to Investigate Subpulse Drifting in Pulsars}
\author[Basu, Mitra \& Melikidze]{Rahul Basu$^{1,2}$, Dipanjan Mitra$^{2,3}$, George I. Melikidze$^{2,4}$ \\
$^{1}$ Inter-University Centre for Astronomy and Astrophysics, Pune, 411007, India; rahulbasu.astro@gmail.com \\
$^{2}$ Janusz Gil Institute of Astronomy, University of Zielona G\'ora, ul. Szafrana 2, 65-516 Zielona G\'ora, Poland \\
$^{3}$ National Centre for Radio Astrophysics, Tata Institute of Fundamental Research, Pune 411007, India \\
$^{4}$ Evgeni Kharadze Georgian National Astrophysical Observatory, 0301, Abastumani, Georgia \\
}
\begin{document}

%\date{Accepted\ldots Received\ldots ; in original form\ldots}

%\pubyear{2017}

\maketitle

\label{firstpage}

\begin{abstract}
Coherent radio emission in pulsars is excited due to instabilities in a 
relativistically streaming non-stationary plasma flow, which is generated from 
sparking discharges in the inner acceleration region (IAR) near the stellar 
surface. A number of detailed works have shown the IAR to be a partially 
screened gap (PSG) dominated by non-dipolar magnetic fields with continuous 
outflow of ions from the surface. The phenomenon of subpulse drifting is 
expected to originate due to variable $\mathbf{E}\times\mathbf{B}$ drift of the
sparks in PSG, where the sparks lag behind corotation velocity of the pulsar. 
Detailed observations show a wide variety of subpulse drifting behaviour where 
subpulses in different components of the profile have different phase 
trajectories. But the drifting periodicity is seen to be constant, within 
measurement errors, across all components of the profile. Using the concept of
sparks lagging behind corotation speed in PSG as well as the different 
orientations of the surface non-dipolar magnetic fields we have simulated the 
expected single pulse behaviour in a representative sample of pulsars. Our 
results show that the different types of drifting phase behaviour can be 
reproduced using these simple assumptions of spark dynamics in a non-dipolar 
IAR.
\end{abstract}


\section{Introduction}
\noindent
One of the most intriguing features seen in the radio emission from pulsars 
involve the phenomenon of subpulse drifting, where systematic periodic shifts 
in individual components of a single pulse, known as subpulses, are seen within
the pulse window \citep{1968Natur.220..231D}. A complete physical understanding
of this phenomenon is still absent, however the most successful explanation is 
provided in the work of \citet[][hereafter RS75]{1975ApJ...196...51R}. 
According to the RS75 model an inner acceleration region (IAR) exists above the
pulsar polar cap, where sparking discharges generate a spark associated 
non-stationary plasma flow, which stream relativistically along the open 
magnetic field lines and leave the pulsar magnetosphere as relativistic pulsar 
wind \citep{1969ApJ...157..869G}. The non-stationary flow is necessary to 
excite the coherent radio emission \citep{1998MNRAS.301...59A,
2000ApJ...544.1081M,2009ApJ...696L.141M,2014ApJ...794..105M,
2018MNRAS.480.4526L}, while  the drift motion of the spark associated plasma 
column results in subpulse drift. In the RS75 model the prototype for the IAR 
is the inner vacuum gap (IVG), which is formed in pulsars with 
$\mathbf{\Omega}\cdot\mathbf{B} <$ 0 above the polar cap, where 
$\mathbf{\Omega} (= 2 \pi/P)$ is the angular velocity of pulsar with period 
$P$, and $\mathbf{B}$ is the magnetic field. It was postulated that due to the 
high binding energy of ions the positive charges cannot escape above the polar 
cap to screen the electric field and hence the IVG forms. The potential drop in 
the IVG is extremely high, about $10^{12}$ V, and several isolated discharges 
mediated via magnetic $e^-e^+$ pair creation is setup in the gap. The electric 
field in the gap separates the pairs and accelerate the charges in opposite 
directions. The $e^-$ travel downwards towards the stellar surface and the 
$e^+$ upwards, producing high energy photons via curvature radiation and/or 
inverse compton scattering, which lead to a pair cascade. This process 
continues till the entire potential drop along the IVG is screened, i.e., when 
the charge density reaches the so called Goldreich-Julian density 
($\rho_{GJ}$). During this interval the discharge also grows in the 
perpendicular direction into adjacent field lines and eventually a full formed 
spark develops. Once the IVG is screened, the spark associated plasma column 
leaves the gap and the electric potential appears once again for the sparking 
process to commence. Several studies have shown that the efficiency of the pair
cascade process requires the presence of strong surface non-dipolar magnetic 
fields (e.g. \citealt{Timokhin2019}) with radius of curvature ($\rho_c$) of 
about $10^5-10^6$ cm. In contrast $\rho_c$ for dipolar magnetic field is around
$10^8$ cm, and hence the presence of highly non-dipolar surface magnetic field 
in the IAR is essential\footnote{There are also indications of the presence of 
non-dipolar magnetic fields near the surface from X-ray observations of normal 
period pulsars \citep[see][]{2019MNRAS.489.4589A,2017JApA...38...46G,
2020MNRAS.493.3770S}.}. When the charge density in sparks reaches $\rho_{GJ}$, 
the force-free condition is achieved and the plasma column corotates with the 
star due to $\mathbf{E}\times\mathbf{B}$ drift. However, during the sparking 
process the charge density in the gap is below $\rho_{GJ}$, resulting in the 
charges lagging behind the corotation motion. This phenomenon is responsible 
for the observed subpulse drifting in observed radio emission (see e.g. RS75, 
\citealt{1985MNRAS.215..111A,2013arXiv1304.4203S,2016ApJ...833...29B,
2020MNRAS.492.2468M}).

Detailed classification studies of the observed drifting behaviour in the 
pulsar population have been carried out in the literature 
\citep{1986ApJ...301..901R,2006A&A...445..243W,2007A&A...469..607W,
2016ApJ...833...29B,2018MNRAS.475.5098B,2019MNRAS.482.3757B}. Subpulse drifting
shows a wide variety of subpulse motion characterised by phase variations in 
fluctuation spectral analysis \citep{1973ApJ...182..245B,1975ApJ...197..481B}. 
The drifting periodicity ($P_3$), the interval at which the subpulses repeat at
any location within the pulse window, is identical across all components in the
pulsar profile despite large phase variations between them. In pulsars where a 
central core emission is seen surrounded by one or two conal pairs, subpulse 
drifting is absent in the central core component and only seen in the 
surrounding cones. The systematic drifting behaviour has been classified by 
\citet{2019MNRAS.482.3757B} into three major categories: 
\begin{enumerate}[label=\emph{\alph*}),align=left, leftmargin=*,topsep=0pt]
\item {\bf Coherent phase-modulated} drifting where the subpulses continuously 
shift from one edge of the pulse window to the other. The phase show large 
monotonic variations across the profile which are usually non-linear.
\item {\bf Switching phase-modulated} drifting is seen in pulsars with more 
than one component in the profile where subpulses show systematic variations 
across each component resulting in large phase variations, but there are sudden
shifts between adjacent components. In certain cases the subpulses show 
opposite sense of variation in different components with slopes of phase 
variations showing opposite signs. This phenomenon is known as bi-drifting 
\citep{2005MNRAS.363..929C,2016A&A...590A.109W,2018MNRAS.475.5098B,
2019MNRAS.486.5216B}. 
\item {\bf Low-mixed phase-modulated} drifting is also seen in pulsars with 
multi-component profiles where the subpulses do not show large shifts during 
drifting. The resulting phase variations are relatively flat across each 
component.
\end{enumerate}
One of the primary challenges is to understand these wide variety seen in the 
drifting behaviour from the perspective of the sparking model, which forms the 
main focus of this work. In section \ref{sec:PSG} we discuss the origin of 
subpulse drifting in the physically consistent Partially Screened Gap 
\citep[PSG,][]{2003A&A...407..315G} model of the IAR. Section 
\ref{sec:singlmod} presents a simplified mechanism to generate single pulses 
exhibiting subpulse drifting, which is subsequently used is section 
\ref{sec:driftsiml} to study the drifting behaviour in different surface 
magnetic field configurations. A short discussion regarding the implications of
these simulations on subpulse drifting as well as their limitations are 
presented in section \ref{sec:disc}.

\section{Subpulse Drifting in the Partially Screened Gap Model}\label{sec:PSG}
\subsection{Partially Screened Gap Model}
Formation of the IVG in the RS75 model requires binding energy at the stellar 
surface to be sufficiently high to prevent positive ions to escape. A number of
subsequent studies have found the above assumptions to be inadequate, 
particularly when the polar cap surface is constantly bombarded by 
back-streaming electrons during the sparking process 
\citep{1980ApJ...235..576C,2003A&A...407..315G}. This can cause the polar cap 
temperatures to rise above $10^6$ K, which is sufficient for a continuous 
outflow of positively charged ions from the stellar surface. 
\citet{2003A&A...407..315G} suggested that the IAR is in fact a partially 
screened gap (PSG), with screening factor $\eta=1-\rho_i/\rho_{GJ}$, where 
$\rho_i$ is the charge density due to production of ions in the IAR. The PSG is
thermally regulated around the critical temperature ($T_i$) of ion free flow 
from the surface, where any drop in temperature below this critical value is 
accompanied by sparking discharges to quickly reheat the surface back to the 
critical level. One of the most important features of the PSG is providing a 
mechanism for the sparking regions to be stable within the IAR. The region 
between sparks in the PSG is screened due to the presence of plasma with charge
density $\rho_{GJ}$, and hence no additional particle acceleration can take 
place. On the other hand in the IVG the region between the sparks is vacuum, 
where unscreened electric field exist and hence can discharge due to pair 
creation. As a result the sparking discharges cannot be confined at any 
location on the surface, but moves continually opposite to the principal normal
of the curvature of the local magnetic field lines \citep{1977ApJ...214..598C,
1980ApJ...235..576C}. In contrast in the PSG the development of pair cascade is
restricted to localised sparking regions where the surface temperature is below
the critical level and is inhibited in the region between the sparks.

The behaviour of subpulse drifting further justifies the requirement of PSG in 
IAR. \citet{2016ApJ...833...29B} found the measured drifting periodicity to be 
anti-correlated with the spin-down energy loss ($\dot{E}$), with an estimated 
dependence of $P_3/P\propto\dot{E}^{-0.6\pm0.1}$. The PSG nature of the IAR 
currently provides the only explanation for this observed dependence. The IAR 
potential is screened by the screening factor which gives an estimate of the 
speed of the sparks in IAR, and consequently the periodicity under certain 
approximations can be estimated as $P_3/P = 1/2 \pi \eta 
\lvert\cos{\alpha_l}\rvert$, where $\alpha_l$ is the angle the local 
non-dipolar magnetic field makes with the rotation axis 
\citep{2013arXiv1304.4203S,2020MNRAS.492.2468M}. It can be shown that in 
typical pulsars $\eta\sim0.1$, and is proportional to the spin-down energy loss
as $\eta\propto\sqrt{\dot{E}}$, which gives the observed dependence \citep[see 
discussions in][]{2016ApJ...833...29B,2020MNRAS.492.2468M}.

\begin{figure*}
\includegraphics[scale=0.9,angle=0.]{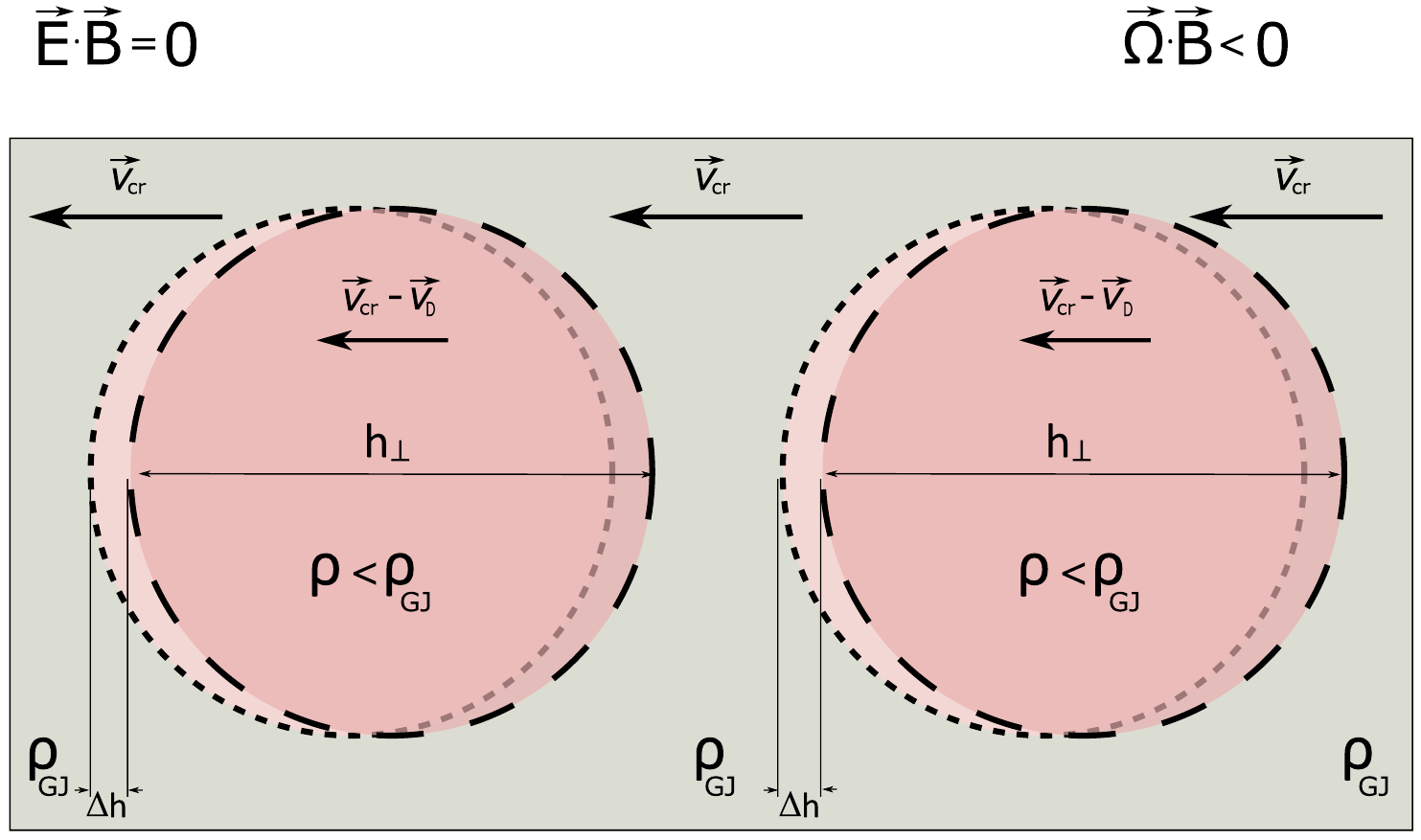} 
\caption{The figure shows a top-down view of the time evolution of the sparking
process from the observer's frame, as proposed in the PSG model. The 
rectangular box corresponds to the polar cap with $\vec{\Omega}\cdot\vec{B} < 
0$, where two distinct sparking regions (circles with lighter shade to darker 
shade of pink) with typical size $h_\perp$ are shown. The direction of rotation
of the pulsar is from right to left indicated by the corotation velocity, 
$\vec{v_{cr}}$. The region outside the polar cap has $\vec{E}\cdot\vec{B} = 0$ 
and also corotates with the pulsar. The pale green region corresponds to the 
region between sparks filled with charge density $\rho_{GJ}$. In the sparking 
region the charge density is $\rho < \rho_{GJ}$. With the progression of time 
the initial spark (lighter shade circle) is replaced by a subsequent spark 
(shown in darker shade) which is shifted by a distance $\Delta h$ opposite to 
the corotation direction and moving with an effective velocity $\vec{v_{cr}} - 
\vec{v_{D}}$.}
\label{fig_sparkgap}
\end{figure*}

In figure~\ref{fig_sparkgap}, the time evolution of the sparking process in the
PSG model is shown. When the surface temperature ($T_S$) in the polar cap 
exceeds $T_i$, the ions steadily flow out from the stellar surface, populating 
the IAR with $\rho_{GJ}$ which screens the electric potential along the gap. 
Sparking commences at localised regions where $T_S$ drops below critical level 
such that the charge density is reduced by a factor $\eta$ and large potential 
difference builds up along the gap. This instigates $e^-e^+$ pair cascades in 
this sparking region, with relativistic back-streaming electrons bombarding the
stellar surface and heating the region below the spark. The sparking process 
continues until $T_s \gsim T_i$ is reached, when the ions can freely flow out 
of the stellar surface and screen the electric potential along the gap with 
$\rho_{GJ}$. In the absence of electric field, intense pair production 
terminates causing the sparking process to stop. The typical timescale over 
which this full spark develops is estimated to be $t_{sp}\sim10~\mu$s (see e.g.
\citealt{2003A&A...407..315G}). The back-streaming electrons during $t_{sp}$ 
lag behind the corotation of the star and hence the peak heating region also 
lags behind along the corotation direction. The sparking discharges grow to 
attain lateral width $h_{\perp}$ whose estimates in the PSG model is given as 
\citep[see eq.4 and discussion above it in][]{2020MNRAS.492.2468M}
\begin{equation}
h_\perp = 2.6 \frac{T_6^2}{\eta b \cos{\alpha_l}}\left(\frac{P}{\dot{P}_{-15}}\right)^{0.5} ~~\textrm{m}.
\label{eq_sprksizePSG}
\end{equation}
Here $T_6$ is the surface temperature in million K and $b=B_s/B_d$ where $B_s$ 
is the surface non-dipolar magnetic field and $B_d$ the equivalent dipolar 
case. For a typical pulsar these parameters can be approximated as $\eta$=0.1, 
$b$=10, $T_6$=$P$=$\dot{P}$=$\lvert\cos{\alpha_l}\rvert$=1, and we have 
$h_\perp\sim2.6$ m. The velocity of lagging behind process in a PSG can be 
estimated as $\eta v_{cr}$, where $v_{cr}$ is the corotation velocity. The 
maximum heated region on the surface is shifted from the center of the spark 
along the corotation direction by $\Delta h_\perp=\eta v_{cr}t_{sp}$. Using 
$\eta$=0.1, $v_{cr}\sim$10$^6$-10$^7$ cm/s (see section \ref{sec:app_IARphy}), 
and $t_{sp}\sim$10$^{-5}$ s, we have $\Delta h_\perp\sim10$~cm. When the spark 
associated plasma column reaches $\rho_{GJ}$, it corotates with the rest of the
star till it leaves the gap region due to inertia. In this process the hot 
region that was formed below the spark lags slightly behind corotation by a 
distance $\Delta h_\perp$. The cooling timescale of this hot region is of the 
order of nanoseconds, while the gap emptying time of the plasma column is 
microseconds which is several times longer. Thus the cooling is most efficient 
in a region shifted by $\Delta h_\perp$ from the spark center, behind the 
corotation direction, where the condition $T_S < T_i$ is satisfied. After the 
plasma column empties a potential drop appears along the IAR causing initiation
of the subsequent sparking discharge. The discharges grows both along the 
horizontal and vertical direction to form a new spark. The locus of this spark 
is shifted by $\Delta h_\perp$ behind the previous one.

As the plasma column from the spark escapes the IAR, they form a secondary 
cloud of pair plasma with high multiplicity 
\citep[$\sim10^5$,][]{1971ApJ...164..529S,Timokhin2019} and typical length 
$ct_{sp}\sim3$~km, along the magnetic field lines. The secondary plasma clouds 
move outwards along the open field lines and generate radio emission at heights
of $\sim$500 km from the surface, where the magnetic field is dipolar 
\citep{2017JApA...38...52M}. The average emission from several 
thousand\footnote{For $P=1$~s and pulse window around 10\% of the period, the 
subpulse usually covers between a fifth and a third of the window making it 
several tens of milliseconds in width. Hence, several thousand secondary plasma 
columns, with typical timescales of tens of microseconds, make up a subpulse.} 
such secondary plasma columns is seen as a subpulse in the pulse window. As the
next spark is shifted behind the corotation direction by $\Delta h_\perp$, the 
subsequent secondary plasma clouds are also formed with equivalent shifts 
across the magnetic field. The physical shift between two consecutive clouds 
are very small, less than 5\% of the lateral size. However, they represent a 
continuous process such that over time the location of the subpulse within the 
pulse window is clearly shifted which is seen as subpulse drifting. 

We consider the IAR to be tightly packed with equidistant sparks 
\citep{2000ApJ...541..351G}, and subsequent sparks are formed at a location 
slightly behind the corotation direction. The maximum number of sparks 
($n_{sp}$) along any diametric cross section of the IAR under such conditions 
can be estimated using the PSG model as \citep{2020MNRAS.492.2468M}
\begin{equation}
n_{sp} \simeq 15~\frac{\eta b^{0.5} \cos{\alpha_l}}{T_6^2}\left(\frac{\dot{P}_{-15}}{P^2}\right)^{0.5}.
\label{eq_nsprkPSG}
\end{equation}
The parameters, $\eta$, $b$, $\alpha_l$ and $T_6$ are not well constrained, but
if we consider typical values $\eta$=0.1, $b$=10, 
$\lvert\cos{\alpha_l}\rvert$=$T_6$=1, we have $n_{sp}\sim$5 for a pulsar with 
$P=1$~s. The pulsar J2144--3933, with $P=8.5$~s and a single component in its 
profile, illustrates the viability of these estimates. If we assume the 
magnetic field configuration to be highly non-dipolar in this case with $b$=40,
then $n_{sp}\sim$1 implying only a single spark can be accommodated within the 
IAR. This concept was used by \citet{2020MNRAS.492.2468M} to predict the death 
line in the $P-\dot{P}$ diagram and provides a tight constrain to the 
population distribution \citep[see figure 6~in][]{2020MNRAS.492.2468M}.

\subsection{Lagging behind co-rotation in Inner Acceleration Region}
The phenomenon of subpulse drifting arising due to sparks lagging behind 
corotation for any arbitrary geometry, has been discussed in earlier studies 
for simpler magnetic field configurations, where surface fields are dominated 
by the radial component \citep{1985MNRAS.215..111A,2013arXiv1304.4203S,
2016ApJ...833...29B, 2020MNRAS.492.2468M}. To further elucidate this concept we
explore the electrodynamics for a general magnetic field structure. Let us 
consider two reference frames primed and unprimed, where the primed frame 
corresponds to the corotating neutron star and the unprimed frame represent the
observer. The relation between the electric fields in the two frames, near the 
polar cap of a slowly rotating pulsar ($\Omega R/c \ll$ 1), can be expressed 
as : 
\begin{equation}\label{eq_coortrans}
\mathbf{E}^{\prime} = \mathbf{E} + \frac{1}{c} \mathbf{v}\times\mathbf{B}
\end{equation}
where $\mathbf{v} = \mathbf{\Omega}\times\mathbf{r}$, is the rotation velocity
of the puslar. 
 
\begin{figure}
\includegraphics[scale=0.37,angle=0.]{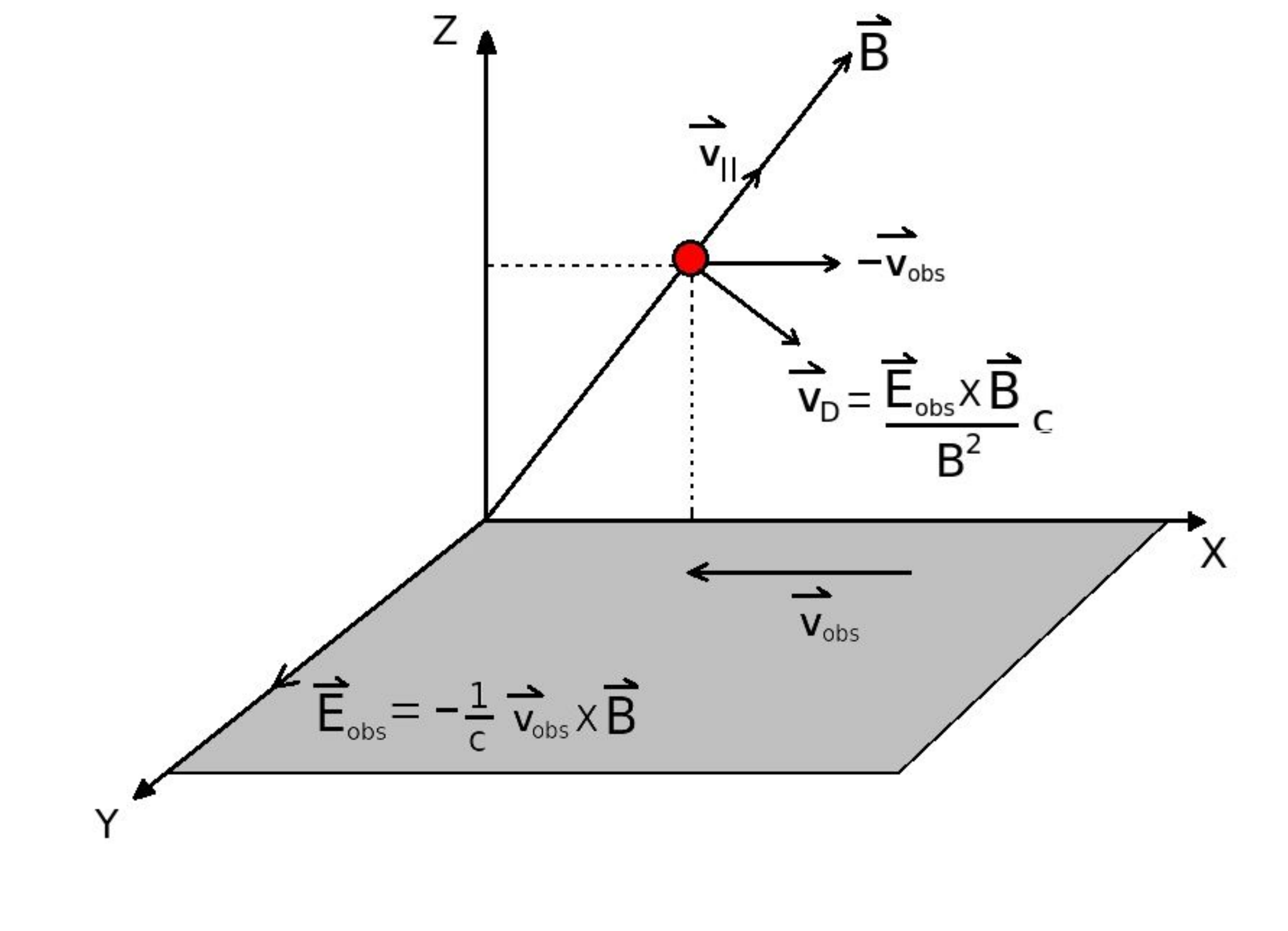}
\caption{The figure presents a simplified schematic of electromagnetic field 
configurations and velocities for a charge particle at rest in a magnetic field
as seen from the perspective of a moving observer. The cartesian coordinate 
system is setup in rest frame of the particle which also contains magnetic 
field $\mathbf{B}$. The particle and $\mathbf{B}$ are setup in the x-z plane 
for simplicity of visualization. There is no electric field in this frame. An 
observer moving with $\mathbf{v}_{obs}$ along the x-axis sees the particle move
along -$\mathbf{v}_{obs}$, with $\mathbf{v}_{D}$ and $\mathbf{v}_{\parallel}$ 
lying in the x-z plane. The electric field $\mathbf{E}_{obs}$ also appears 
along y-axis to the observer due to frame transformation.}
\label{fig_corot}
\end{figure}

A simple schematic to explain the relation between the fields and velocities in
a moving frame and rest frame is shown in figure \ref{fig_corot}. In this 
figure a test charge particle is considered to be at rest in the x-z plane, and
the magnetic field $\mathbf{B} = (B_x, 0, B_z)$ is confined in the x-z plane.
Now if we consider an observer moving along the negative x-axis with velocity 
$\mathbf{v}_o = -v_o \hat{x}$, the test particle will appear to move in the 
opposite direction in the observer's frame. An electric field $\mathbf{E}_o$ 
appears in this frame along the y-axis given by, 
\begin{equation}\label{eq_elecdemo}
\begin{split}
\mathbf{E}_o & = -\frac{1}{c} \mathbf{v}_o\times\mathbf{B} \\
             & = \frac{1}{c} v_o B_z/c ~\hat{y}
\end{split}
\end{equation}
In the observer's frame the velocity of the charged particle can be expressed 
in terms of two perpendicular components, the drift velocity $\mathbf{v}_{D}$ 
perpendicular to the magnetic field and $\mathbf{v}_{\parallel}$ along magnetic
field, such that --$\mathbf{v}_o$ = $\mathbf{v}_{D} + \mathbf{v}_{\parallel}$.
The $\mathbf{E}\times\mathbf{B}$ drift velocity of particle $\mathbf{v}_{D}$ is
given as : 
\begin{equation}\label{eq_veldemo}
\begin{split}
\mathbf{v}_{D} & = c\frac{\mathbf{E}_o\times\mathbf{B}}{B^2} \\
              & = \frac{v_o}{B^2}\left(-B_z^2 ~\hat{x} + B_x B_z ~\hat{z}\right).
\end{split}
\end{equation}

An analogous behaviour is seen in pulsars. When full corotational charge 
density $\rho_{GJ}$ exist in the pulsar magnetosphere the corotational electric
field $\mathbf{E}^{\prime} = 0$ in the rest frame of the pulsar. It follows 
from eq.~(\ref{eq_coortrans}) that in the observer's frame the corotation 
electric field is $\mathbf{E_c} = 
-\frac{1}{c}(\mathbf{\Omega}\times\mathbf{r})\times\mathbf{B}$. We choose a 
coordinate system centered on the neutron star, and consider a slowly rotating 
pulsar with angular velocity $\mathbf{\Omega}$ aligned along the z-axis, such 
that $\mathbf{\Omega}$ = $\Omega\hat{z}$, where $\hat{z} = \cos{\theta}\hat{r} 
- \sin{\theta}\hat{\theta}$, and magnetic field given by $\mathbf{B} = (B_r, 
B_{\theta}, B_{\phi})$. The corotation electric field $\mathbf{E}_c = (E^c_r, 
E^c_{\theta}, E^c_{\phi})$ in the above coordinate system at a point 
$\mathbf{r} = (r, \theta, \phi)$ is given as :
\begin{equation}\label{eq_corotelec}
\begin{split}
E^c_r & = \frac{\Omega r \sin{\theta}B_{\theta}}{c}, \\
E^c_{\theta} & = -\frac{\Omega r \sin{\theta}B_{r}}{c}, \\
E^c_{\phi} & = 0.
\end{split}
\end{equation}
The electric field in eq.(\ref{eq_corotelec}) is a more generalised version of
eq.(\ref{eq_elecdemo}) and is perpendicular to the magnetic field. 
Thus the drift velocity ($\mathbf{v}^c_D$), which is the component of the 
corotation velocity ($\mathbf{\Omega}\times\mathbf{r}$) perpendicular to the 
local magnetic field, is given as $\mathbf{v}^c_D = 
c(\mathbf{E}\times\mathbf{B})/B^2$. Using eq.(\ref{eq_corotelec}) we obtain 
$\mathbf{v}^c_D = (v^c_{D,r}, v^c_{D,\theta}, v^c_{D,\phi})$ to be : 
\begin{equation}\label{eq_corotvel}
\begin{split}
v^c_{D,r} & = - \frac{\Omega r B_r B_{\phi} \sin{\theta}}{B^2}, \\
v^c_{D,\theta} & = - \frac{\Omega r B_{\theta} B_{\phi} \sin{\theta}}{B^2}, \\
v^c_{D,\phi} & = \frac{\Omega r (B_r^2 + B_{\theta}^2) \sin{\theta}}{B^2},
\end{split}
\end{equation}
which is a more generalised form of eq.(\ref{eq_veldemo}). 

The steady state condition where the magnetosphere is filled with corotational
charge density $\rho_{GJ}$, breaks down in the IAR where a continuous outflow  
of plasma is setup, leaving behind a gap. As the charge density empties in the
gap region an electric field $\mathbf{E}_{\parallel}$ along the magnetic field 
appears in the pulsar frame in addition to perpendicular component of the 
electric field --$\mathbf{E}_c$ which is opposite to the corotation field 
derived above. It is often reported in the literature that the charge particles
in the IAR would see an electric field such that they rotate within the IAR 
around the magnetic axis \citep{1975ApJ...196...51R,2012ApJ...752..155V,
2017ApJ...845...95S}. However, in the absence of any external electric field, a
charge in IAR would lag behind corotation in the observer's frame, and rotate 
opposite to the corotation direction in the pulsar frame due to the 
--$\mathbf{E}_c$ field. It is clear from eq.(\ref{eq_corotvel}) that the 
$v^c_{D,\phi}$ has the same sign throughout the polar cap\footnote{except in 
rare cases when magnetic inclination angle and hence $\theta$ is close to 
0\degr or 180\degr} and hence the charges cannot rotate around the magnetic 
axis.

There are two primary features of the plasma generation process in the IAR 
discussed above. Firstly, the sparking is associated with a continuous lagging 
behind corotation motion, where every subsequent spark is formed slightly 
behind the corotation direction. Secondly, the surface of the neutron star is 
dominated by non-dipolar magnetic field while the magnetic field in emission 
region higher up in the magnetosphere is purely dipolar in nature. As a 
consequence, the sparks lagging behind in the IAR will have different paths in 
the emission region. Using these basic assumptions we demonstrate in this work 
that the majority of observed subpulse drifting behaviour in pulsars, outlined 
earlier, can be reproduced by considering different manifestations of surface 
non-dipolar fields.

\section{Modelling single pulse sequence}\label{sec:singlmod}

\begin{figure*}
\includegraphics[scale=1.1,angle=0.]{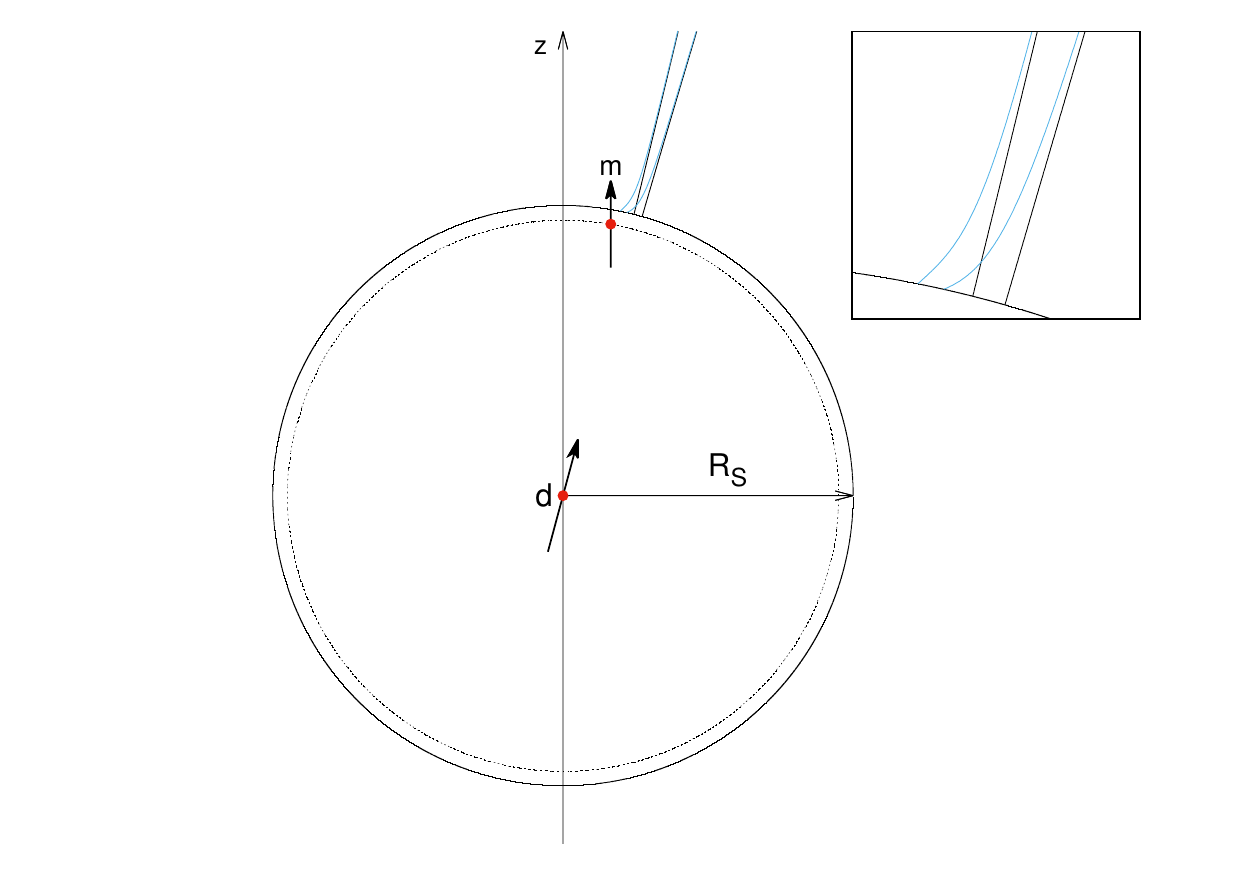}
\caption{The figure shows a representation of the magnetic configuration 
comprising of a star centered dipole $\mathbf{d}$ = ($d$, 15\degr, 0\degr) 
inclined from the rotation axis, and a second dipole $\mathbf{m}_s$ = 
(0.001$d$, 0\degr, 0\degr) anchored in the crust at $\mathbf{r}_s$ = 
(0.95$R_S$, 10\degr, 0\degr). The resultant non-dipolar polar cap (the inset 
figure) is shifted from the dipolar polar cap but at heights of few times the 
stellar radius the dipolar field dominates.}
\label{fig_field}
\end{figure*}

\begin{figure*}
\includegraphics[scale=1.1,angle=0.]{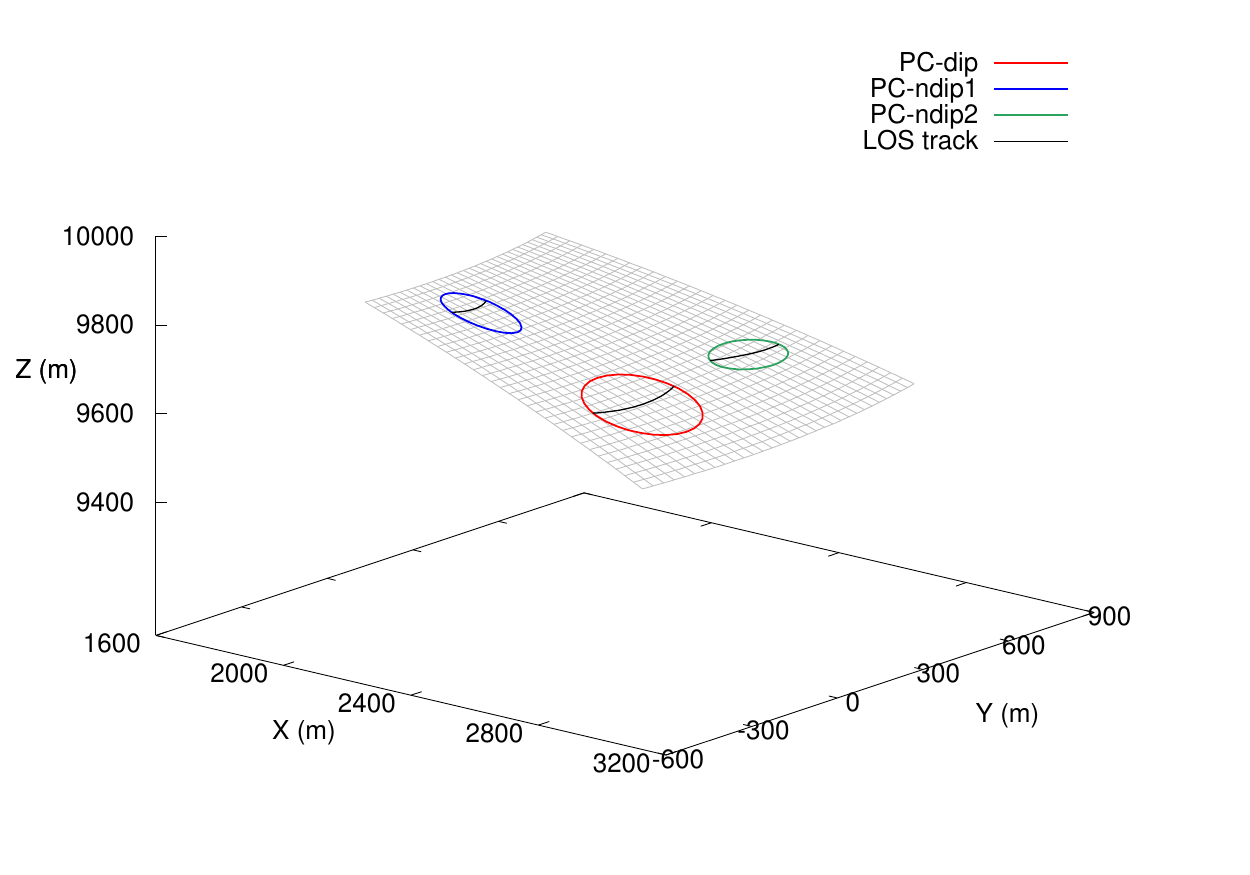}
\caption{The figure shows three different polar cap orientation for a pulsar 
with $P$ = 1 sec and the star centered dipole $\mathbf{d}$ = ($d$, 15\degr, 
0\degr). The dipolar polar cap is shown in addition to two different 
realization of non dipolar polar caps. In each of the non-dipolar polar caps 
one single surface dipole oriented along the z-axis $\mathbf{m}_s$ = 
(0.001$d$, 0\degr, 0\degr) is located at $\mathbf{r}_s$ = (0.95$R_S$, 
10\degr, 0\degr) and (0.95$R_S$, 15.8\degr, 18.7\degr) respectively. In 
addition the projections of the line of sight (LOS), corresponding to $\beta$ =
0\degr, from the emission region to the polar cap in each of the non dipolar 
cases are also shown.}
\label{fig_polcap}
\end{figure*}

In order to generate a single pulse sequence there are two primary inputs that 
are required, firstly, the nature of the magnetic field configuration, and, 
secondly, the nature of spark motion in IAR. We describe below our simplified 
model for the magnetic field configuration as well as the spark motion.

\subsection{Magnetic Field Structure}
We have used a simple configuration consisting of a star centered dipole 
($\mathbf{d}$) along with one or more surface dipoles ($\mathbf{m}_i$, with
$i = 1, 2, ..., N$), anchored on the crust near the IAR, as our model for the 
non-dipolar magnetic field \citep{2002A&A...388..235G}. In figure 
\ref{fig_field} a schematic of a configuration consisting of star centered 
dipole and one surface dipole is shown. The simulations are carried out in the 
rest frame of the pulsar which implies that observer's line of sight evolves 
with time and periodically cuts across the emission region, while the magnetic 
configuration remains fixed during the temporal evolution of the pulse 
sequence. The coordinate system is oriented such that the global dipole is 
located at the origin and along the x-z plane, $\mathbf{d}$ = ($d$, $\theta_d$,
0\degr). The surface field on the other hand is more complicated and requires 
two sets of variables for each dipole, one specifying the location 
$\mathbf{r}_i$ = ($r_s^i$, $\theta_s^i$, $\phi_s^i$) and the other the 
magnetic field orientation $\mathbf{m}_i$ = ($m^i$, $\theta_m^i$, $\phi_m^i$). 
The detailed calculations of the resultant magnetic fields, in spherical 
coordinates, for the general configuration is shown in appendix 
\ref{app:magcon}. For these simulations we have used a complete three 
dimensional solution of the magnetic line of force. This requires numerical 
solution of the system of differential equations in spherical coordinates,
\begin{equation}\label{eq_field}
\begin{split}
\frac{\mathrm{d}\theta}{\mathrm{d}r} & = \frac{B_{\theta}^d + \sum\limits^N_{i=1}B_{\theta}^i}{r\left(B_r^d + \sum\limits^N_{i=1}B_r^i\right)} \\
\frac{\mathrm{d}\phi}{\mathrm{d}r} & = \frac{B_{\phi}^d + \sum\limits^N_{i=1}B_{\phi}^i}{r\left(B_r^d + \sum\limits^N_{i=1}B_r^i\right)\sin{\theta}}
\end{split}
\end{equation}
In accordance with \citet{2002A&A...388..235G} we used typical parameters $m^i$
= 0.001-0.05$d$ and $r_s^i$ = 0.95$R_S$, where $R_S$ = 10 km, the radius of 
neutron star. For the above parameters the surface dipole contributions are 
$\approx$ 0 at heights of 50$R_S$, which was used as the initial condition to 
trace the field lines to the stellar surface. We have varied the remaining 
parameters, i.e, the location of the crust dipoles specified by $\theta_s^i$, 
$\phi_s^i$, as well as the dipole orientations using $\theta_m^i$, $\phi_m^i$, 
to generate different configurations of the surface field as required. 

The pulsar magnetosphere is separated into the open and closed field line 
regions which are bound by the light cylinder radius $R_{LC}$ (=$cP/2\pi$). The
radio emission originates at heights of $R_E\sim$50$R_S$ 
The pulsar magnetosphere is separated into the open and closed field line
regions which are bound by the light cylinder radius $R_{LC}$ (=$cP/2\pi$). The
radio emission originates at heights of $R_E\sim$50$R_S$
\citep{1997MNRAS.288..631K,2002ApJ...577..322M,2003A&A...397..969K,
2004A&A...421..215M,2009MNRAS.393.1617K,2017JApA...38...52M} along the open
field lines. The opening angle of the open field line region is estimated from 
equation of dipolar fields as $\theta_o^E = 
\sin^{-1}\left(\sqrt{R_E/R_{LC}}\right)$. The entire open field line region at 
the emission height is also referred to as emission beam. Using the boundary 
value of $\theta=\theta_o^E$ at $r=R_E$, and varying $\phi$, the boundary of 
the non-dipolar polar cap is estimated from the solutions of the field line 
equations (eq. \ref{eq_field}). In figure \ref{fig_polcap} the polar cap 
boundaries for two different configuration of the non-dipolar fields are shown 
in addition to the purely dipolar polar cap. In each case the surface dipole is
located 5\degr~offset from the center of the dipolar axis, the first case along
$\theta$-axis and second along the $\phi$-axis. The figure shows that the polar
cap is elongated and smaller in area compared to the dipolar polar cap which is
circularly symmetric. 

The pulsed emission is seen when the line of sight (LOS) traverses the open 
field lines in the emission region during the pulsar rotation. The LOS is 
characterized by the angle $\beta$ which corresponds to the angular separation 
between the LOS and the axis of the star centered dipole during their closest 
approach \citep{1984A&A...132..312G}. Consequently, the profile shape and the 
observed subpulse motion is determined by the LOS traverse on the emission 
region as well as its corresponding track on the IAR. The track of the LOS on 
the IAR can be estimated by varying the $\phi$ coordinate across the open field
line in the emission region, $r=R_E$, for a constant $\theta=\theta_d+\beta$, 
and for each field line estimating the equivalent location in the IAR. In 
figure \ref{fig_polcap} we also show the LOS in the IAR for the two non-dipolar
configurations where $\beta$ = 0\degr, i.e., the LOS cuts the emission beam 
centrally. As seen in the figure the LOS on the IAR are not centrally located 
due to the asymmetry of the non-dipolar fields.

\subsection{Spark Motion in Inner Acceleration Region}
The next step in generating single pulse sequence is establishing a dynamical 
sparking system in the IAR and associating the sparking process with subpulse 
variation. The sparks lag behind the star in the corotation direction, and move 
around the rotation axis with a drifting periodicity longer than the pulsar 
period. There are no studies explaining the detailed evolution of sparking 
process in IAR, which is also beyond the scope of this work. However, under the
assumption that the IAR is tightly packed with sparks 
\citep{2000ApJ...541..351G}, the PSG model suggests a typical pulsar can 
accommodate a maximum of 5 sparks across any diametric cross section (see eq. 
\ref{eq_nsprkPSG} and discussion below it). Beyond this basic understanding 
there is no detailed modelling which connects the spark width with the lateral 
size of the secondary plasma clouds above the IAR, as well as their relation to
the subpulse width in the emission region. In a detailed work involving a large
set of pulsars studied in the MSPES survey \citep{2016ApJ...833...28M}, 
\citet{2018ApJ...854..162S} showed that a lower boundary in the distribution of
component widths is present in pulsars, with a $P^{-0.5}$ dependence. The 
opening angle of the dipolar beam in the radio emission region also scales as 
$P^{-0.5}$, which suggests the presence of an upper limit for the number of 
components within the pulse window as well. Additionally, detailed studies of 
the emission beam suggests that a maximum of five components can be 
accommodated within the emission beam \citep{1993ApJ...405..285R,
1999A&A...346..906M} which is consistent with the estimations of the PSG model.
In this work we have assumed the ratio of the diameter of the spark ($D$) and 
polar cap radius ($r_p$) is $D/r_p \sim 0.2$, for a maximum of 5 sparks present
across the diameter of the polar cap. We have also assumed a direct one is to 
one correspondence between the sparks in the IAR and the subpulses in the 
emission region. This is obviously a simplistic approximation for reproducing 
the subpulse emission features, but is adequate for investigating their 
drifting behaviour. 

We have setup a system of sparks of equal size and separation between adjacent 
sparks given as $\theta_{sp}$=0.2$\theta_p$, where $\theta_p$ is the angular 
radius of the non-dipolar polar cap. The sparks resemble a Gaussian shape, 
\begin{equation}
\label{eq_sprksize}
I_{sp} = I_0~{\rm exp}\left(-4r^2/a_{sp}^2\right),
\end{equation} 
where $I_0$ is the maximum intensity, $a_{sp}$ = $R_S\theta_{sp}$ and $r$ is 
given as 
\begin{equation}
\label{eq_sprkpos}
r = \sqrt{d^2 + d_i^2 -2d_id(\sin\theta\sin\theta_i\cos(\phi-\phi_i)+\cos\theta\cos\theta_i)}.
\end{equation}
Here ($d$, $\theta$, $\phi$) specifies the reference point and ($d_i$, 
$\theta_i$, $\phi_i$) the center of the $i^{th}$ spark. All points considered 
are on the surface of the neutron star, hence, $d$=$d_i$=$R_S$ and 
eq.(\ref{eq_sprkpos}) simplifies to the form
\begin{equation}
\label{eq_sprkpos1}
r = 1.4R_S\sqrt{1 - (\sin\theta\sin\theta_i\cos(\phi-\phi_i)+\cos\theta\cos\theta_i)}.
\end{equation}
In the dynamic sparking
system a number of spark tracks are setup within the IAR separated by 
$\theta_{sp}$. In each of these tracks the sparks move around the rotation axis
(implying constant $\theta$, since rotation axis is aligned with the z-axis) 
with the drifting periodicity $P_3$. The track locations are identified with 
$\theta^n$, $n$ representing the track number, which are setup such that a 
central track passes through the middle of the IAR, $\theta^0$, as well as 
tracks on either side, $\theta^n$ = $\theta^0$+2$n\theta_{sp}$, $n$ = 0, 
$\pm$1, $\pm$2. At any given time, $t$, the center of the $i^{th}$ spark is 
located at ($R_S$, $\theta^n$, $\phi_i^n(t)$),  
\begin{equation}
\label{eq_sprktrk}
\phi_i^n(t) = 2\pi (t-t_0)/NP_3 + 2\pi i/N.
\end{equation}
Here $N = \pi\sin\theta^n/\theta_{sp}$; $i = 0, 1, ..., N-1$ and $t_0$ is any 
arbitrary start time. Using the above setup the time evolution of the sparking 
process in the IAR from the lagging behind model can be simulated as 
\begin{equation}
\label{eq_sprkmodl}
I(t) = \sum_{n} B_n(\phi) \sum_{i=0}^{N-1} I_{sp}(t,n,i).
\end{equation}
$B_n(\phi)$ is a box function with $B_n(\phi)$ = 1 for $\phi_{min}^n < \phi < 
\phi_{max}^n$ and 0 otherwise. Here $\phi_{min}^n$ and $\phi_{max}^n$ are the 
boundaries of the polar cap along the $n^{th}$ spark track. 

\begin{figure*}
\begin{tabular}{@{}cr@{}}
{\mbox{\includegraphics[scale=0.7,angle=0.]{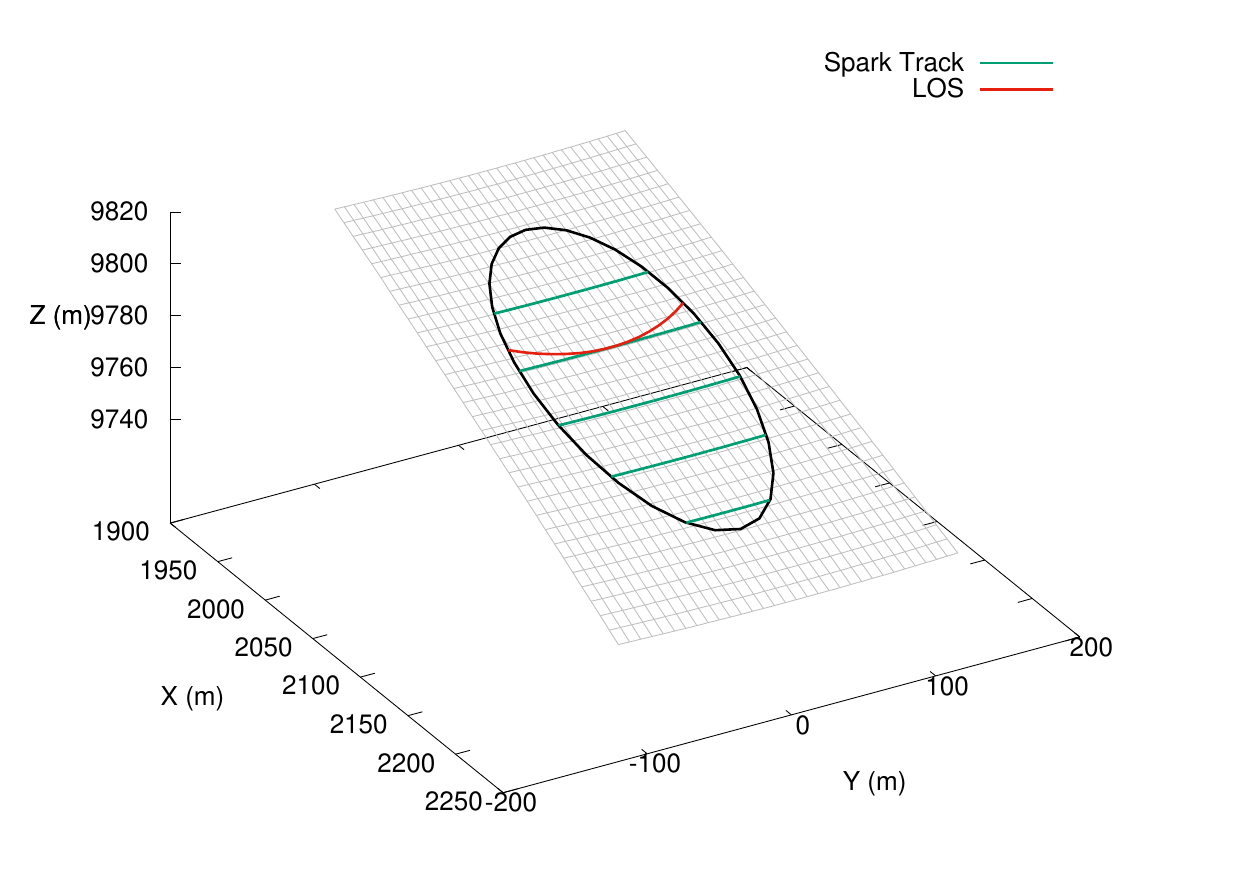}}} &
{\mbox{\includegraphics[scale=0.7,angle=0.]{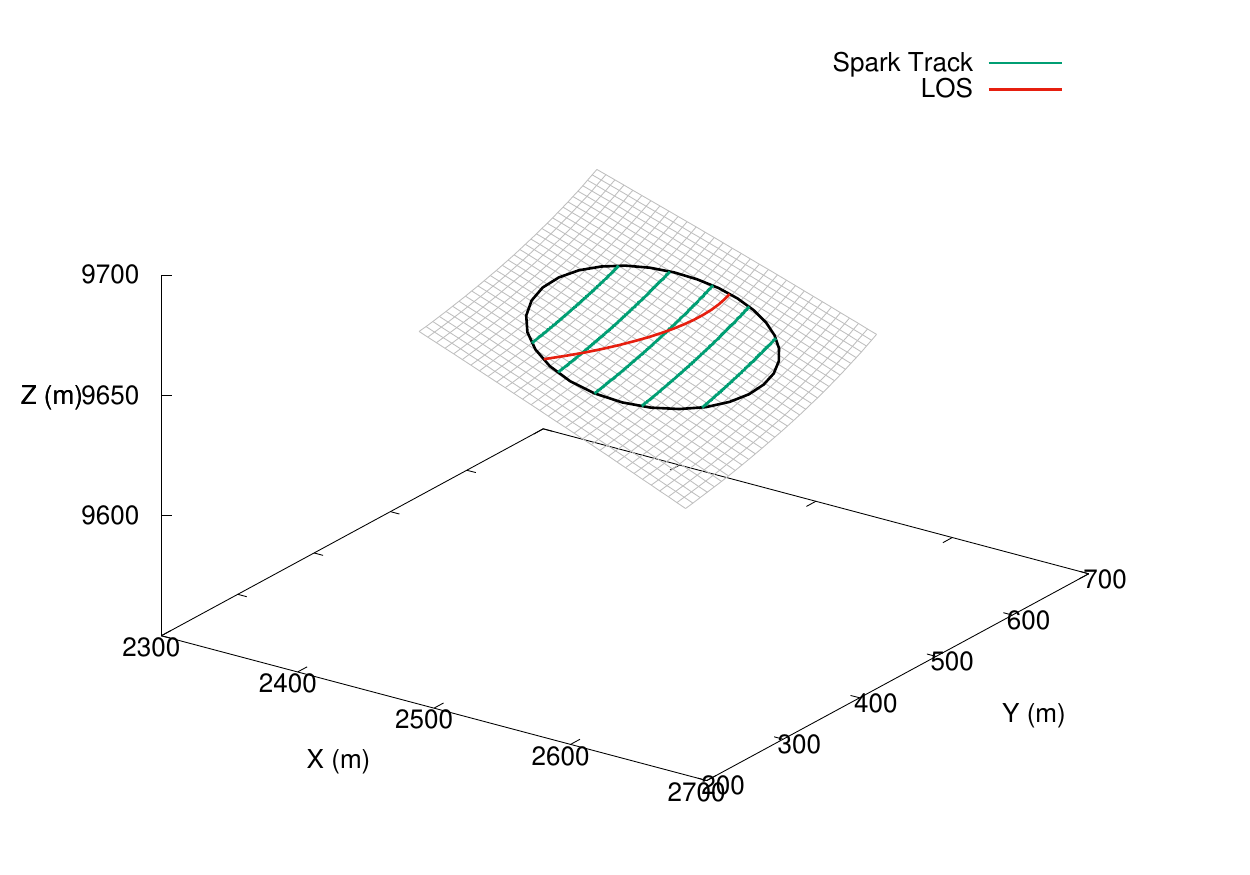}}} \\
\end{tabular}
\caption{The figure shows track of the sparks which are lagging behind 
corotation in the Inner Acceleration Region (IAR) of pulsars. In addition the 
transition of the line of sight (LOS) from the emission region to the polar cap
is also shown. The magnetic field structure is governed by two dipoles, the 
first located at the star center with $\mathbf{d}$ = ($d$, 15\degr, 0\degr) 
and the second near the crust with $\mathbf{m}_s$ = (0.001$d$, 0\degr, 
0\degr) and located at $\mathbf{r}_s$ = (0.95$R_S$, 10\degr, 0\degr) for the 
left panel and $\mathbf{r}_s$ = (0.95$R_S$, 15.8\degr, 18.7\degr) for the 
right panel. In the right panel it can be seen that the LOS cuts across 
multiple tracks of spark motion. This is due to the fact that the non-dipolar 
IAR is asymmetrically displaced from the purely dipolar polar cap.}
\label{fig_sparktrack}
\end{figure*}

In figure \ref{fig_sparktrack} the tracks of the spark motion in the 
non-dipolar polar caps for two different configurations of the magnetic field 
(corresponding to figure \ref{fig_polcap}) are shown. Additionally, the locus 
of the LOS variation on these polar caps corresponding to $\beta$= 0\degr~are 
also shown in figure \ref{fig_polcap}. Depending on the orientation of the 
surface fields the polar cap is shifted from the corresponding dipolar case. 
Thus the LOS can traverse the spark tracks at different angles, sometimes 
across multiple tracks as seen in the right panel of figure 
\ref{fig_sparktrack}. This will lead to different phase behaviours in subpulse 
drifting.

\section{Simulating Pulse sequence for Subpulse Drifting}\label{sec:driftsiml}
In this section we demonstrate that the simplified algorithm presented above 
can be used to explain the different drifting classes seen in pulsars. To show 
this in detail we have explored different orientations of the surface dipole to
form diverse realisations of the non-dipolar polar cap. In each of these 
magnetic field configurations we have generated a series of single pulses from 
sparks lagging behind the corotation speed. This is achieved by rotating the 
LOS around the star such that $\theta_E$= $\theta_d+\beta$ and $\phi_E$= 2$\pi 
t/P$. When the LOS encounters the open field line region a corresponding 
translation is made to the IAR by solving the field line equations 
(eq.\ref{eq_field}) with initial condition $\mathbf{r}_E=(50R_S, \theta_E,
\phi_E)$ to obtain the corresponding location $\mathbf{r}_s=(R_S, \theta,
\phi)$ on the polar cap. 

Next the dynamic structure of the spark motion in IAR is estimated using 
eq.(\ref{eq_sprkmodl}) and the reference point $\mathbf{r}_s$ to obtain the 
subpulse behaviour. This process is repeated for a number of cycles of the LOS 
rotation, each rotation corresponding to a single pulse, and the relevant pulse
sequence is generated. Finally, we have carried out fluctuation spectral 
studies on the simulated pulse sequence to explore the phase behaviour of 
subpulse drifting. In all IAR configurations explored here we show a detailed 
calculation of the different physical parameters like surface magnetic field 
behaviour, the ratio $b$, between the non-dipolar and equivalent dipolar field,
corotation electric field, $\rho_{GJ}$, $\cos{\alpha_l}$, drift speed due to 
corotation, etc., in section \ref{sec:app_IARphy}. 

Note that in our analysis, we show simulations for radio emission for a given 
observing frequency arising from the same emission height across the line of 
sight. While this is a reasonably good assumption, however if the emission 
originated from a range of heights (e.g. \citealt{2004ApJ...609..335G}), then 
one expects to see minor changes in the phase behaviour of the drifting 
subpulse, although the drifting periodicity i.e. $P_3$ would remain the same. 
This effect of phase change arises because for a given pulsar the line of sight
remains constant, however for different emission heights the observer cuts 
different sets of magnetic field lines thus sampling slightly different part of
the spark motion in the polar cap. Since this is a purely geometrical effect, 
there is no change in the repetition time i.e. $P_3$. Similar effects of phase
change and constant $P_3$ would also be seen in observations at different 
frequencies, since according to radius to frequency mapping progressively 
higher and higher frequencies arises closer to the neutron star (see e.g. 
\citealt{2002ApJ...577..322M}). Additionally, there is possibility of around 
10\% variation in $v_D$ and $\cos{\alpha_l}$ across the LOS in certain magnetic
field configurations (see section \ref{sec:app_IARphy}), which will likely 
cause $P_3$ to vary along the LOS. It is possible to carry out detailed 
estimations of the variations of $P_3$ along the LOS, but these involve 
detailed calculations and any estimated variations are expected to be well 
within the measurement errors in $P_3$. As a result we have considered a 
constant value of $P_3$ across the LOS. There is the possibility of certain 
magnetic field configurations resulting in large change of both $v_D$ and 
$\cos{\alpha_l}$ across the polar cap and as a result the drifting speed is 
also expected to show large variations across it. However, most observations 
suggest that $P_3$ is mostly constant across the pulsar profile and show very 
little variations \citep{2006A&A...445..243W,2016ApJ...833...29B, 
2019MNRAS.482.3757B}. Hence, any surface field configuration showing large 
variations in $v_D$ and $\cos{\alpha_l}$ are not viable candidates for these 
drifting studies.

\subsection{Coherent phase-modulated Drifting}\label{sec:cohdrift}
\begin{figure*}
\begin{tabular}{@{}cr@{}}
{\mbox{\includegraphics[scale=0.4,angle=0.]{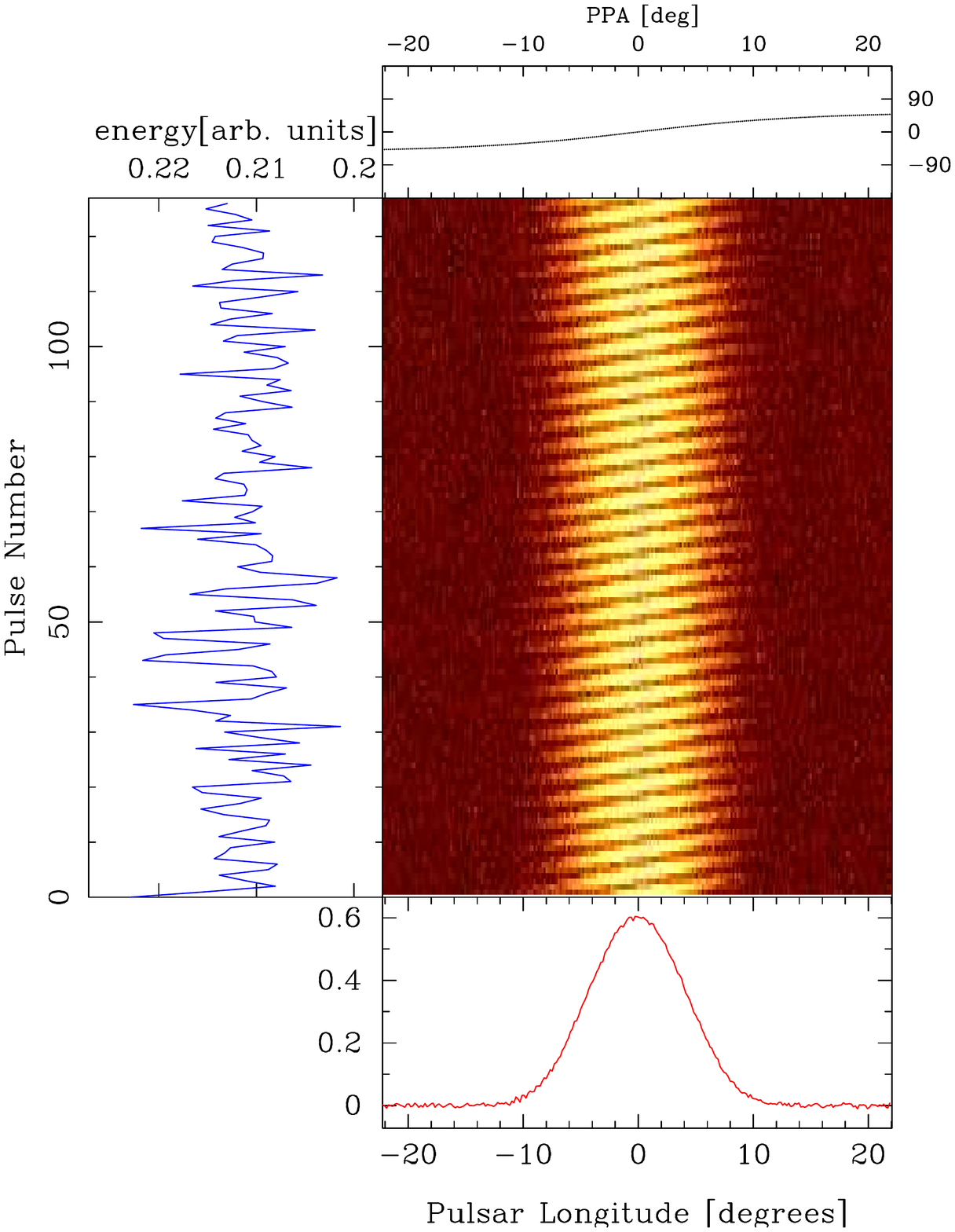}}} &
{\mbox{\includegraphics[scale=0.4,angle=0.]{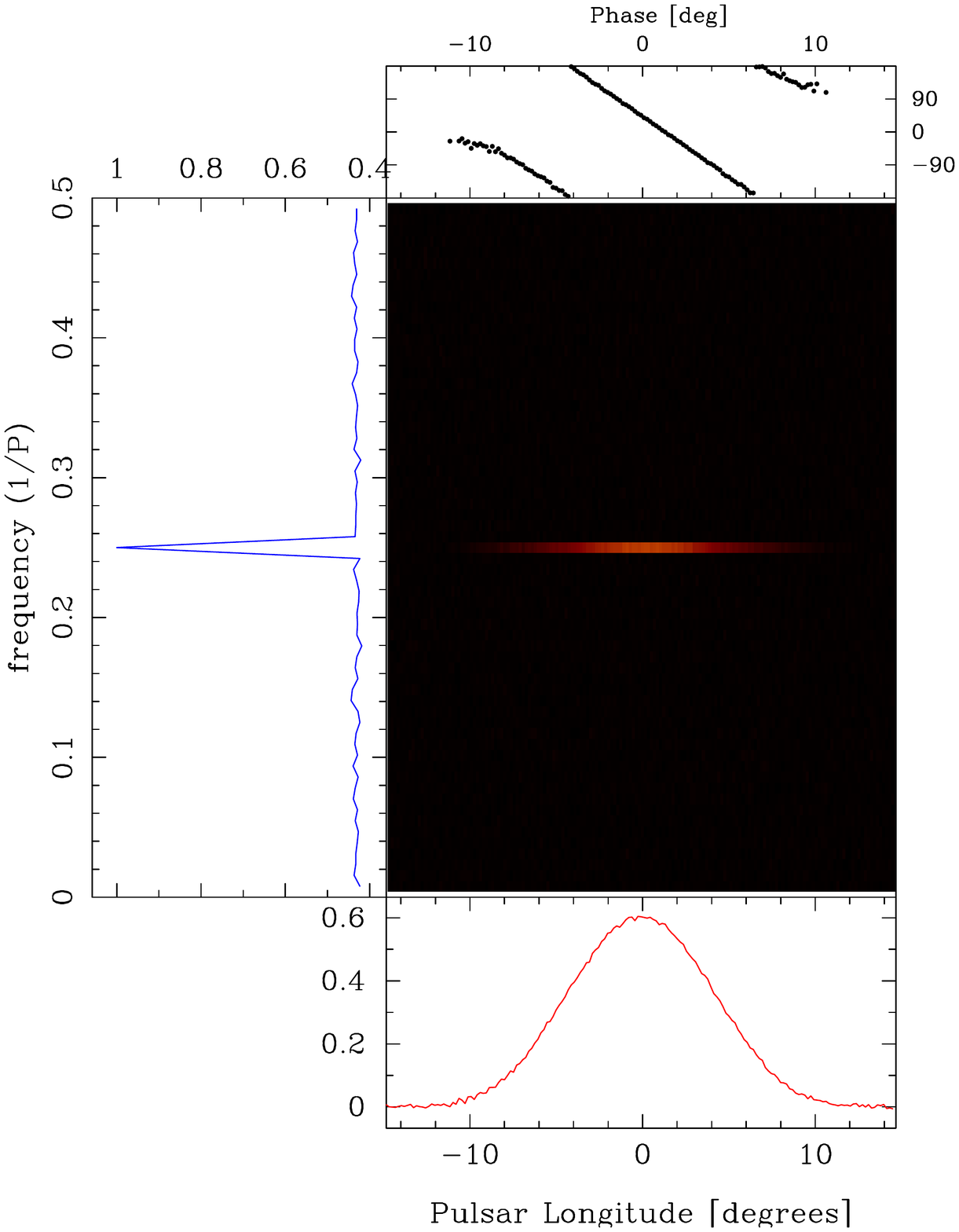}}} \\
\end{tabular}
\caption{The figure shows a simulation of of a sequence of pulses exhibiting
coherent phase-modulated positive drifting. The left panel shows 128 
consecutive single pulses where the subpulses in subsequent pulses appear at 
later longitudes across the pulse window. The right panel shows the LRFS for 
this pulse sequence which exhibits a peak at $f_p$ = 0.25 cycles/$P$ and the 
corresponding phase changes showing a negative slope from the leading to the 
trailing edge of the window. The pulse sequence was simulated using $P_3$ = 
1.33$P$. The surface magnetic fields were specified as $\mathbf{r}_s$ = 
(0.95$R_S$, 10\degr, 0\degr) and $\mathbf{m}_s$ = (0.001$d$, 0\degr, 0\degr).
The inclination of star centered dipole was $\theta_d$ = 15\degr~and the line 
of sight inclination angle $\beta$ = 4\degr.}
\label{fig_posdrift}
\end{figure*}

\begin{figure*}
\begin{tabular}{@{}cr@{}}
{\mbox{\includegraphics[scale=0.72,angle=0.]{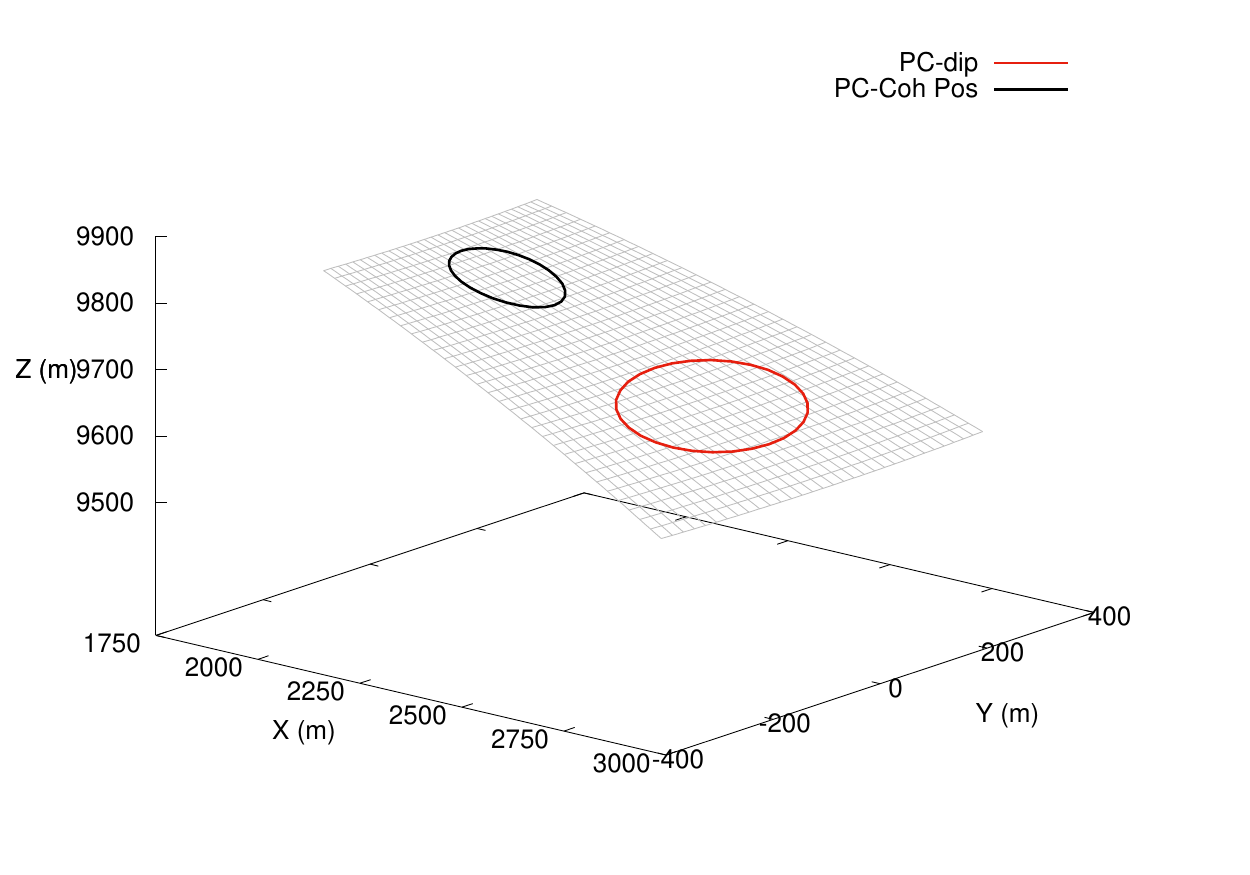}}} &
{\mbox{\includegraphics[scale=0.72,angle=0.]{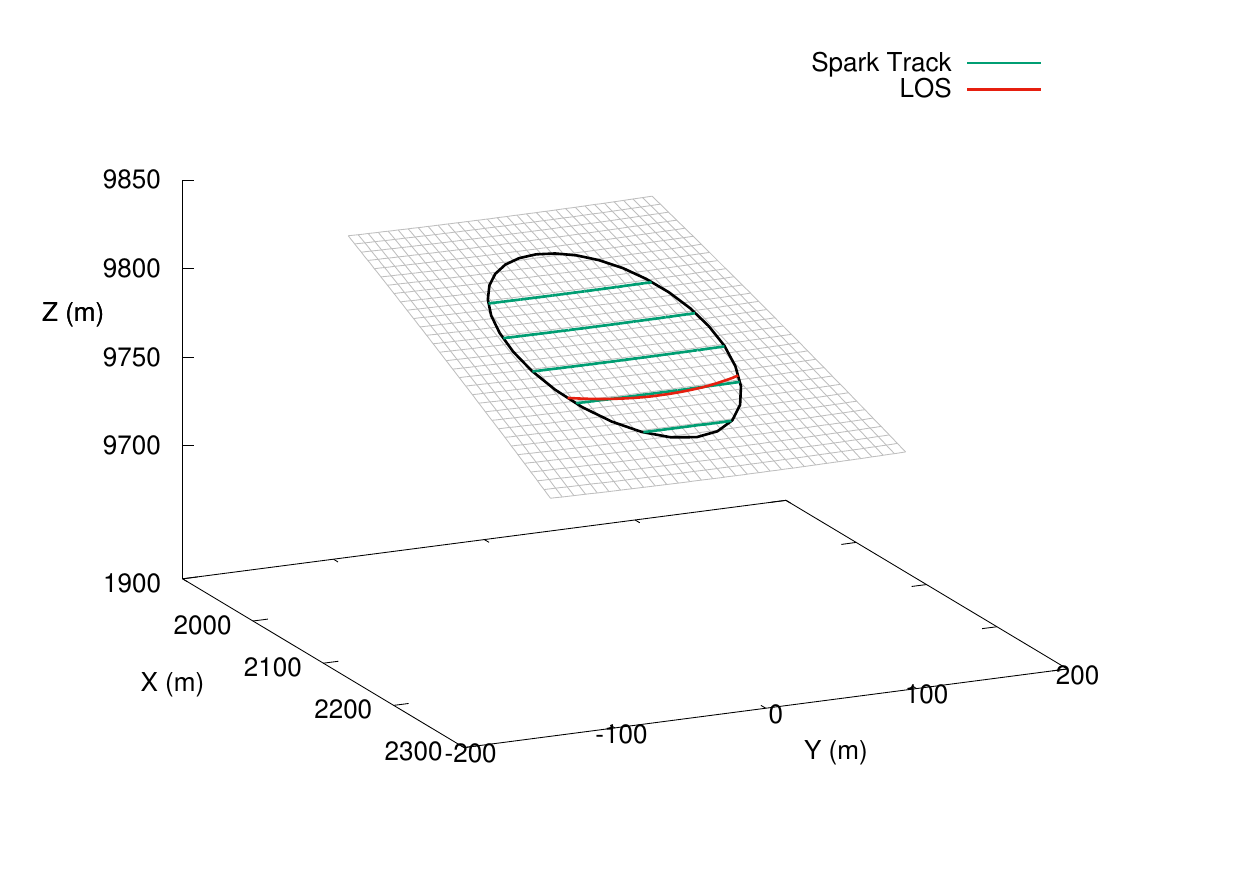}}} \\
\end{tabular}
\caption{The figure shows location of the non-dipolar polar cap with respect to
an equivalent dipolar case (left panel) and the tracks of the sparks (right 
panel) which are lagging behind corotation in the Inner Acceleration Region 
(IAR). In addition the transition of the line of sight (LOS) from the emission 
region to the polar cap is also shown. The magnetic field is characterised by 
two dipoles, one located at the star center with $\mathbf{d}$ = ($d$, 
15\degr, 0\degr) and the second near the crust with $\mathbf{m}_s$ = 
(0.001$d$, 0\degr, 0\degr) located at $\mathbf{r}_s$ = (0.95$R_S$, 10\degr, 
0\degr). The LOS corresponds to $\beta$= 4\degr~and closely follows the spark 
track.}
\label{fig_posspark}
\end{figure*}

\begin{figure*}
\begin{tabular}{@{}cr@{}}
{\mbox{\includegraphics[scale=0.4,angle=0.]{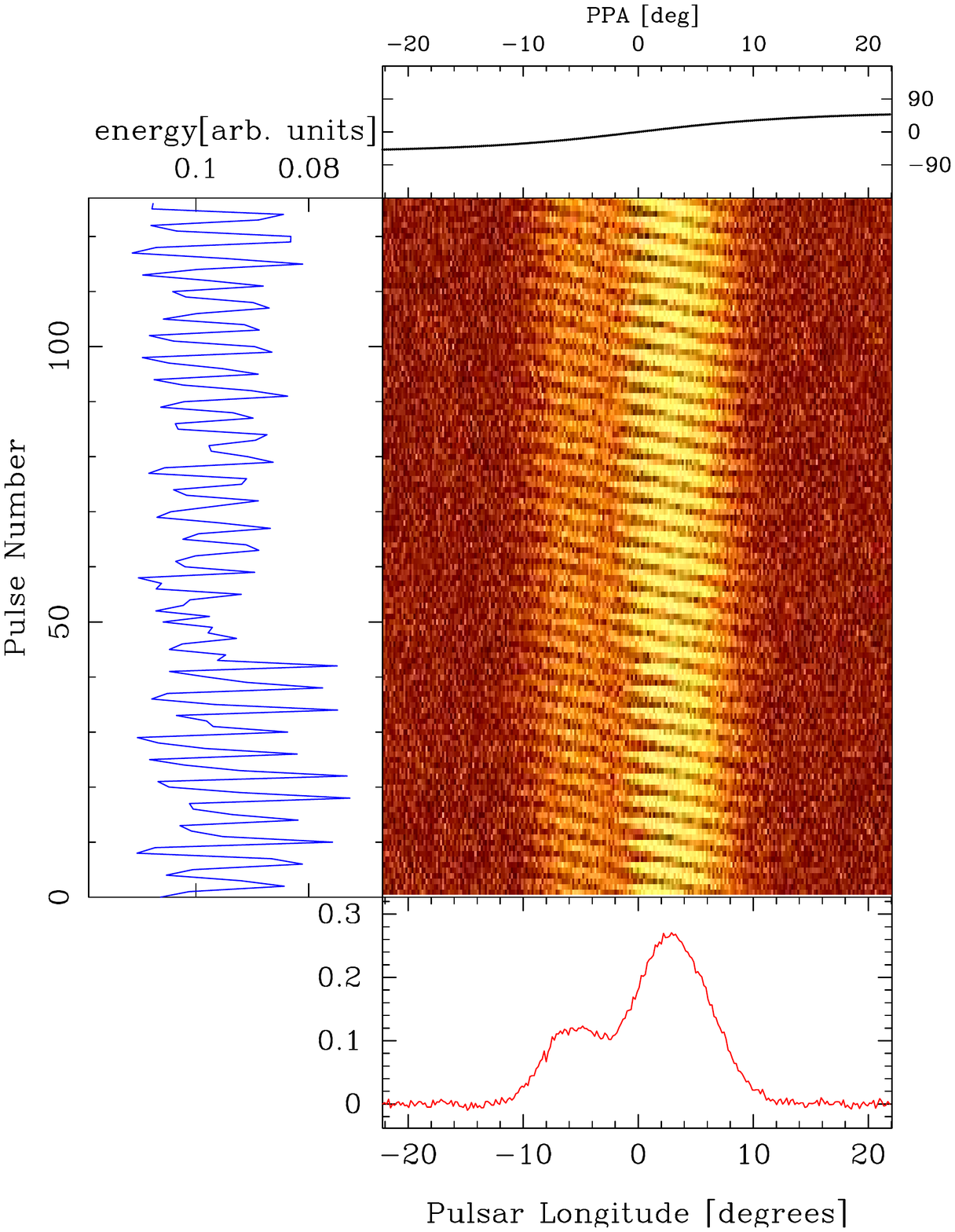}}} &
{\mbox{\includegraphics[scale=0.4,angle=0.]{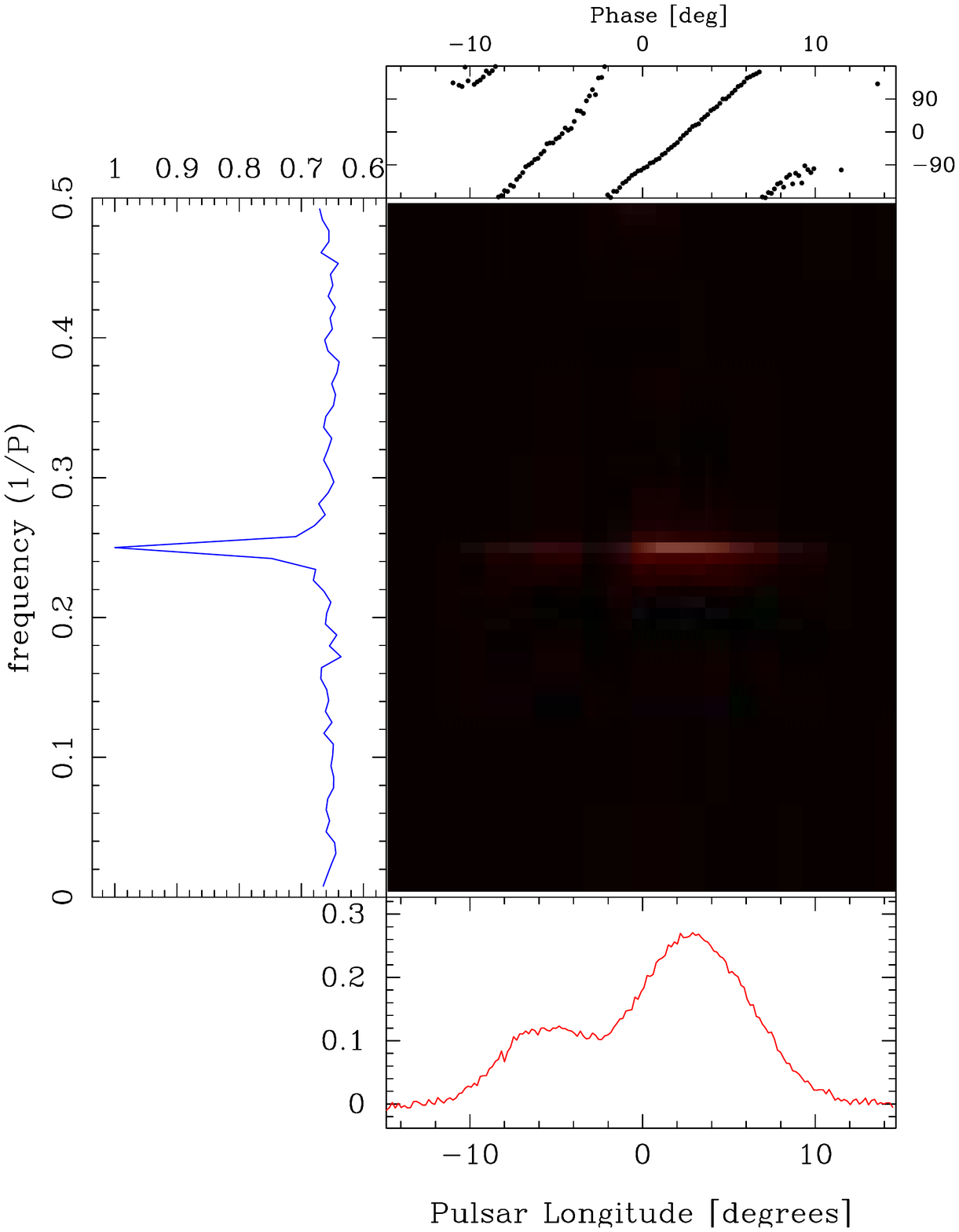}}} \\
\end{tabular}
\caption{The figure shows a simulation of a sequence of pulses exhibiting 
coherent phase modulated negative drifting. The left panel shows 128 
consecutive single pulses where the subpulses in subsequent pulses appear at 
earlier longitudes across the pulse window. The right panel shows the LRFS for 
this pulse sequence which exhibits a peak at $f_p$ = 0.25 cycles/$P$ and the 
corresponding phase changes showing a positive slope from the leading to the 
trailing edge of the window. The pulse sequence was simulated using $P_3$ = 
4$P$. The surface magnetic fields were specified as $\mathbf{r}_s$ = 
(0.95$R_S$, 18.86\degr, 10.99\degr) and $\mathbf{m}_s$ = (0.001$d$, 0\degr, 
0\degr). The inclination of star centered dipole was $\theta_d$ = 15\degr~and 
the line of sight inclination angle $\beta$ = 4\degr.}
\label{fig_negdrift}
\end{figure*}

\begin{figure*}
\begin{tabular}{@{}cr@{}}
{\mbox{\includegraphics[scale=0.72,angle=0.]{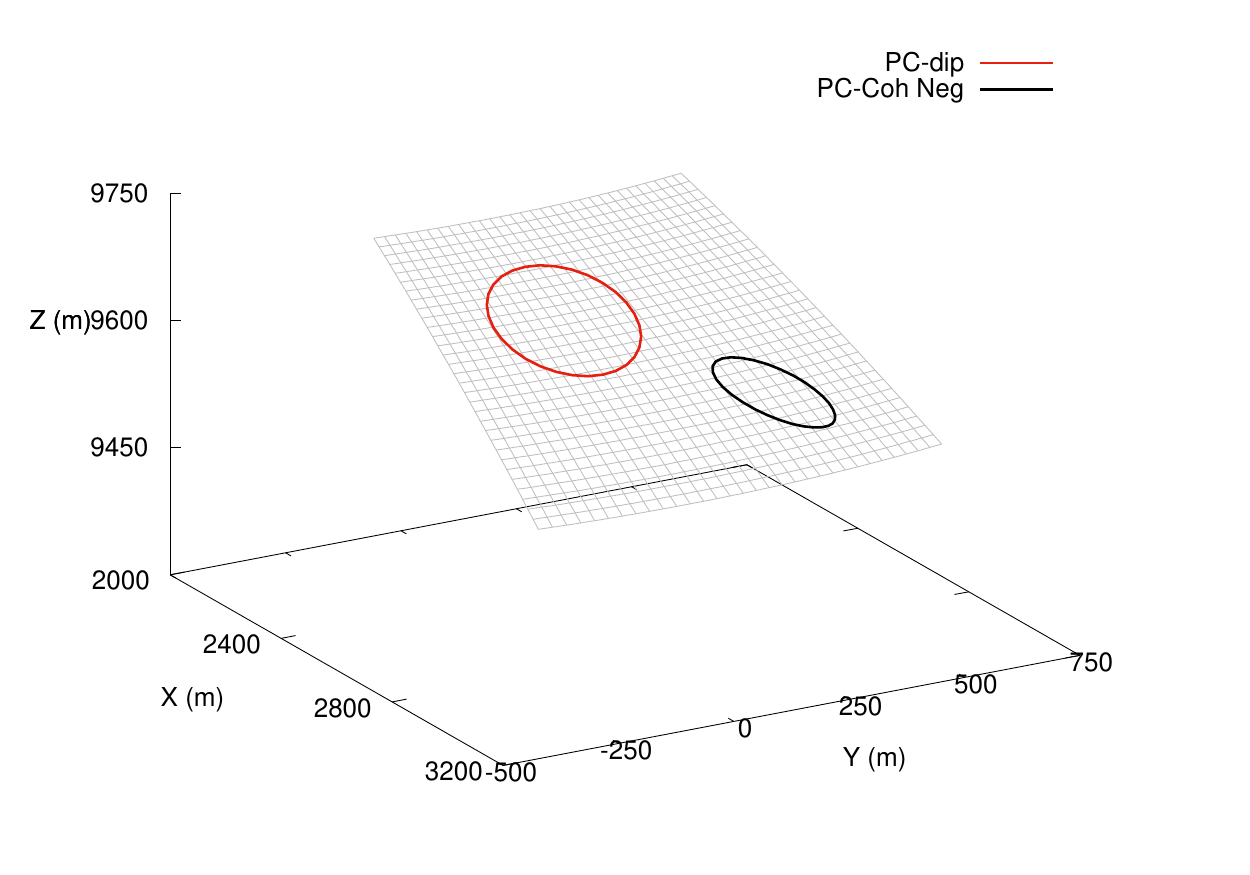}}} &
{\mbox{\includegraphics[scale=0.72,angle=0.]{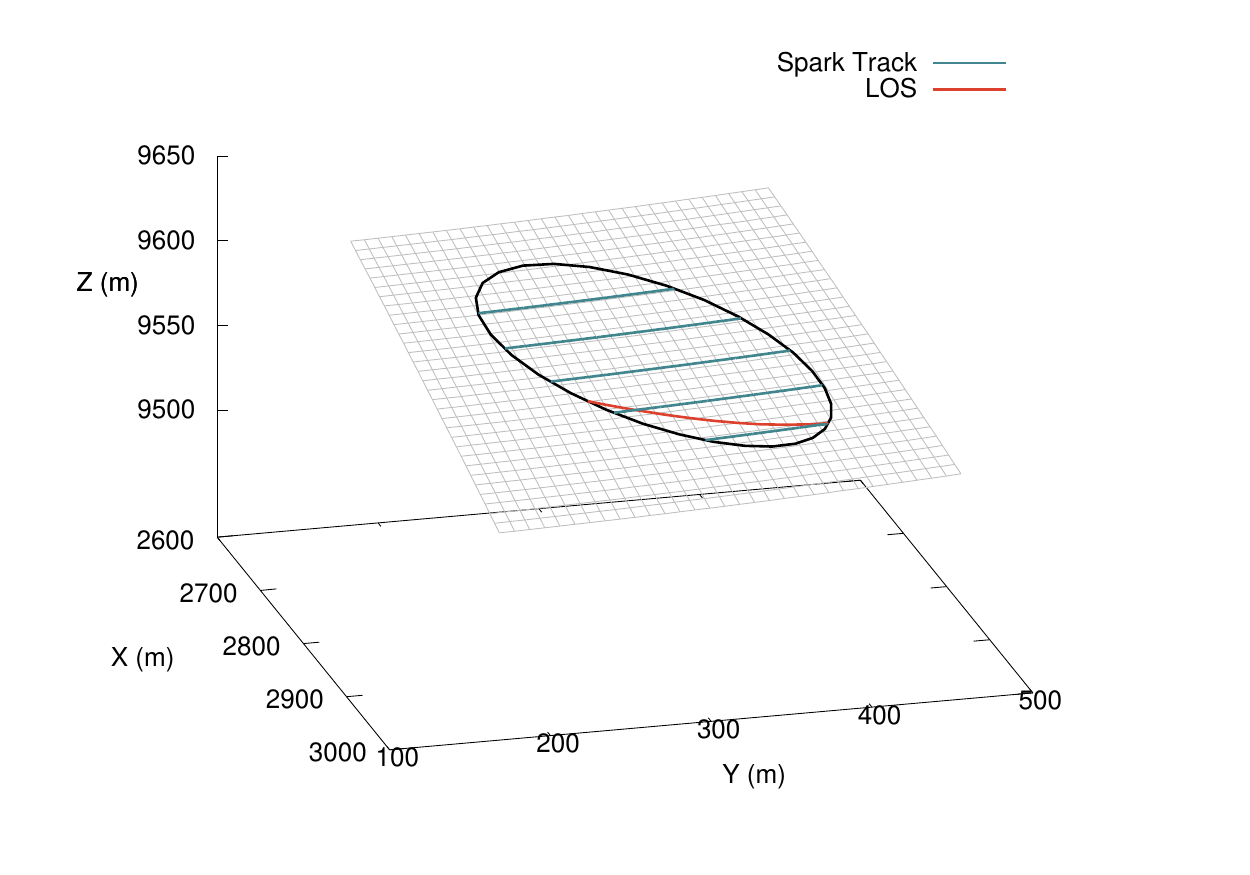}}} \\
\end{tabular}
\caption{Equivalent to figure \ref{fig_posspark} for the magnetic field 
configuration $\mathbf{d}$ = ($d$, 15\degr, 0\degr) and $\mathbf{m}_s$ = 
(0.001$d$, 0\degr, 0\degr) located at $\mathbf{r}_s$ = (0.95$R_S$, 18.86\degr, 
10.99\degr).}
\label{fig_negspark}
\end{figure*}

The coherent phase-modulated drifting is the most prominent drifting class 
which shows regular drift bands in the single pulse sequence. This can be 
further divided into two distinct types, positive and negative drifting. The 
subpulses in such cases show systematic shift in position across the pulse 
window, with the shift being towards the leading part in case of positive 
drifting and towards the trailing part in negative drifting. They are also seen 
as two distinct phase behaviours associated with the peak frequencies in the 
longitude resolved fluctuation spectral analysis 
\citep[LRFS,][]{1973ApJ...182..245B}. In case of positive drifting the phases 
exhibit a systematic variation with a negative gradient from the leading to the
trailing edge while the slope is reversed in the case of negative drifting. 
\citet{2016ApJ...833...29B} argued that the positive and negative drifting 
classes are due to aliasing of $P_3$ around 2$P$. The sparks in the IAR lag 
behind corotation speed which is characterised by the period of pulsar 
rotation, i.e, $\Omega = 2\pi/P$. Hence, within the pulse window the sparks 
move from the trailing to the leading edge as the LOS cuts the emission beam 
from the leading to the trailing side. Our observations are limited by the fact
that the emission is only seen once every period which implies that we can only
see features below $f_p <$ 0.5 cycles/$P$ in the LRFS. So, if the drifting 
periodicity $P_3 >$ 2$P$ the subpulses in subsequent periods appear at earlier 
longitudes which is the negative drifting. However, if the drifting periodicity
is $P < P_3 <$ 2$P$, the subpulses in every subsequent period would appear at 
later longitudes and correspond to the positive drifting. The measured peak 
frequency in the LRFS in this case is aliased and corresponds to $f_p$ = 
1-$P$/$P_3$. Here, the underlying assumption is $P_3 < P$ is not a viable 
solution, hence the frequency peaks are not associated with higher order 
aliases.

In figure \ref{fig_posdrift} we show one realisation of positive drifting. The 
left panel shows 128 single pulses where the subpulses shift towards the 
trailing part of the profile. The right panel shows the LRFS corresponding to 
this pulse sequence. The magnetic field configuration comprises of a star 
centered dipole with $\theta_d$ = 15\degr~and the surface dipole with 
$\mathbf{m}_s$ = (0.001$d$, 0\degr, 0\degr) located at $\mathbf{r}_s$ = 
(0.95$R_S$, 10\degr, 0\degr). We have used $\beta$ = 4\degr~for simulating the 
single pulses with the average profile showing a single component. The polar 
cap structure is represented in figure \ref{fig_posspark}, which shows the 
different spark tracks in the IAR as well as the LOS traverse across them. The 
magnetic field configuration in this case is symmetric around the y-axis. This 
is reflected in the LOS traverse which closely follow a single spark track. 
This is also seen in the phase variations associated with the drifting peak 
which is linear, apart from the edges which show a slight flattening due to the
LOS being more curved than the spark tracks. We have used $P_3$ = 1.33$P$ for 
generating the subpulse motion which corresponds to positive drifting and is 
seen as the drifting peak at $f_p$ = 0.25 cycles/$P$ in the LRFS. The 
corresponding negative drifting for this configuration can be produced using 
$P_3$ = 4$P$ which will give identical $f_p$ with the sign of the phase slope 
reversed.

Another realisation of a single pulse sequence with prominent drift bands is 
shown in figure \ref{fig_negdrift}. In this case we have reproduced negative
drifting with the subpulses shifting towards the leading part of the profile 
(left panel). We have used $P_3$ = 4$P$ for the spark motion in the IAR which 
once again gives a peak around $f_p$ = 0.25 cycles/$P$ in the LRFS (right 
panel). But contrary to the previous case the magnetic field configuration is 
more asymmetric with the surface dipole once again given as $\mathbf{m}_s$ = 
(0.001$d$, 0\degr, 0\degr) but located at $\mathbf{r}_s$ = (0.95$R_S$, 
18.86\degr, 10.99\degr). We have once again used $\beta$ = 4\degr~for 
simulating the single pulses. The polar cap structure for this configuration is
shown in figure \ref{fig_negspark} where the asymmetry is reflected in the 
elongated polar cap shape as well as the LOS traverse. The LOS cuts across two 
spark tracks resulting in a barely resolved double peaked structure. It should 
also be noted that the phase variations associated with the drifting (figure 
\ref{fig_negdrift}, right panel, top window) show distinct non-linear 
behaviour. This is more representative of the phase behaviour seen in pulsar 
observations \citep{2016ApJ...833...29B}.

\subsection{Low-mixed phase-modulated Drifting}\label{sec:lowmixdrift}
\begin{figure*}
\begin{tabular}{@{}cr@{}}
{\mbox{\includegraphics[scale=0.4,angle=0.]{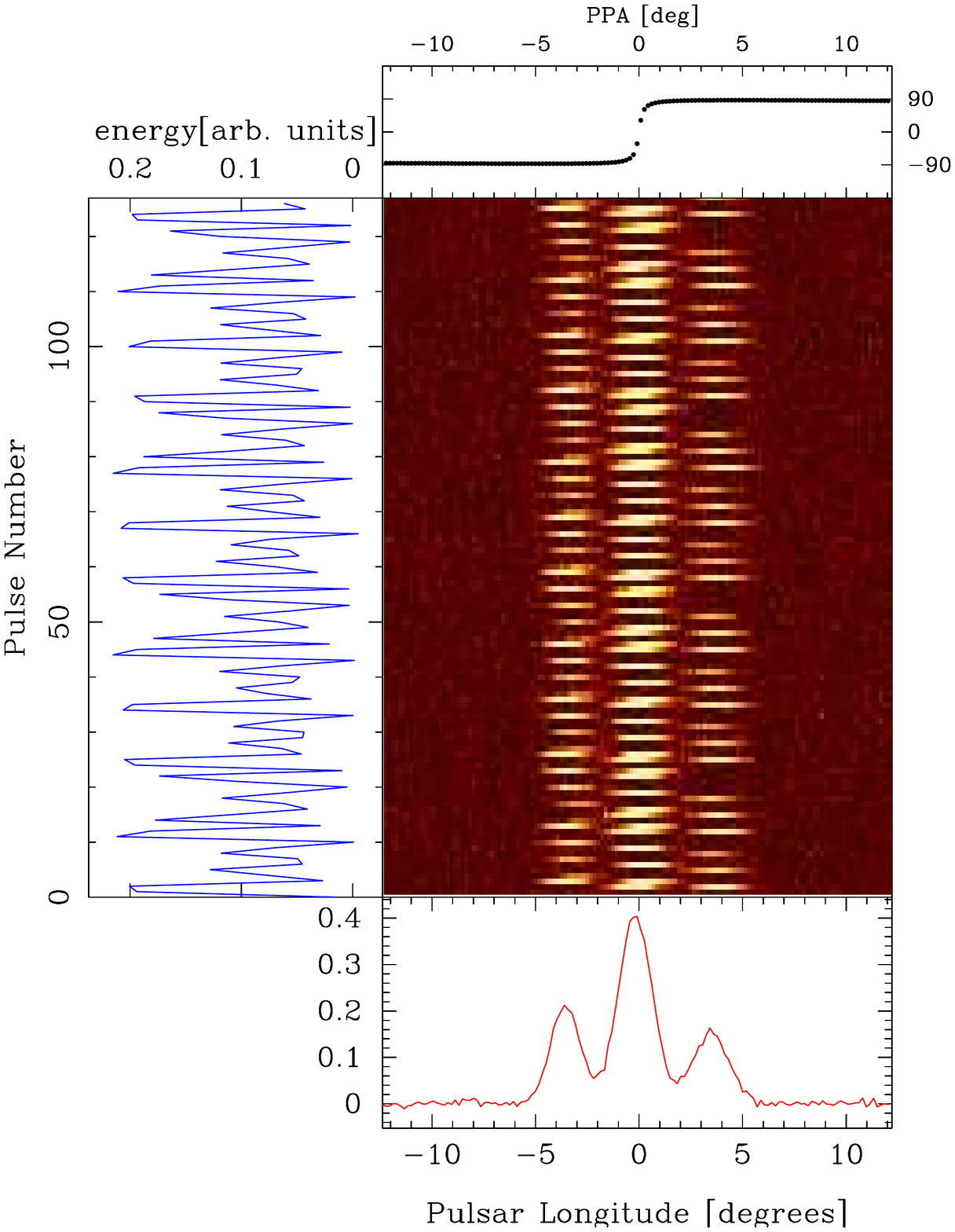}}} &
{\mbox{\includegraphics[scale=0.4,angle=0.]{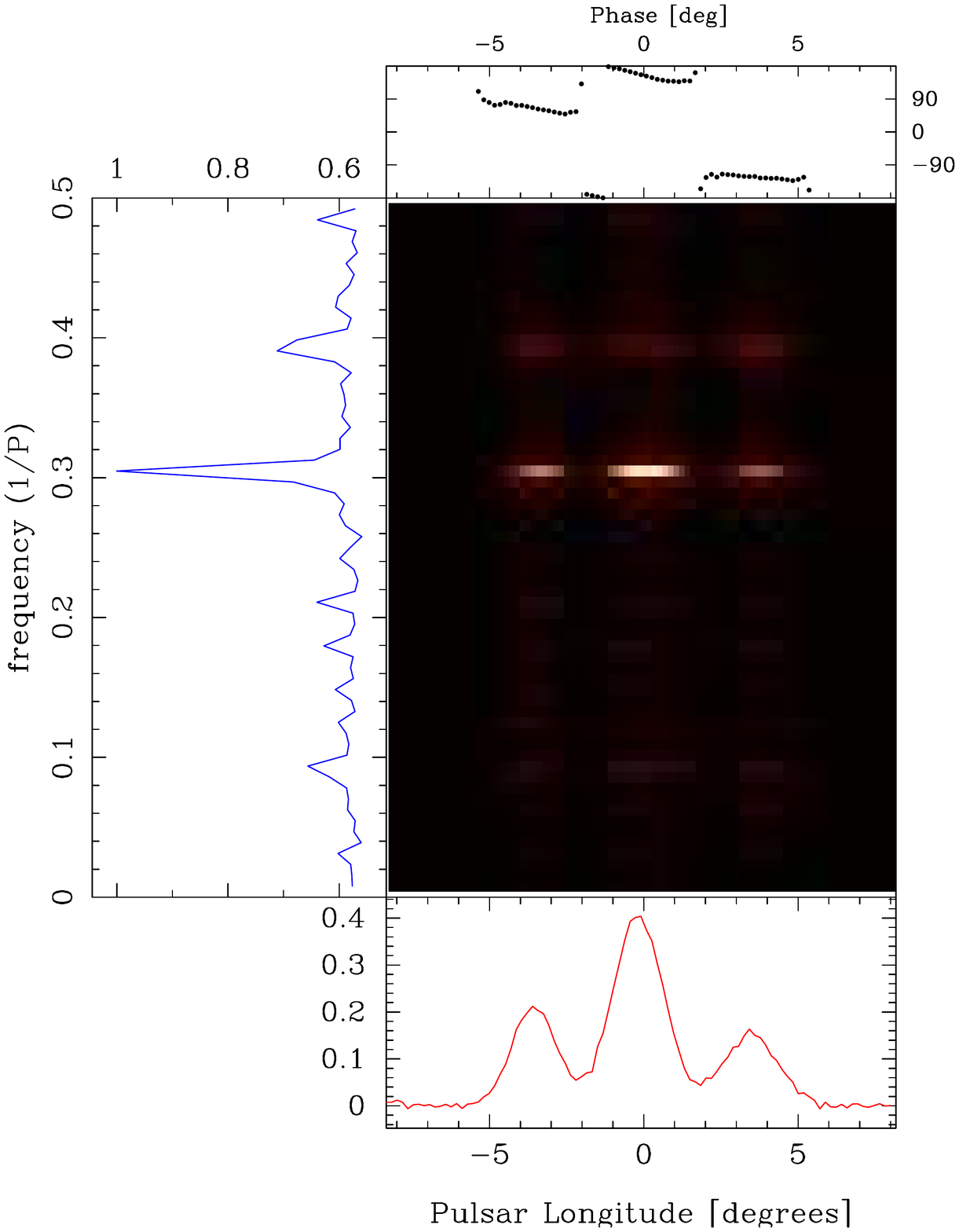}}} \\
\end{tabular}
\caption{The figure shows a simulation of a sequence of pulses exhibiting 
low-mixed phase-modulated drifting. The left panel shows 128 consecutive single
pulses where the subpulses in subsequent pulses do not move steadily across the
pulse window but vary periodically in intensity. The right panel shows the LRFS
for this pulse sequence which exhibits a peak at $f_p$ = 0.3 cycles/$P$ and the
corresponding phase changes across each component are relatively flat. The 
pulse sequence was simulated using $P_3$ = 3.33$P$. The surface magnetic fields
were specified as $\mathbf{r}_s$ = (0.95$R_S$, 57.08\degr, 20.66\degr) and 
$\mathbf{m}_s$ = (0.05$d$, 0\degr, 0\degr). The inclination of star centered 
dipole was $\theta_d$ = 45\degr~and the line of sight inclination angle $\beta$
= 0.1\degr.}
\label{fig_lowmixdrift}
\end{figure*}

\begin{figure*}
\begin{tabular}{@{}cr@{}}
{\mbox{\includegraphics[scale=0.72,angle=0.]{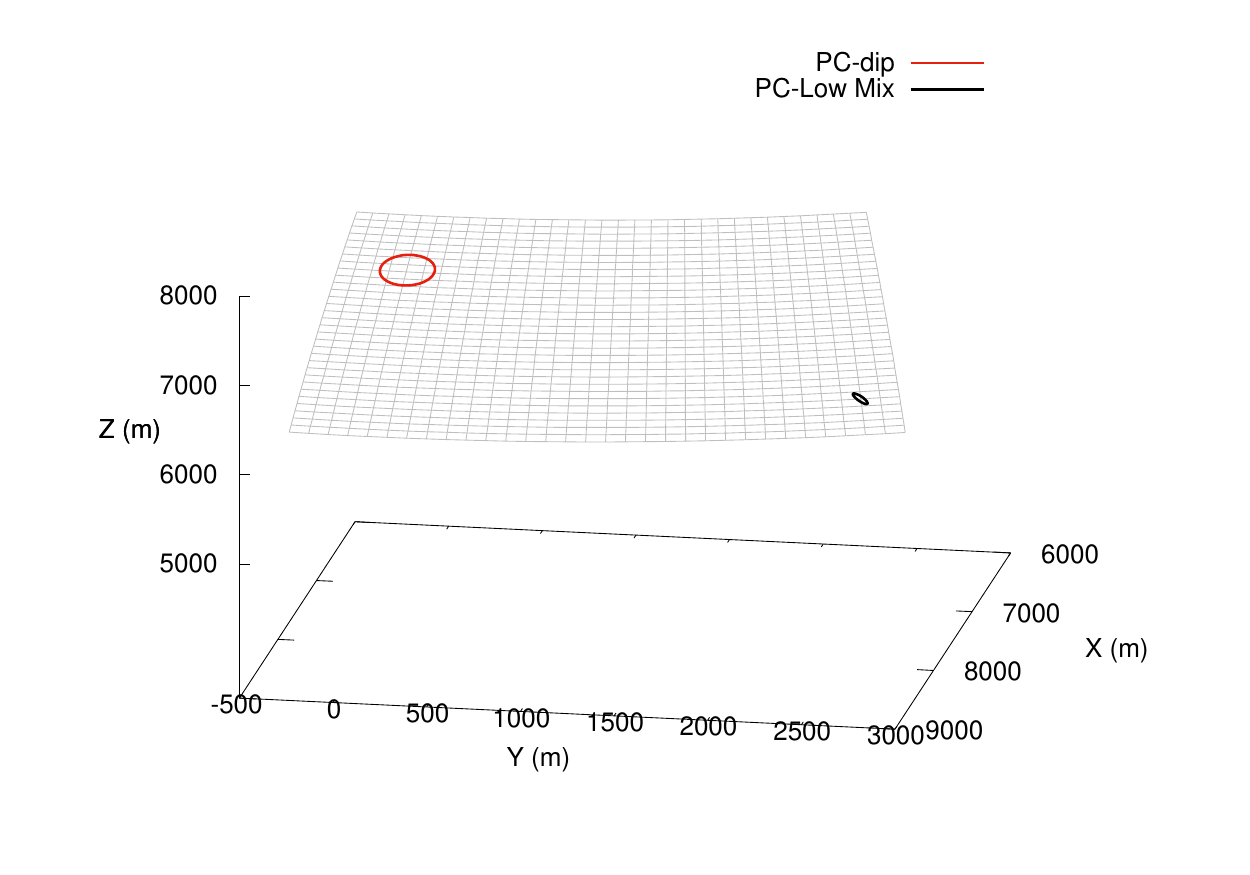}}}
{\mbox{\includegraphics[scale=0.72,angle=0.]{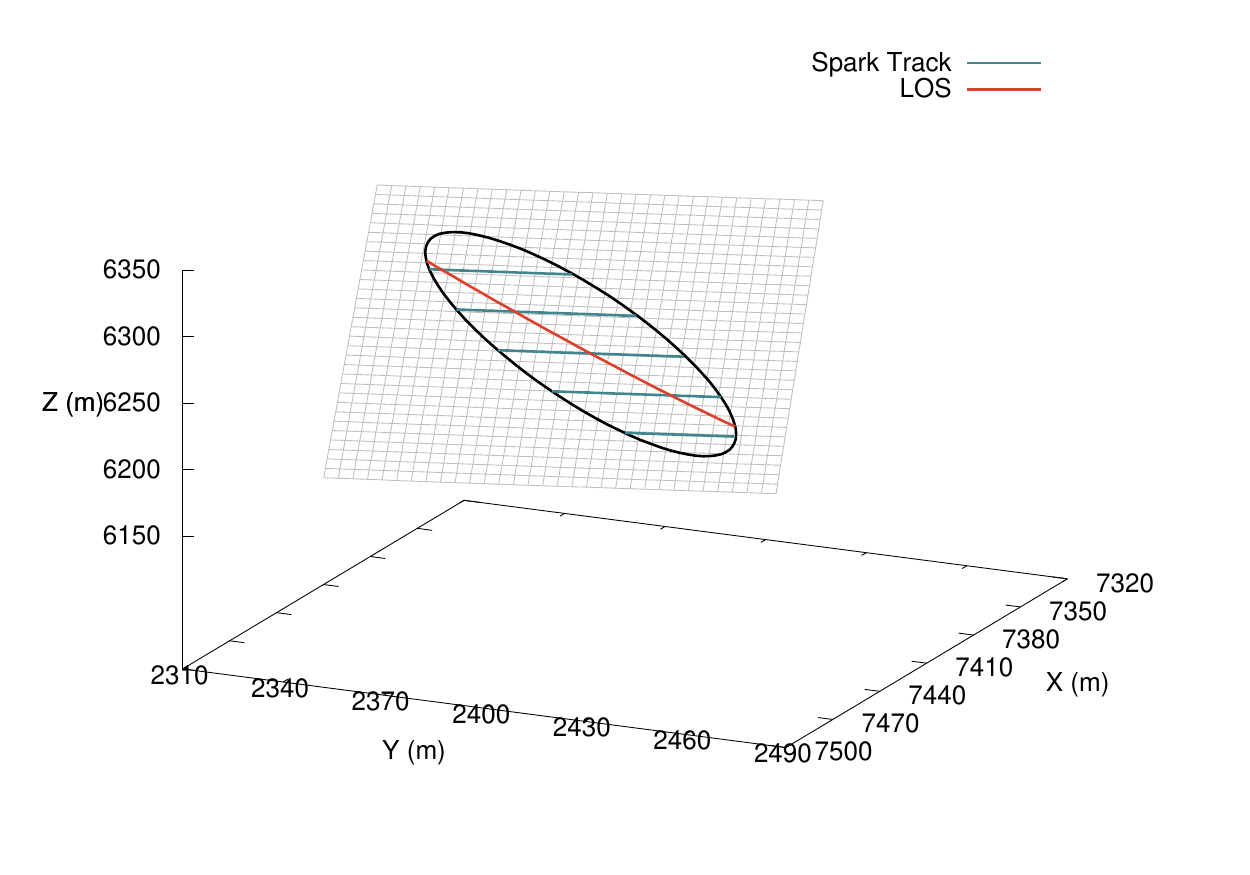}}}
\end{tabular}
\caption{Equivalent to figure \ref{fig_posspark} for the magnetic field 
configuration $\mathbf{d}$ = ($d$, 45\degr, 0\degr) and $\mathbf{m}_s$ = 
(0.05$d$, 0\degr, 0\degr) located at $\mathbf{r}_s$ = (0.95$R_S$, 57.08\degr, 
20.66\degr). The LOS corresponds to $\beta$= 0.1\degr and cuts across the spark
tracks.}
\label{fig_lowmixspark}
\end{figure*}

The second category of subpulse drifting investigated here belongs to the 
low-mixed phase-modulated drifting, where the subpulses do not show much change
in their relative location within the pulse window but periodically vary in 
intensity. This is usually associated with profiles which show multiple 
components, implying that the LOS cuts the emission beam more centrally with 
low $\beta$ \citep{2019MNRAS.482.3757B}. We have reproduced an example of this 
phenomenon as shown in figure \ref{fig_lowmixdrift}. The left panel shows 128 
consecutive single pulses with three subpulses which do not move across the 
profile but periodically varies in intensity, forming a three component average
profile. We have used a magnetic configuration specified by a star centered 
dipole with $\theta_d$ = 45\degr~and another surface dipole with $\mathbf{m}_s$
= (0.05$d$, 0\degr, 0\degr) located at $\mathbf{r}_s$ = (0.95$R_S$, 57.08\degr,
20.66\degr). The $\beta$ used for LOS estimates was 0.1\degr. The right panel 
of the figure shows the LRFS for this pulse sequence. We have use $P_3$ = 
3.33$P$ for spark motion in IAR which is reflected as the frequency peak $f_p$ 
= 0.3 cycles/$P$ in the LRFS. The phase variations in the top window of the 
right panel show very little change across each component, but exhibit jumps 
between adjacent components which is consistent with observations of low-mixed 
phase-modulated drifting \citep{2019MNRAS.482.3757B}. Figure 
\ref{fig_lowmixspark} shows the conditions in the polar cap corresponding to 
the magnetic configuration. The small shifts of the subpulse position suggests 
that the LOS should traverse across the paths of spark motion. This is clearly 
seen in the right panel of the figure where the LOS is indeed cutting across 
the tracks. To achieve this arrangement the non-dipolar polar cap had to be 
shifted significantly away from the dipolar counterpart as seen in the left 
panel of the figure.

\subsection{Switching phase-modulated Drifting}\label{sec:swiphsdrift}
\begin{figure*}
\begin{tabular}{@{}cr@{}}
{\mbox{\includegraphics[scale=0.4,angle=0.]{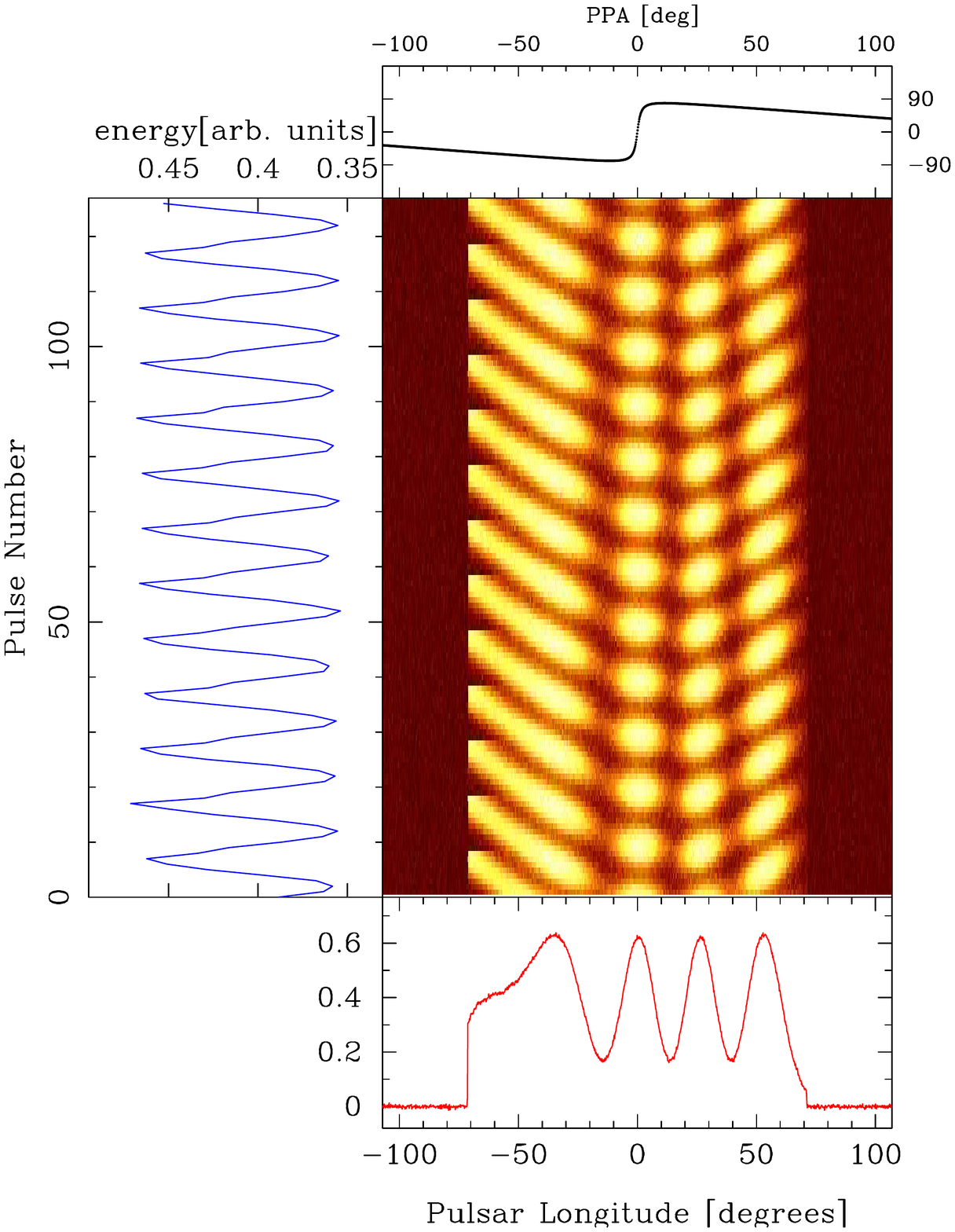}}} &
{\mbox{\includegraphics[scale=0.4,angle=0.]{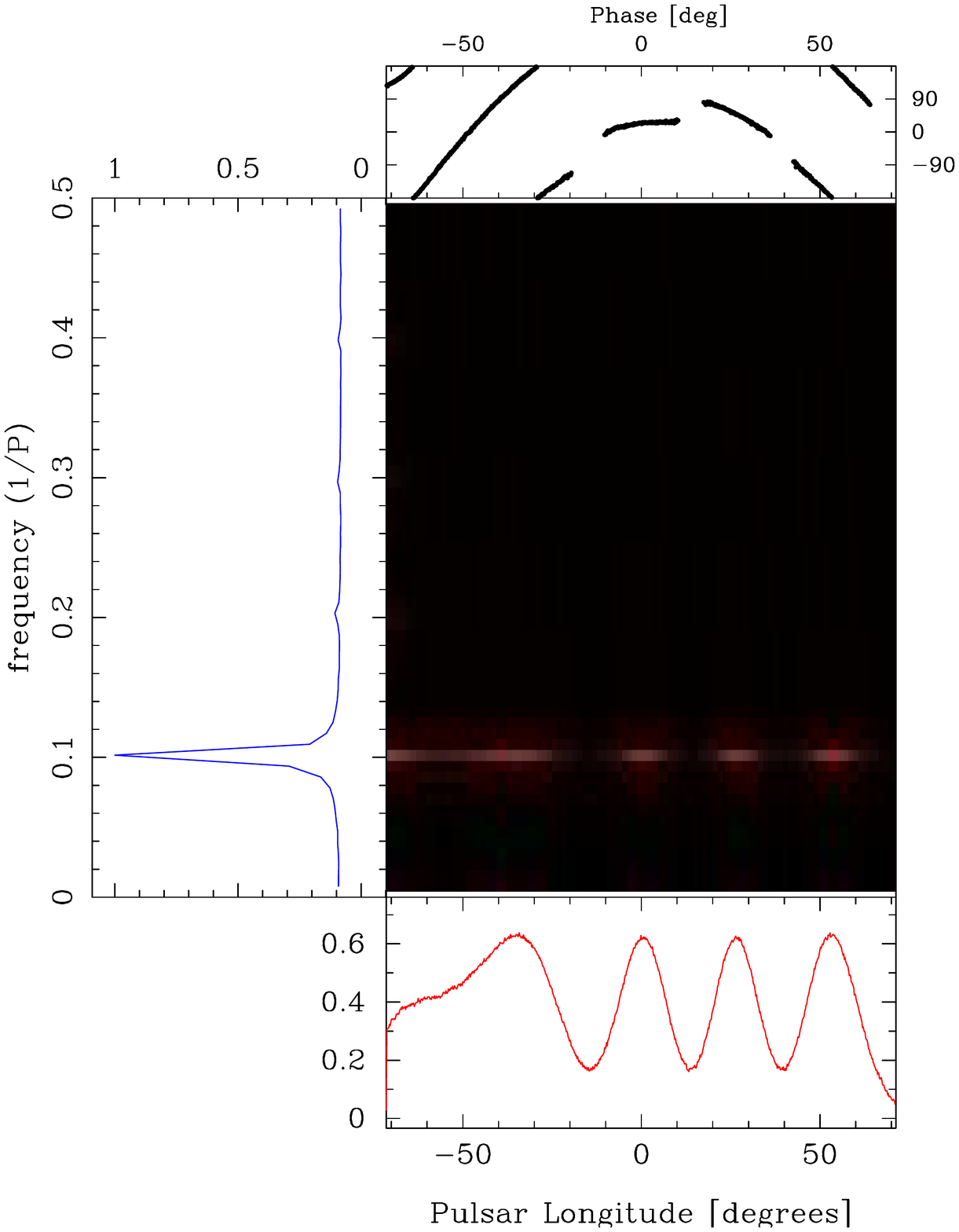}}} \\
\end{tabular}
\caption{The figure shows a simulation a sequence of pulses exhibiting 
switching phase-modulated drifting where the drift direction is opposite in 
different parts of the profile. The left panel shows 128 consecutive single 
pulses where the subpulses in the leading part of the profile appear at earlier
longitudes while the subpulses near the trailing edge appear at later 
longitudes in subsequent pulses. The right panel shows the LRFS for this pulse 
sequence which exhibits a peak at $f_p$ = 0.1 cycles/$P$ and the corresponding 
phases show a positive slope in the leading part of the profile which becomes 
flatter near the middle and changes to negative slope in the trailing part. The
pulse sequence was simulated using $P_3$ = 10$P$. The surface magnetic fields 
were specified as $\mathbf{r}_s$ = (0.95$R_S$, 5\degr, 120\degr) and
$\mathbf{m}_s$ = (0.005$d$, 0\degr, 0\degr). The inclination of star centered
dipole was $\theta_d$ = 5\degr~and the line of sight inclination angle $\beta$ 
= 0.1\degr.}
\label{fig_swiphsdrift}
\end{figure*}

\begin{figure*}
\begin{tabular}{@{}cr@{}}
{\mbox{\includegraphics[scale=0.68,angle=0.]{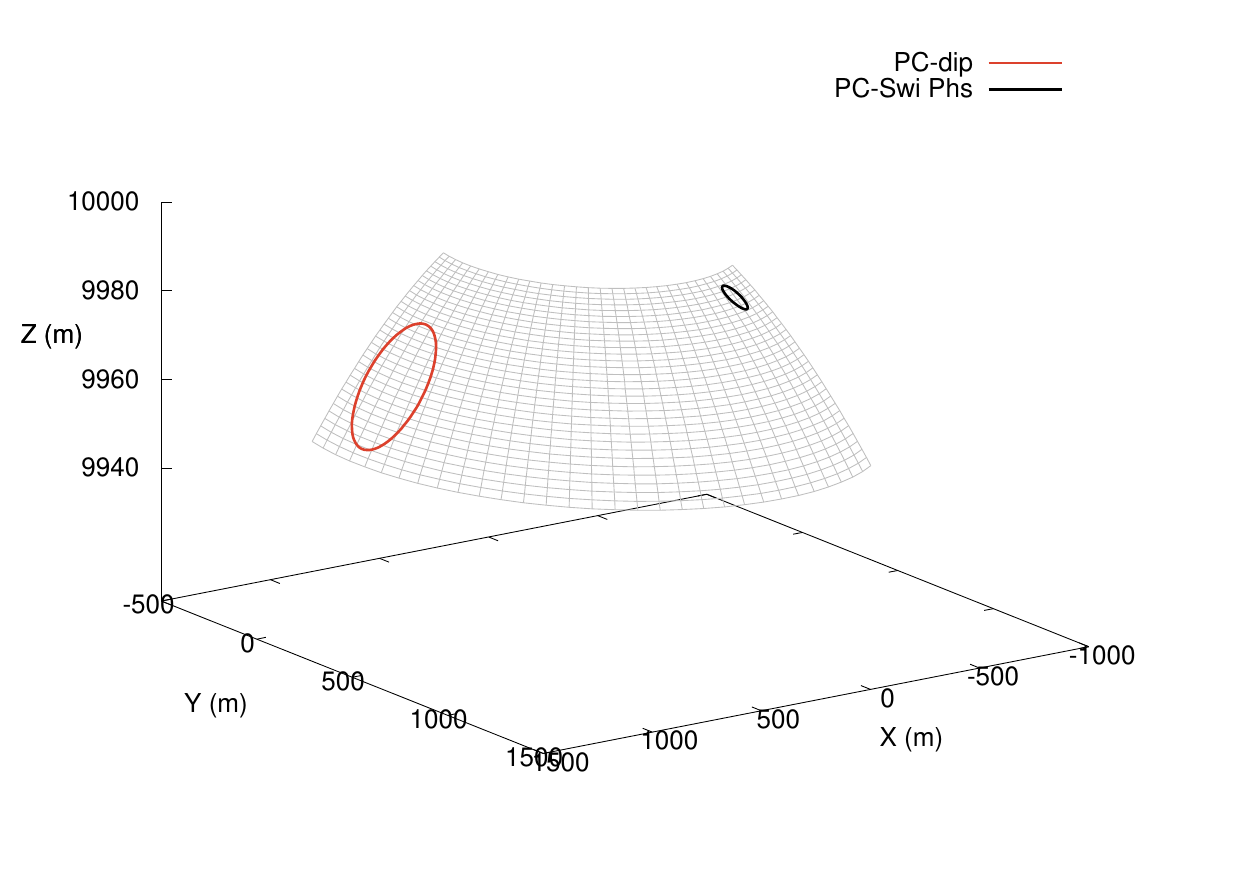}}} &
{\mbox{\includegraphics[scale=0.68,angle=0.]{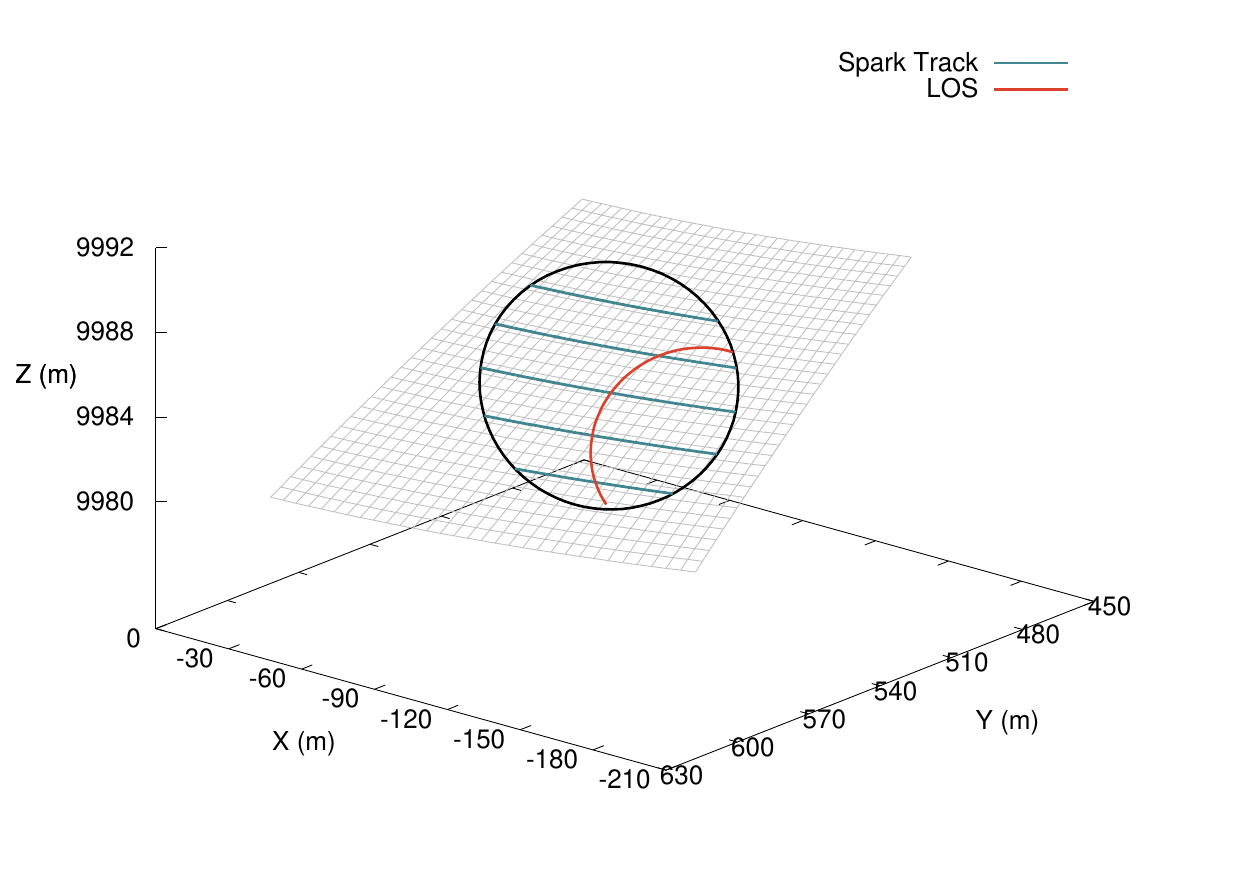}}} \\
\end{tabular}
\caption{Equivalent to figure \ref{fig_posspark} for the magnetic field
configuration $\mathbf{d}$ = ($d$, 5\degr, 0\degr) and $\mathbf{m}_s$ = 
(0.005$d$, 0\degr, 0\degr) located at $\mathbf{r}_s$ = (0.95$R_S$, 5\degr, 
120\degr). The LOS corresponds to $\beta$= 0.1\degr~and cuts the leading and 
trailing parts at opposite directions of spark motion leading to the 
bi-drifting phenomenon.}
\label{fig_swiphsspark}
\end{figure*}

The switching phase-modulated drifting is seen in a small group of pulsars 
where the subpulses belonging to different components of the profile show large
shifts and reversals in phase directions. This implies that both positive and 
negative drifting are seen at different regions of the pulsar profile. In 
figure \ref{fig_swiphsdrift} we show an example of the switching 
phase-modulated drifting behaviour using sparks lagging behind corotation 
speed. The left panel shows 128 single pulses where subpulses show the gradual 
reversal in phase from the first to the last component. The pulsar profile 
consists of four components with significant subpulse variation across each of 
them. The leading component has the most variation and shows negative drifting 
with shifts towards the leading part of the profile. The two central components
has lower variations in phases with the central left component showing slight 
negative drifting and the central right component small positive drifting. The 
trailing component shows positive drifting with the subpulses shifting towards 
the trailing edge. The magnetic configuration used in these simulations 
consists of a star centered dipole with $\theta_d$ = 5\degr~and another surface
dipole with $\mathbf{m}_s$ = (0.005$d$, 0\degr, 0\degr) located at 
$\mathbf{r}_s$ = (0.95$R_S$, 5\degr, 120\degr). We have used $\beta$ = 
0.1\degr~for the LOS. The right panel of the figure shows the LRFS 
corresponding to the single pulse behaviour. We have used $P_3$ = 10$P$ for the
spark motion in the IAR which is reflected as the peak frequency $f_p$ = 0.1 
cycles/$P$ in the LRFS. The top window also shows the phase behaviour which 
indicates the bi-drifting behaviour. 

Figure \ref{fig_swiphsspark} shows the conditions in the IAR which leads to
switching phase-modulated drifting. The LOS makes a curved traverse across the 
different spark paths. Initially the LOS moves in the opposite direction of the
spark motion which results in negative drifting. For the two central tracks the
LOS cuts across them resulting in very little variations across the components.
Finally, towards the trailing side the LOS reverses direction and moves in the 
same direction of the sparks which results in positive drifting. The left panel
in the figure shows that the non-dipolar polar cap has to be highly asymmetric 
compared to the equivalent dipolar polar cap in order to see the bi-drifting 
behaviour. The polar cap is shifted to the other side of the neutron star 
rotation axis (negative x-direction) resulting in highly curved nature of the 
LOS in this example. This further explains the rarity of this phenomenon in the
pulsar population. 

Also note that the $\theta_d$ = 5\degr~is the lowest amongst the different 
configurations explored in this work. This has important observational 
implications. In all pulsars exhibiting reversals in phases the profile width 
is much larger than the general pulsar population \citep[$>$ 50\degr,
see][]{2019MNRAS.482.3757B,2019MNRAS.486.5216B}, which suggests a low value of 
magnetic inclination angle, $\theta_d$. In our simulations the reversals in 
phase is possible because the non-dipolar polar cap is rotated to the other 
side of the neutron star. In case the inclination angle is large the surface 
field needs to be significantly larger on the other side for this realization. 
This is difficult to achieve in a realistic pulsar where there are likely to be
multiple surface anomalies resulting in strong local fields, and the nearest 
maximum to the dipolar polar cap dominating the eventual IAR. Hence, the 
relatively wide profiles in bi-drifting pulsars provide indirect evidence for 
the sparks to lag behind corotation speed in IAR dominated by non-dipolar 
fields.

\section{Discussion}\label{sec:disc}
The different drifting classes can be associated with the differential 
orientation of the non-dipolar polar cap compared to the corresponding dipolar 
case. The systematic drift bands corresponding to coherent phase-modulated 
drifting are seen when the polar cap orientation is close to the dipolar case, 
i.e, the dominating surface dipoles are close to the magnetic axis. As the 
surface dipoles are oriented further away from the axis the systematic drift 
bands give way to non-linearity in the phase variations till phase stationary 
behaviour is seen when the non-dipolar polar cap is on the sides of the neutron
star relative to the dipolar case, i.e. roughly 90\degr away. Finally, the 
reversals in phase behaviour is seen when the surface magnetic field is highly 
curved and the LOS reverses direction as it traverses the polar cap. This is 
usually seen when the non-dipolar polar cap is rotated behind the dipolar polar
cap, i.e. almost 180~\degr~away. There is increasing evidence for the presence 
of non-dipolar polar caps in normal pulsars ($P>0.1$ s) from simultaneous 
observations at radio and X-ray frequencies. The thermal X-ray peak, which 
originates from the heated polar cap, and the radio emission peak, arising from
the dipolar field lines are significantly misaligned 
\citep{2019MNRAS.489.4589A,2020MNRAS.491...80P}. This is only possible when the
polar cap is highly non-dipolar in nature and located far away from the 
corresponding dipolar case. Additionally, other observed correlations, like 
bi-drifting behaviour being associated with relatively wide profile pulsars, 
the constant drifting periodicity across all components of the profile (see 
section \ref{sec:app_IARphy}), also find simple interpretation in this proposed
mechanism. There is no requirement for the sparks to lag behind corotation 
speed in certain parts of the polar cap and exceed corotation speed in other 
parts, in order to reproduce the different drifting features.

\begin{figure*}
\begin{tabular}{@{}cr@{}}
{\mbox{\includegraphics[scale=0.42,angle=0.]{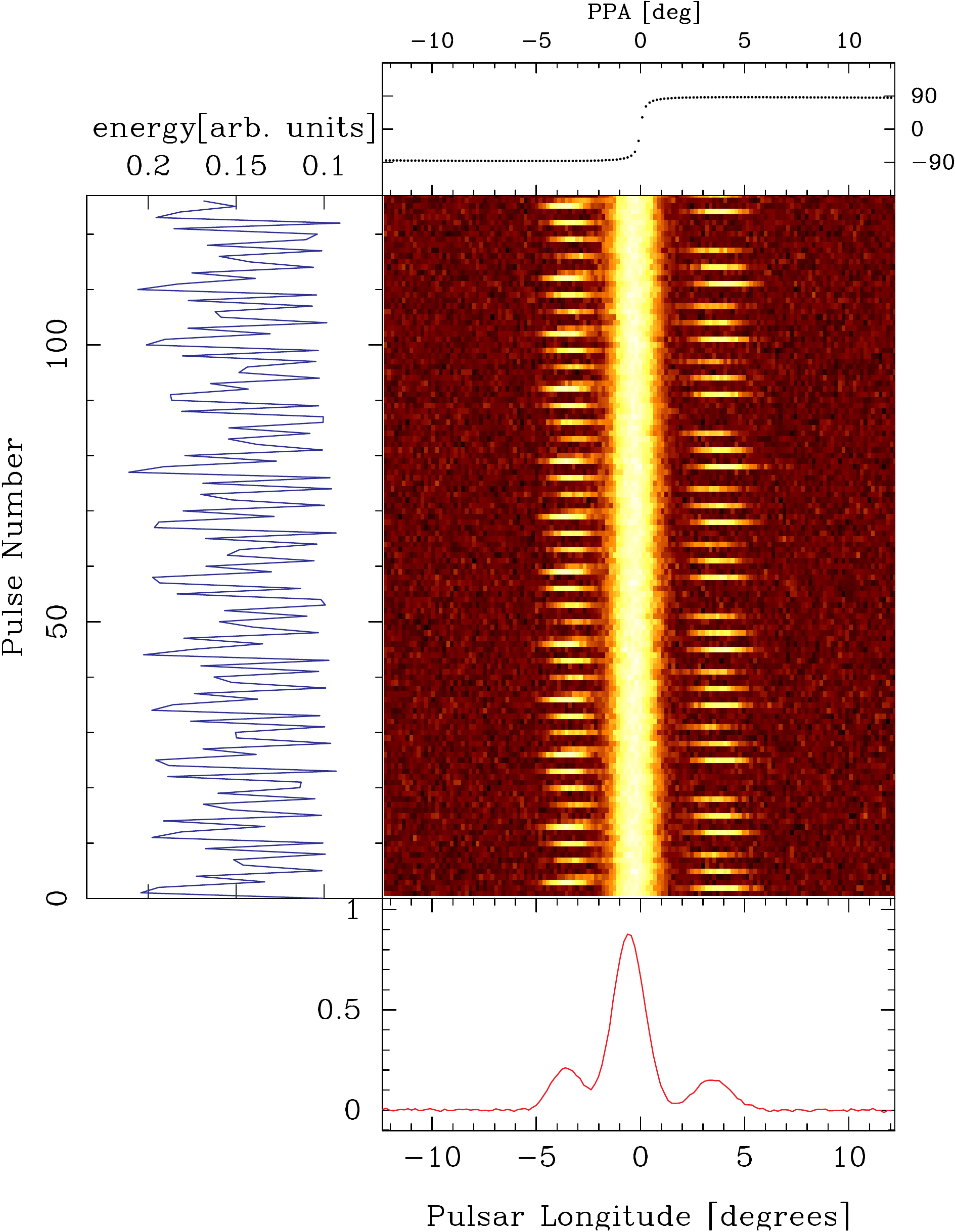}}} &
{\mbox{\includegraphics[scale=0.42,angle=0.]{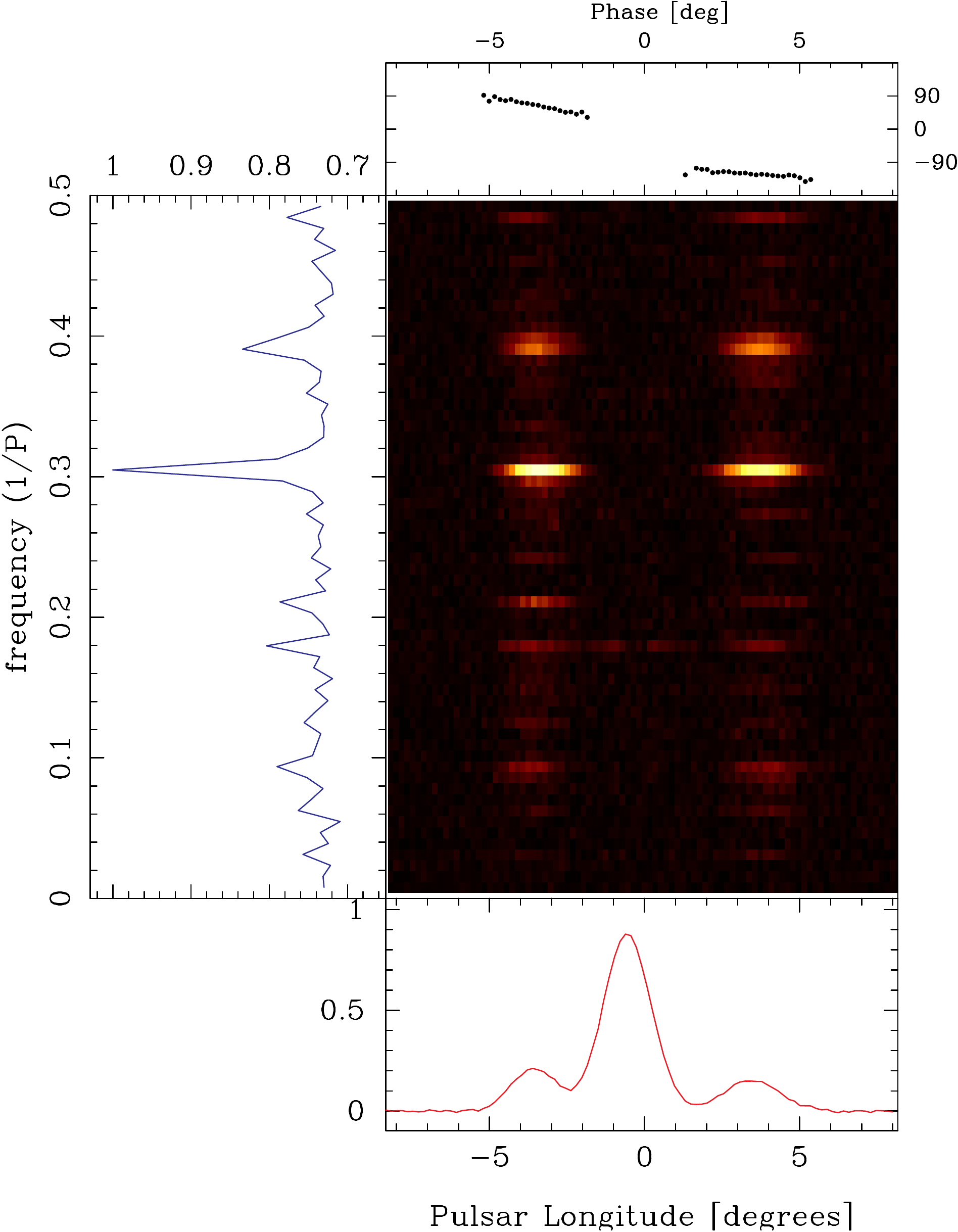}}} \\
\end{tabular}
\caption{The figure shows the single pulse behaviour for the modified spark 
motion in the IAR corresponding to the magnetic field configuration shown in 
figure \ref{fig_lowmixspark} which results in the low-mixed phase-modulated 
drifting. In this modified figure the sparks are assumed to be stationary in
the central track. The presence of a boundary around the polar cap and tightly 
packed sparks in the IAR inhibit the spark motion, particularly near the 
center. The single pulse sequence for 128 pulses is shown in the left panel,
where the central component is unchanging while the outer components show 
periodic modulation in intensity. The right panel shows the LRFS for this pulse
sequence which exhibits a peak at $f_p$ = 0.3 cycles/$P$ in the two outer 
components with relatively flat phase variations. No periodic fluctuation is 
seen in the central component.}
\label{fig_lowmixcore}
\end{figure*}

\begin{figure*}
\begin{tabular}{@{}cr@{}}
{\mbox{\includegraphics[scale=0.4,angle=0.]{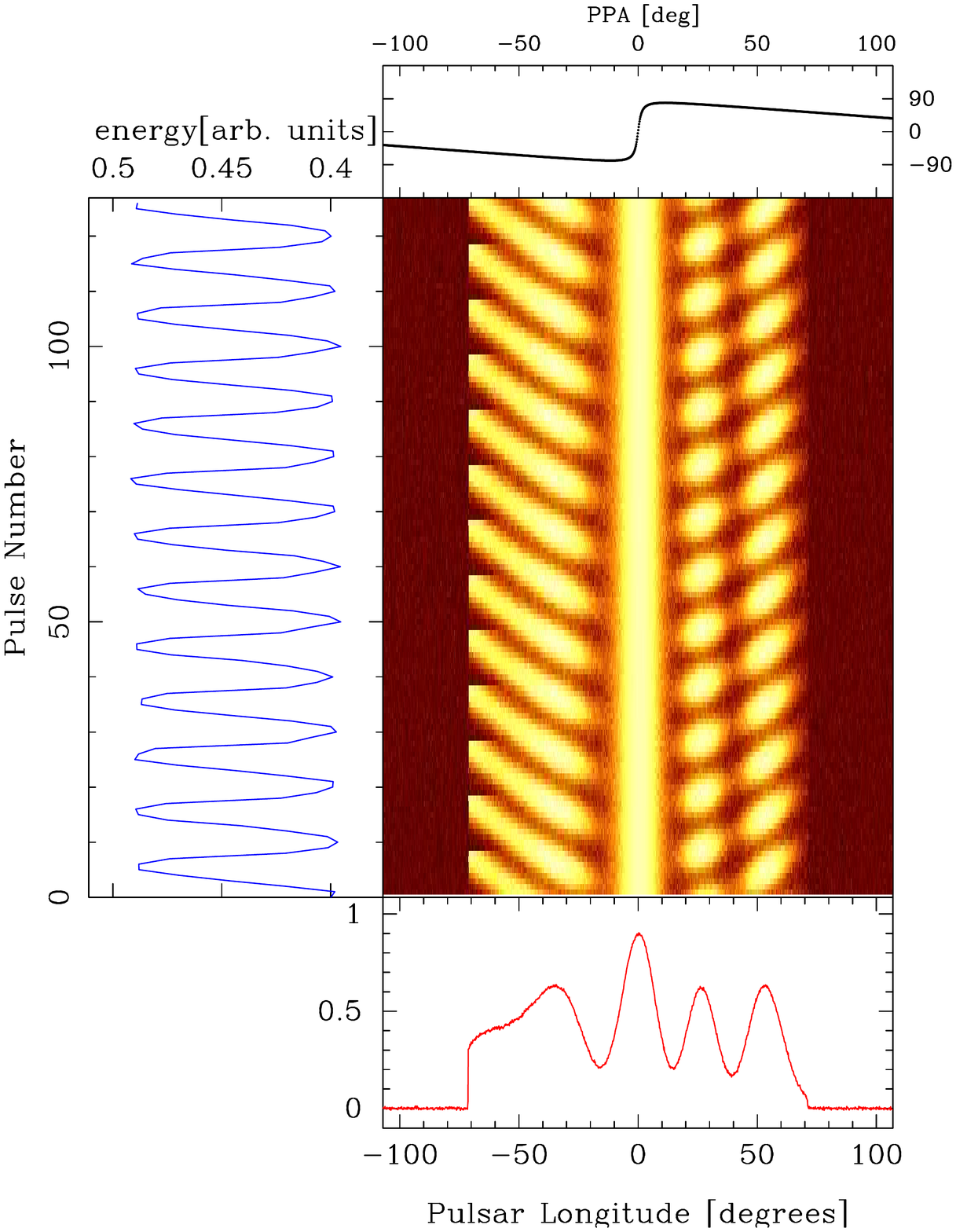}}} &
{\mbox{\includegraphics[scale=0.4,angle=0.]{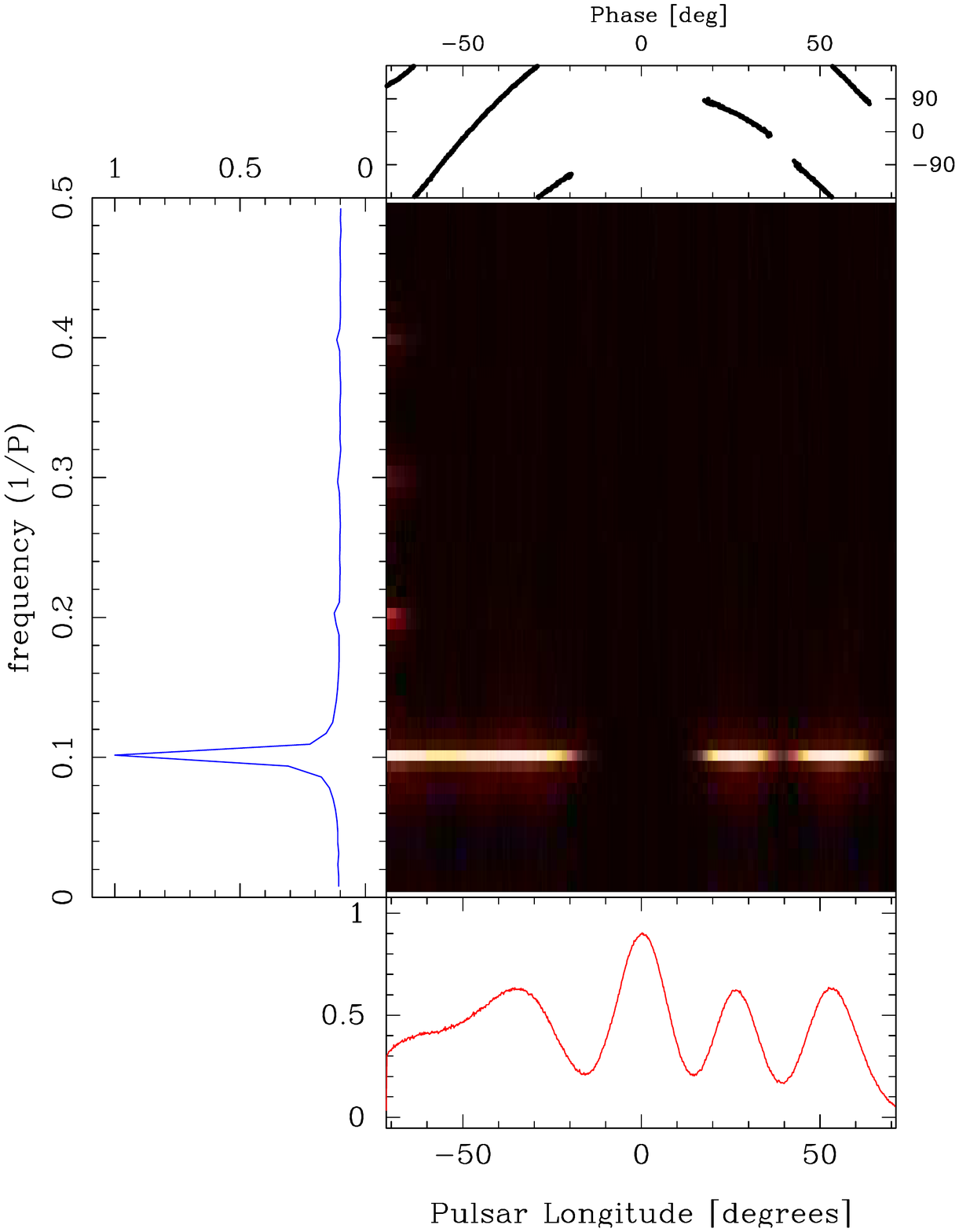}}} \\
\end{tabular}
\caption{The figure shows the single pulse behaviour for the modified spark
motion in the IAR corresponding to the magnetic field configuration shown in
figure \ref{fig_swiphsspark} which results in the switching phase-modulated
drifting. In this modified figure the sparks are assumed to be stationary in
the central track. The presence of a boundary around the polar cap and tightly
packed sparks in the IAR inhibit the spark motion, particularly near the
center. The single pulse sequence for 128 pulses is shown in the left panel,
where the central component is unchanging while in the outer components the 
subpulses in the leading part of the profile appear at earlier longitudes while
the subpulses near the trailing edge appear at later longitudes in subsequent 
pulses. The right panel shows the LRFS for this pulse sequence which exhibits a
peak at $f_p$ = 0.1 cycles/$P$ in the outer components and the corresponding
phases show a positive slope in the leading part of the profile which 
transforms to negative slope in the trailing part. No periodic fluctuation is
seen in the central component.}
\label{fig_swiphscore}
\end{figure*}

There is also an underlying assumption of steady evolution of sparks, where 
they freely lag behind the corotation speed throughout the IAR. However, this 
idealized lagging behind scenario is restricted in reality by the presence of 
polar cap boundaries beyond which no spark can be formed in the closed field 
line region. As the the pair production process is inhibited in the closed 
field line region a continuous sparking discharge is expected to happen near 
the edge of the polar cap for the thermal regulation in the PSG model. If a 
spark is not present near the boundary, since there is no heating from the 
other side, the temperature drops quickly inhibiting the ion flow from the 
surface. A large potential drop along the IAR develops resulting in sparking 
discharge to commence immediately near the boundary. Hence, the most likely 
dynamics of the plasma flow in IAR would involve the sparks to lag behind 
corotation, but constrained to move along the boundary of the polar cap, such 
that there is always presence of a spark near the edge. In addition, the PSG 
model also demands the IAR to be tightly packed with sparks for effective 
thermal regulation throughout the polar cap. As a result of the boundary effect
one expects the spark motion to be increasingly constrained as one moves away 
from the edge and the sparks are expected to be completely stationary near the 
central region. A detailed study incorporating the boundary effect on the 
sparks and their natural tendency to lag behind corotation speed will be 
explored in a future work. The possibility of stationary sparks near the center
has important observational implications. The pulsar emission is characterised 
by a stationary core which does not show drifting \citep{1986ApJ...301..901R,
2019MNRAS.482.3757B}. In addition, there are also a large number of cases where
no subpulse drifting is seen in the single pulse emission 
\citep{2006A&A...445..243W,2007A&A...469..607W,2016ApJ...833...29B,
2019MNRAS.482.3757B}. In the absence of detailed estimates a simplified 
application of the effect of stationary sparks in the center of the IAR can be 
used for the single pulse simulations shown in section \ref{sec:driftsiml}. The
sparks in the central tracks used in all cases can be considered to be 
stationary with no time evolution, i.e., in eq.(\ref{eq_sprktrk}), $\phi_i^0(t)
= 2\pi i/N$. There is no change in the observed drifting behaviour for the 
coherent phase-modulated drifting cases in section \ref{sec:cohdrift}, where 
the LOS traverses the IAR peripherally without cutting across the central spark
track. The modified single pulse behaviour corresponding to low-mixed 
phase-modulated drifting case is shown in figure \ref{fig_lowmixcore}, and 
switching phase-modulated drifting is shown in figure \ref{fig_swiphscore}. The
magnetic field configurations and the drifting periodicities used in these 
plots are identical to section \ref{sec:lowmixdrift} and \ref{sec:swiphsdrift},
respectively. The central component, resembling the core emission, is 
stationary in both cases without any periodic behaviour.

\subsection{Physical parameters in non-dipolar Inner Acceleration Region}\label{sec:app_IARphy}
\begin{figure*}
\begin{tabular}{@{}cr@{}}
{\mbox{\includegraphics[scale=0.65,angle=0.]{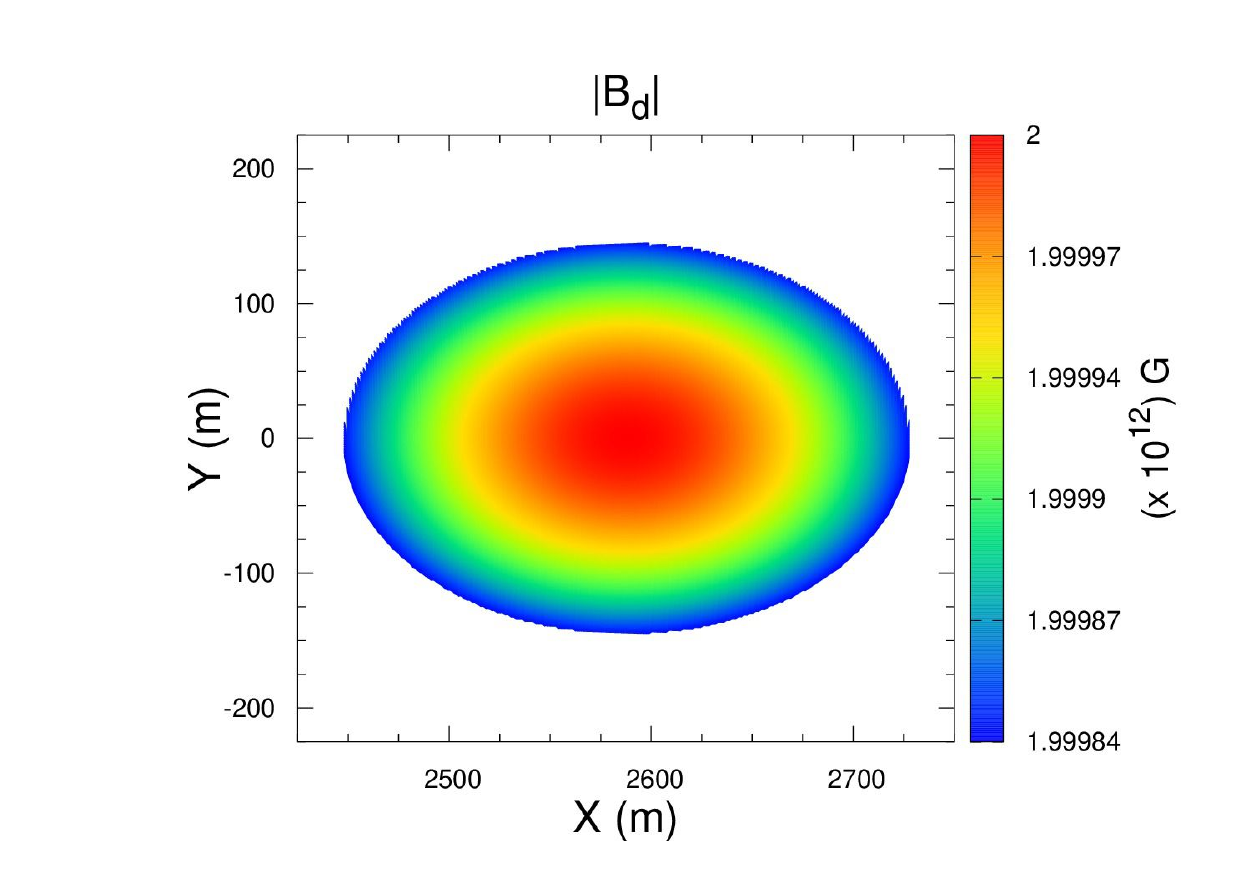}}} &
{\mbox{\includegraphics[scale=0.65,angle=0.]{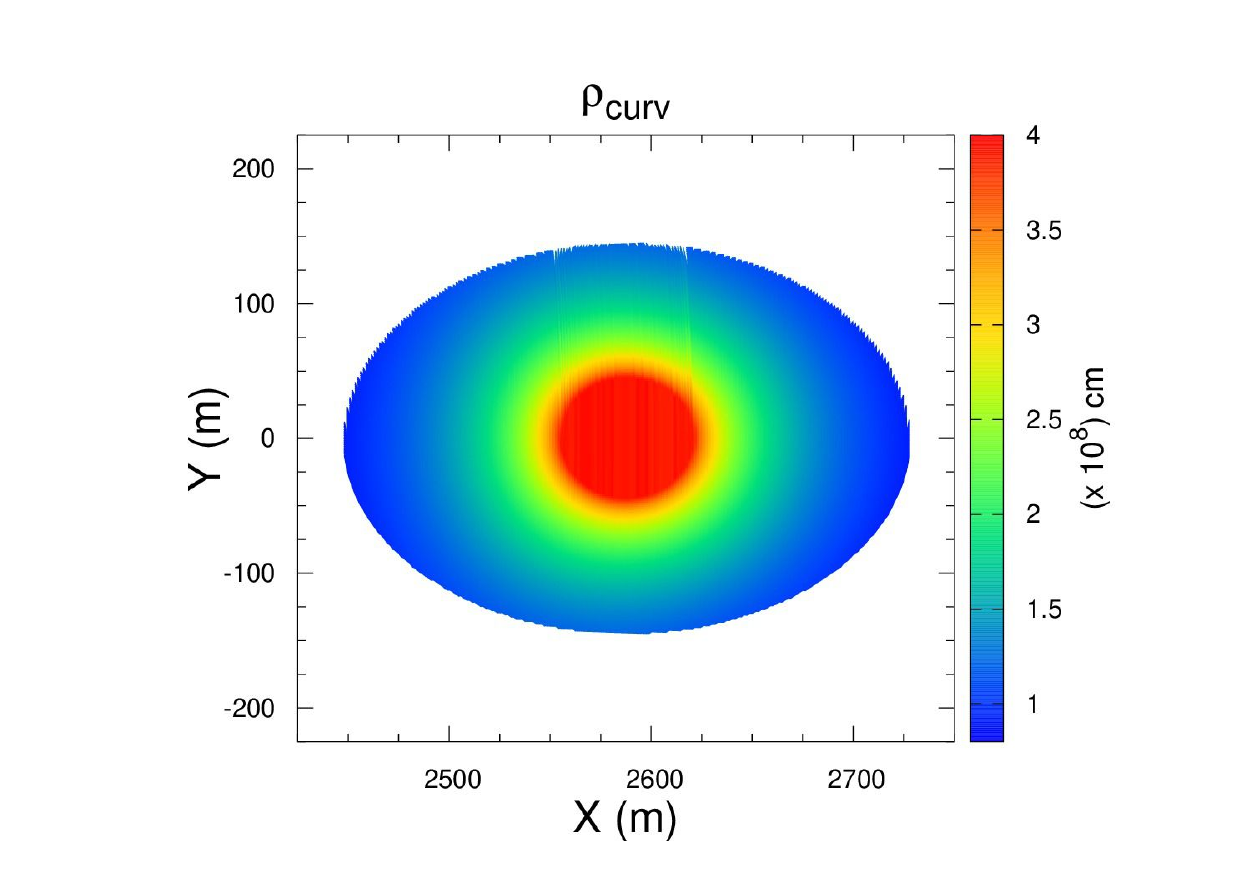}}} \\
{\mbox{\includegraphics[scale=0.65,angle=0.]{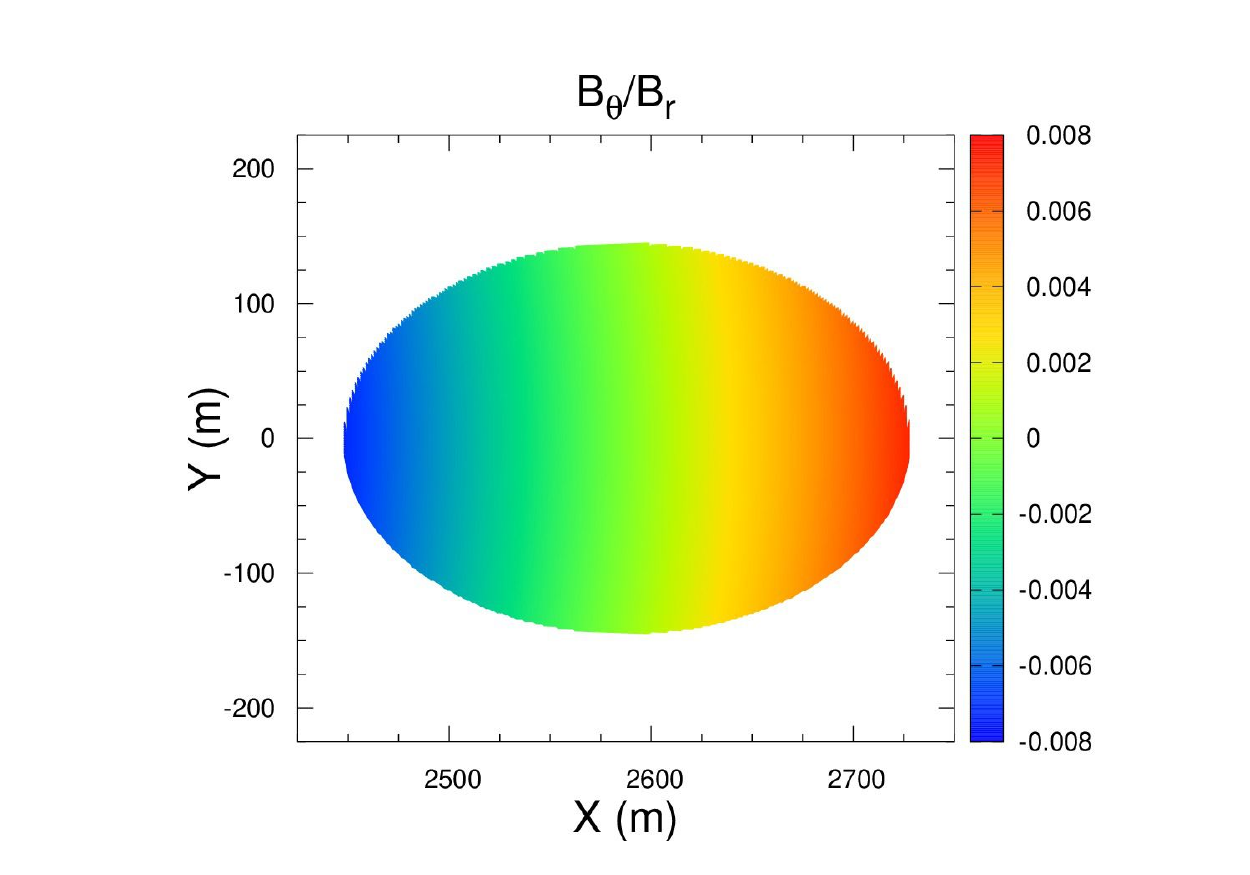}}} &
{\mbox{\includegraphics[scale=0.65,angle=0.]{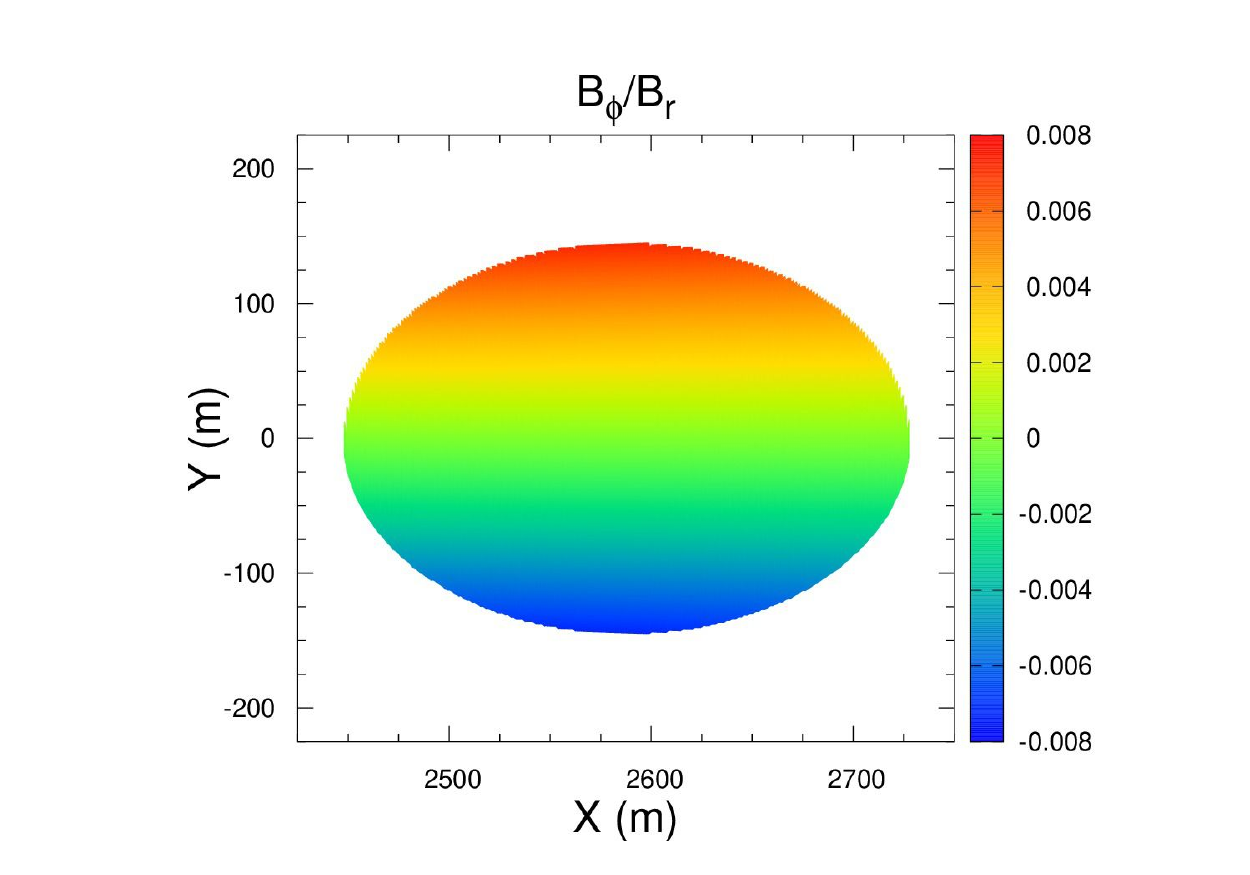}}} \\
{\mbox{\includegraphics[scale=0.65,angle=0.]{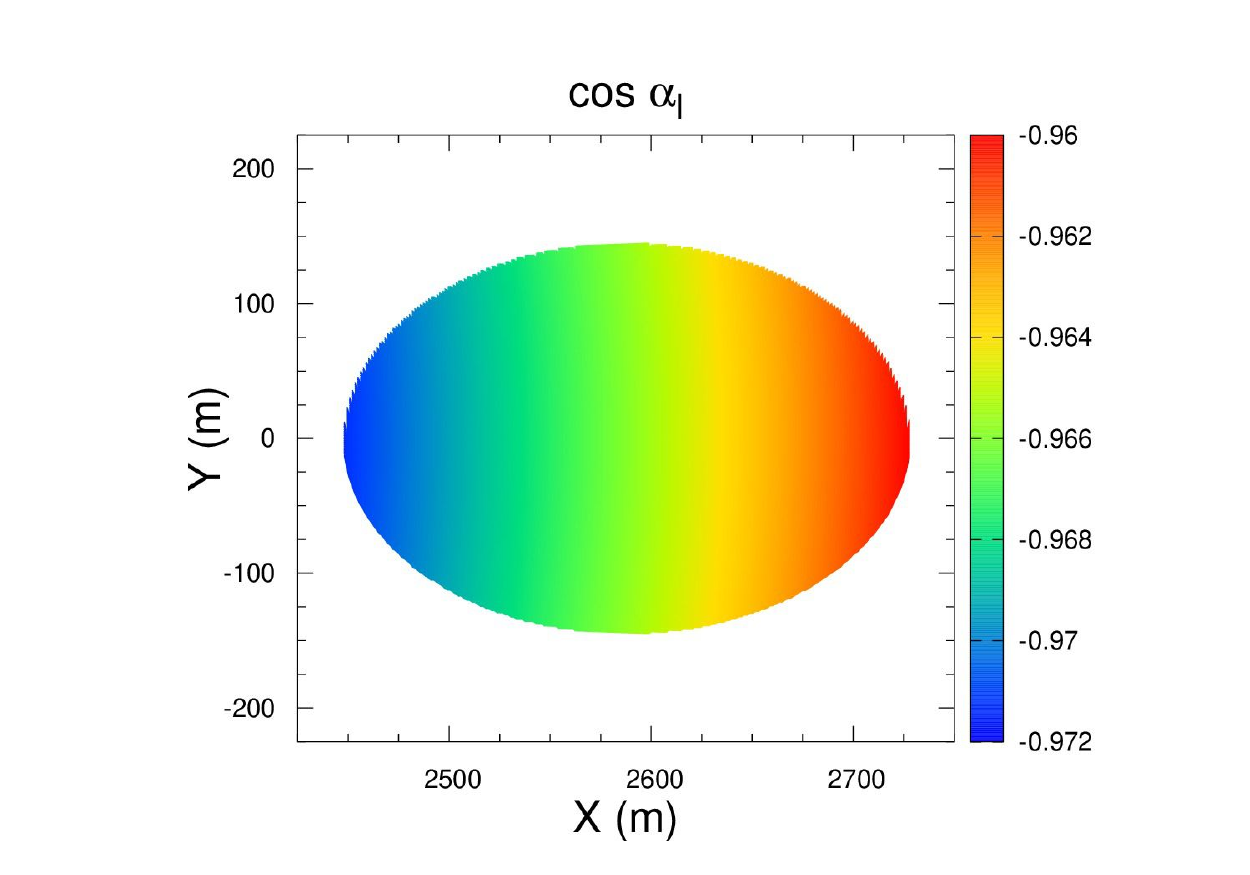}}} &
{\mbox{\includegraphics[scale=0.65,angle=0.]{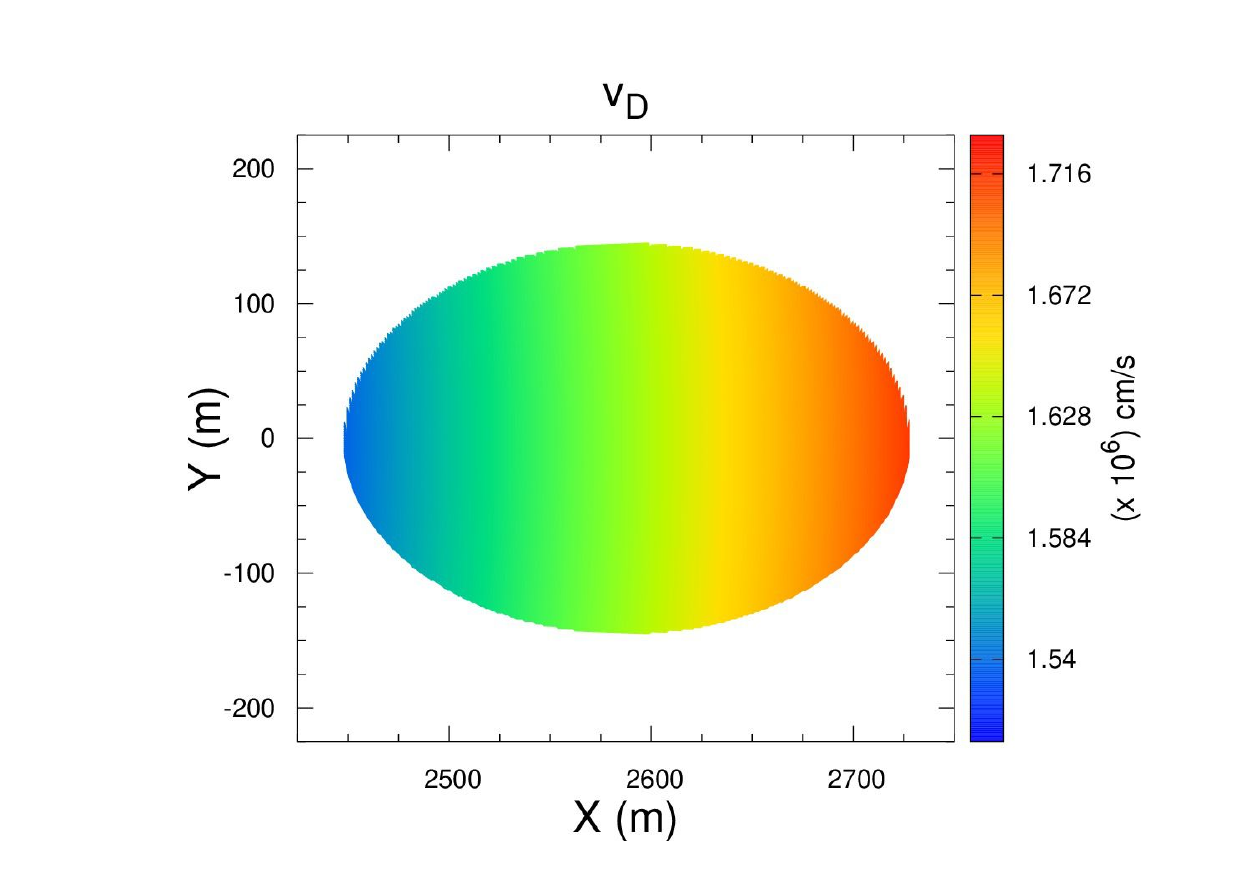}}} \\
\end{tabular}
\caption{The figure shows the estimates of different physical parameters, 
magnetic field intensity ($B_d$, top left), the radius of curvature of the 
field lines ($\rho_{curv}$, top right), the ratio between the components of the 
magnetic field in spherical coordinates ($B_{\theta}/B_r$, middle left and 
$B_{\phi}/B_r$, middle right), the inclination angle of surface magnetic field
with the rotation axis ($\cos{\alpha_l}$, bottom left) and the drift velocity 
of charges in presence of corotation electric field ($v_D$, bottom right) in a 
dipolar polar cap. The magnetic moment has an inclination angle $\theta_d$ = 
15\degr~and all quantities are estimated  at the surface of the neutron star, 
$R_S$ = 10 km, and the variations across the polar are represented by a colour 
scheme defined in each plot.}
\label{fig_appdip}
\end{figure*}

\begin{figure*}
\begin{tabular}{@{}cr@{}}
{\mbox{\includegraphics[scale=0.65,angle=0.]{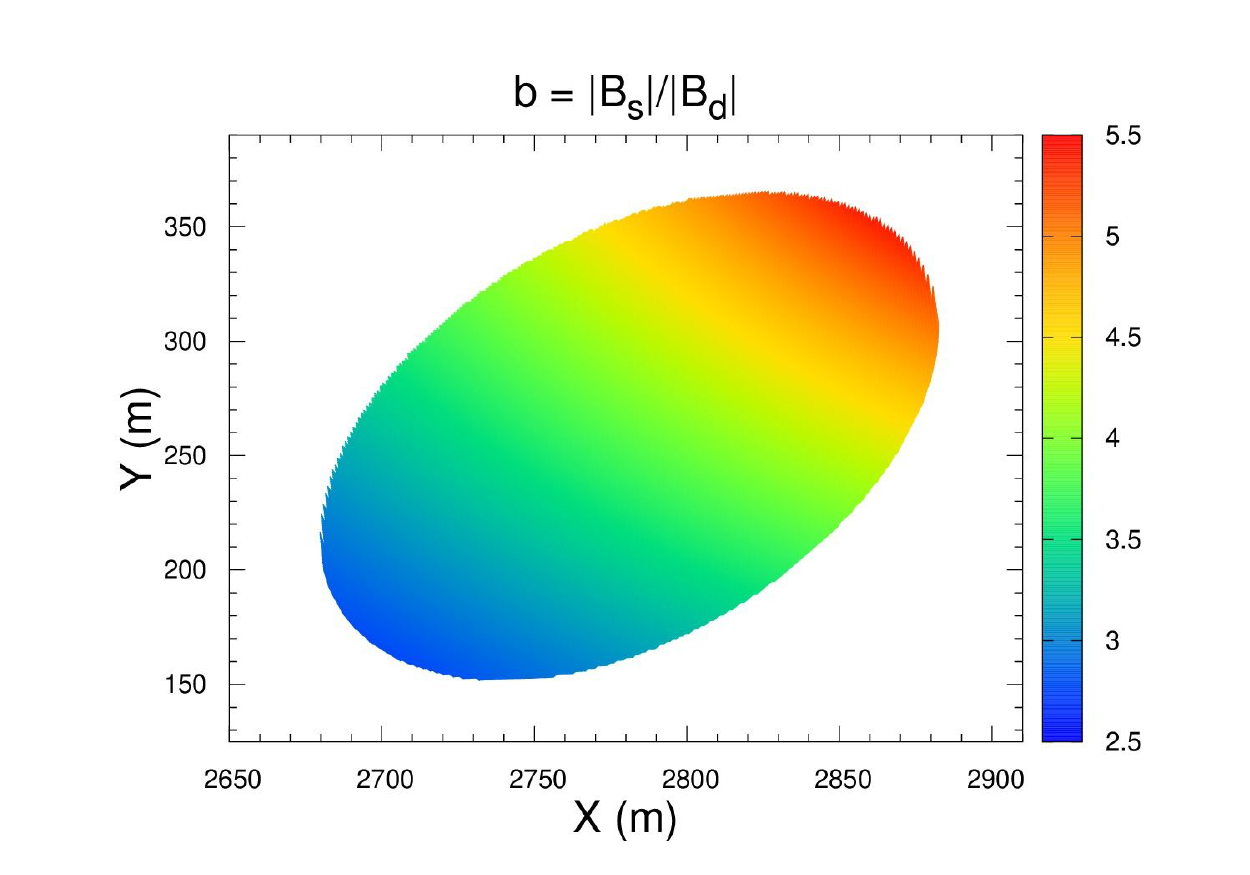}}} &
{\mbox{\includegraphics[scale=0.65,angle=0.]{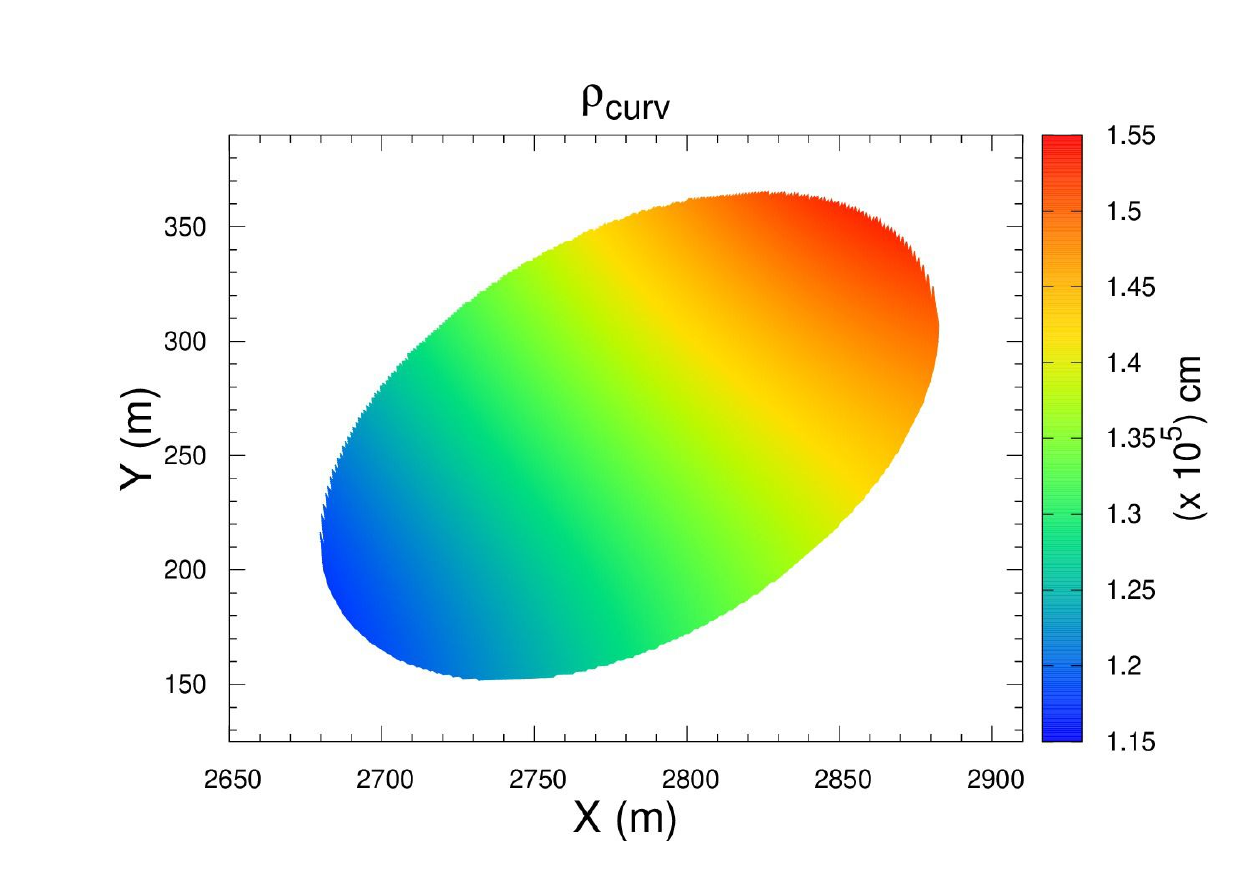}}} \\ 
{\mbox{\includegraphics[scale=0.65,angle=0.]{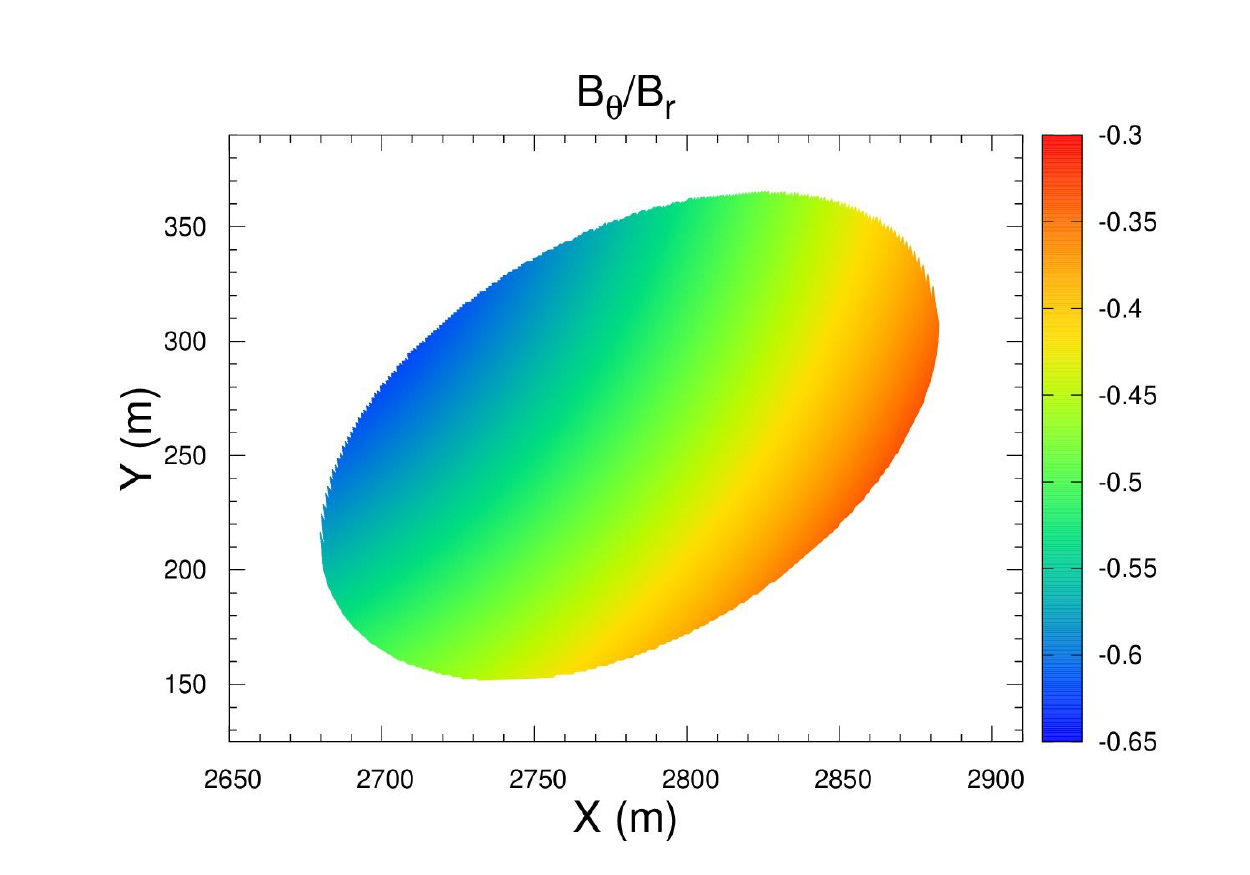}}} &
{\mbox{\includegraphics[scale=0.65,angle=0.]{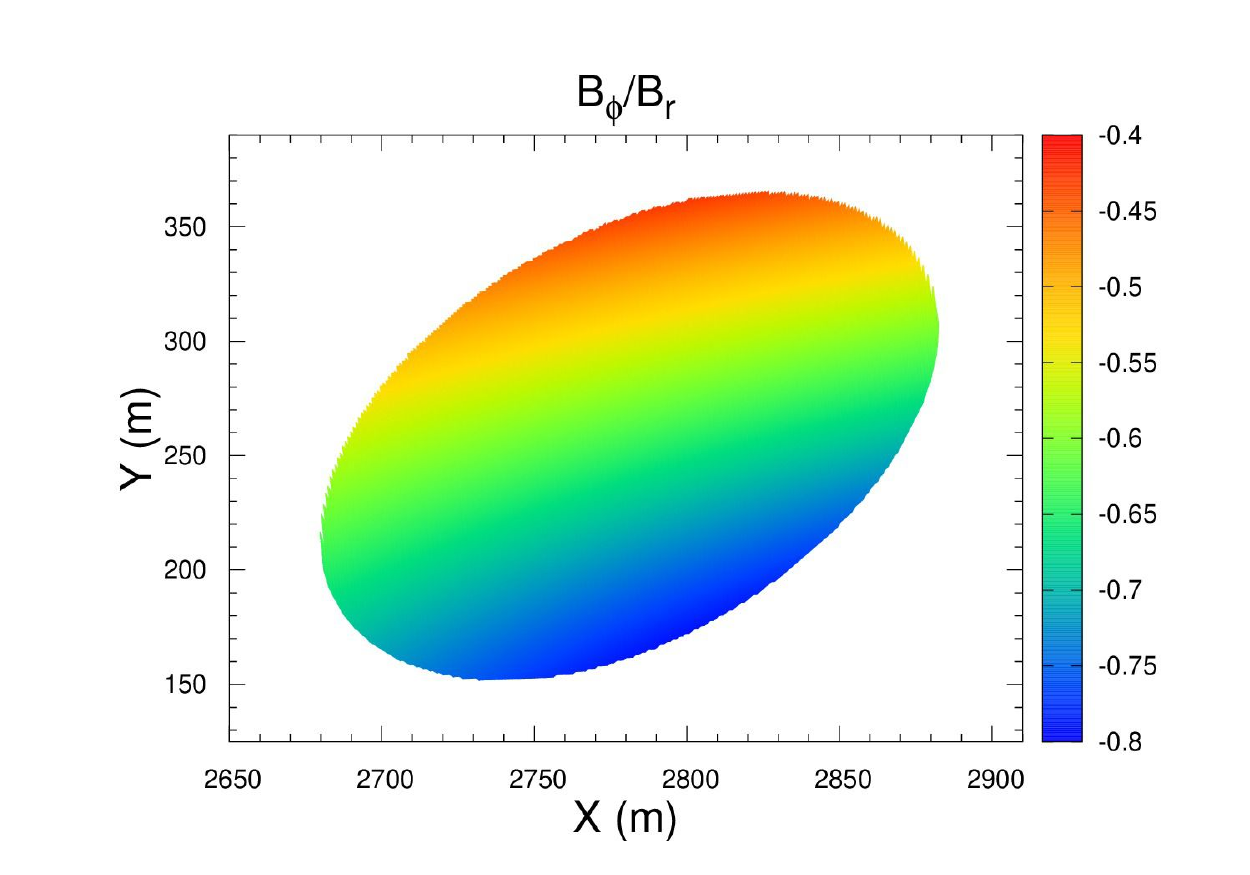}}} \\
{\mbox{\includegraphics[scale=0.65,angle=0.]{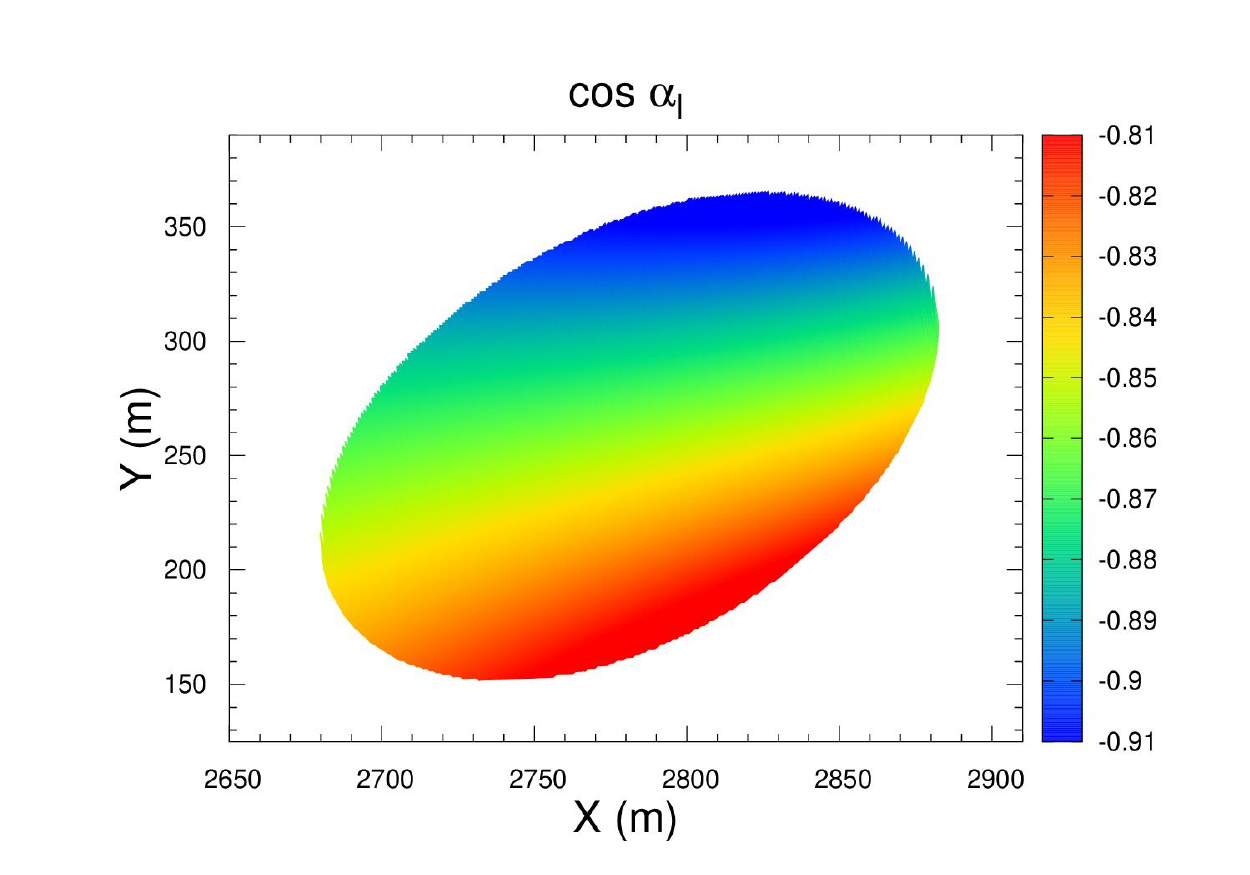}}} &
{\mbox{\includegraphics[scale=0.65,angle=0.]{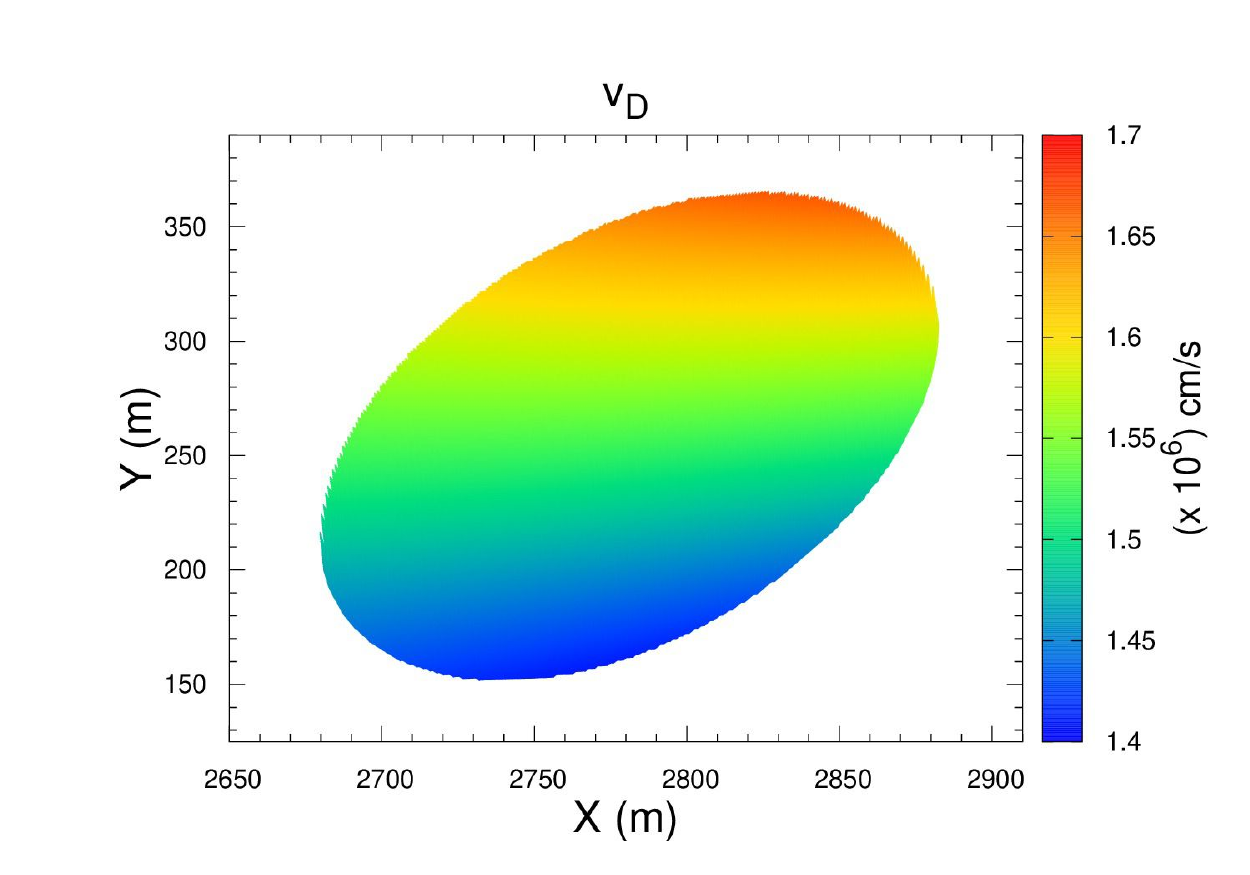}}} \\
\end{tabular}
\caption{The figure shows the estimates of different physical parameters in a 
non-dipolar polar cap consisting of a combination of a star centered dipole 
with inclination angle $\theta_d$ = 15\degr, and a dipole near the surface 
located at $\mathbf{r}_s$ = (0.95$R_S$, 18.86\degr, 10.99\degr) with magnetic 
moment $\mathbf{m}_s$ = (0.001$d$, 0\degr, 0\degr). The above magnetic field 
configuration results in coherent phase-modulated drifting. The different 
parameters are same as figure \ref{fig_appdip}, with the only exception being 
the top right panel which shows ratio between non-dipolar magnetic field 
intensity and equivalent dipolar field (b=$B_{s}/B_d$).}
\label{fig_appcoh}
\end{figure*}

\begin{figure*}
\begin{tabular}{@{}cr@{}}
{\mbox{\includegraphics[scale=0.65,angle=0.]{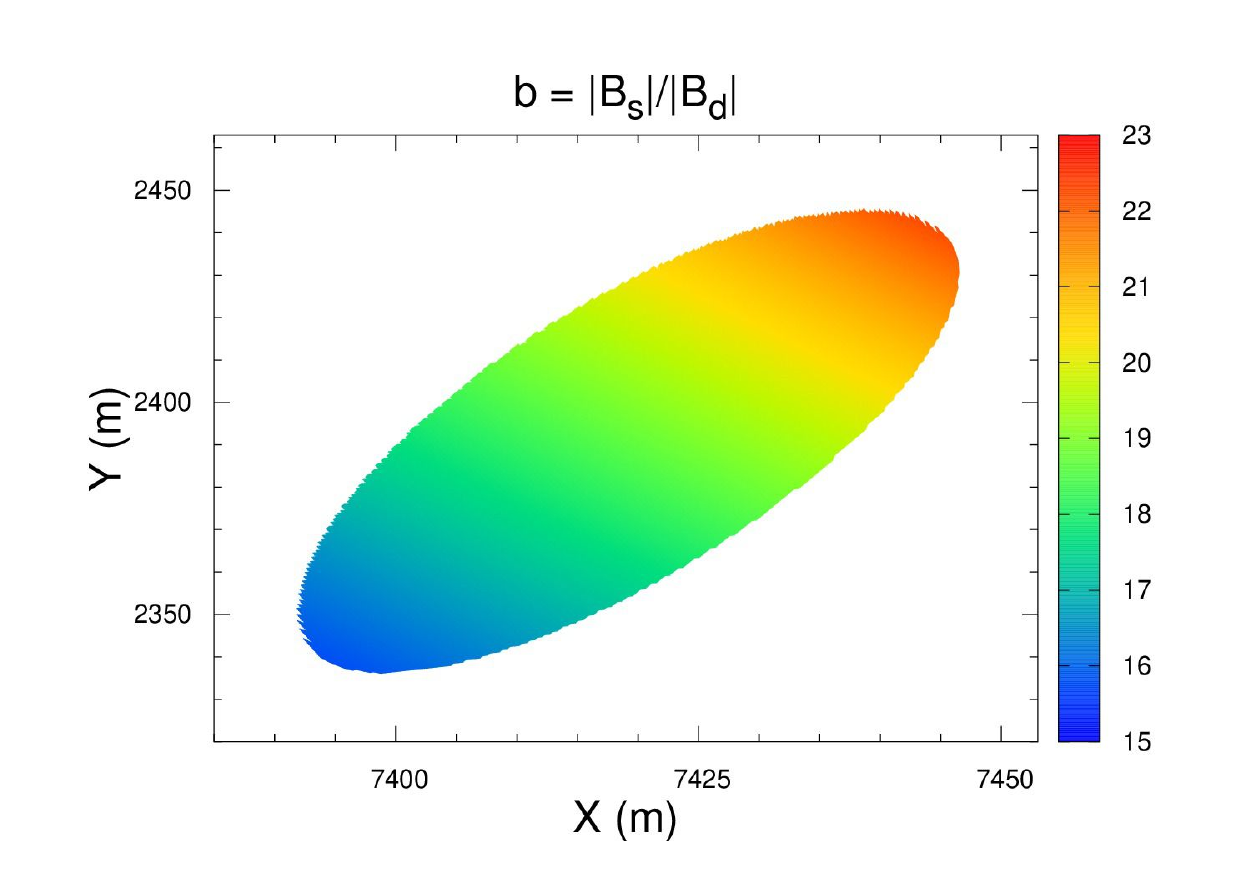}}} &
{\mbox{\includegraphics[scale=0.65,angle=0.]{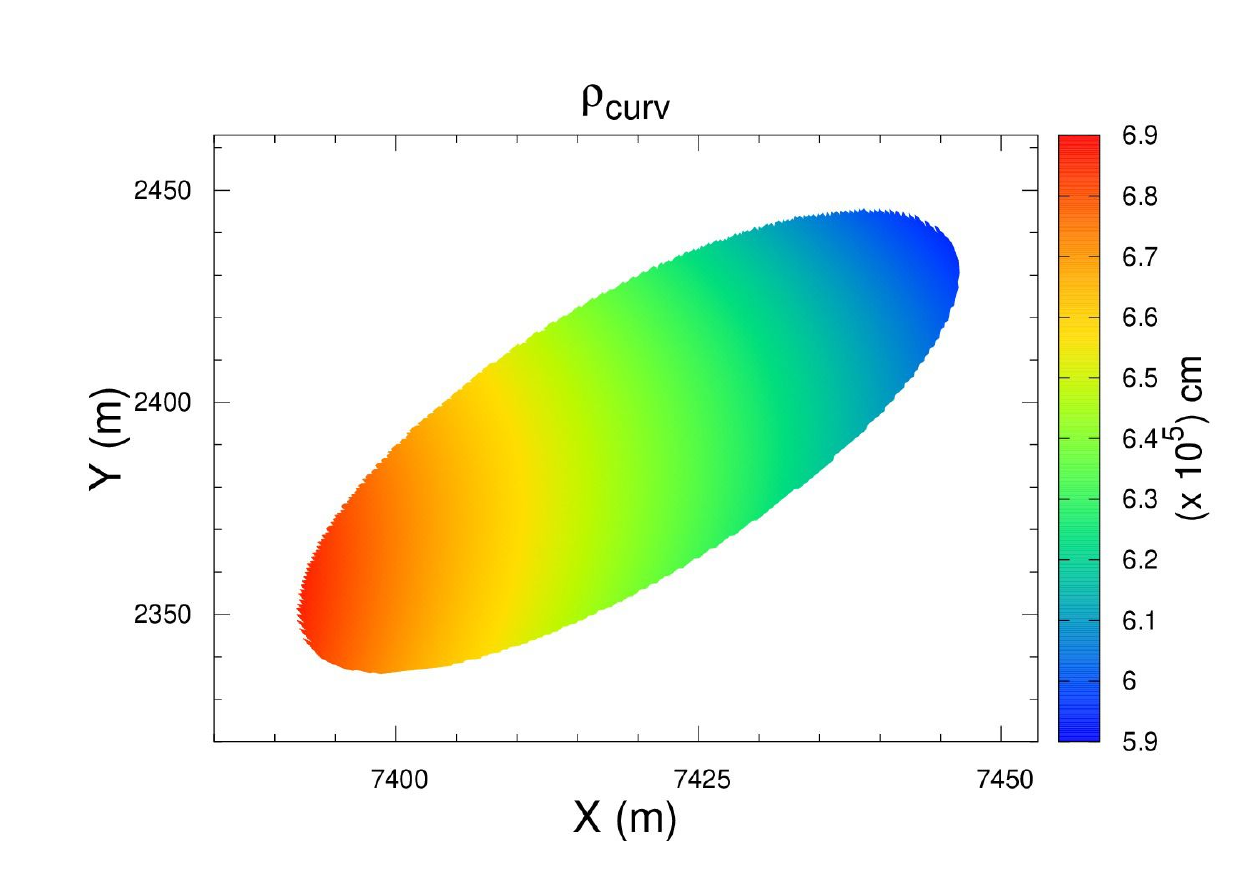}}} \\
{\mbox{\includegraphics[scale=0.65,angle=0.]{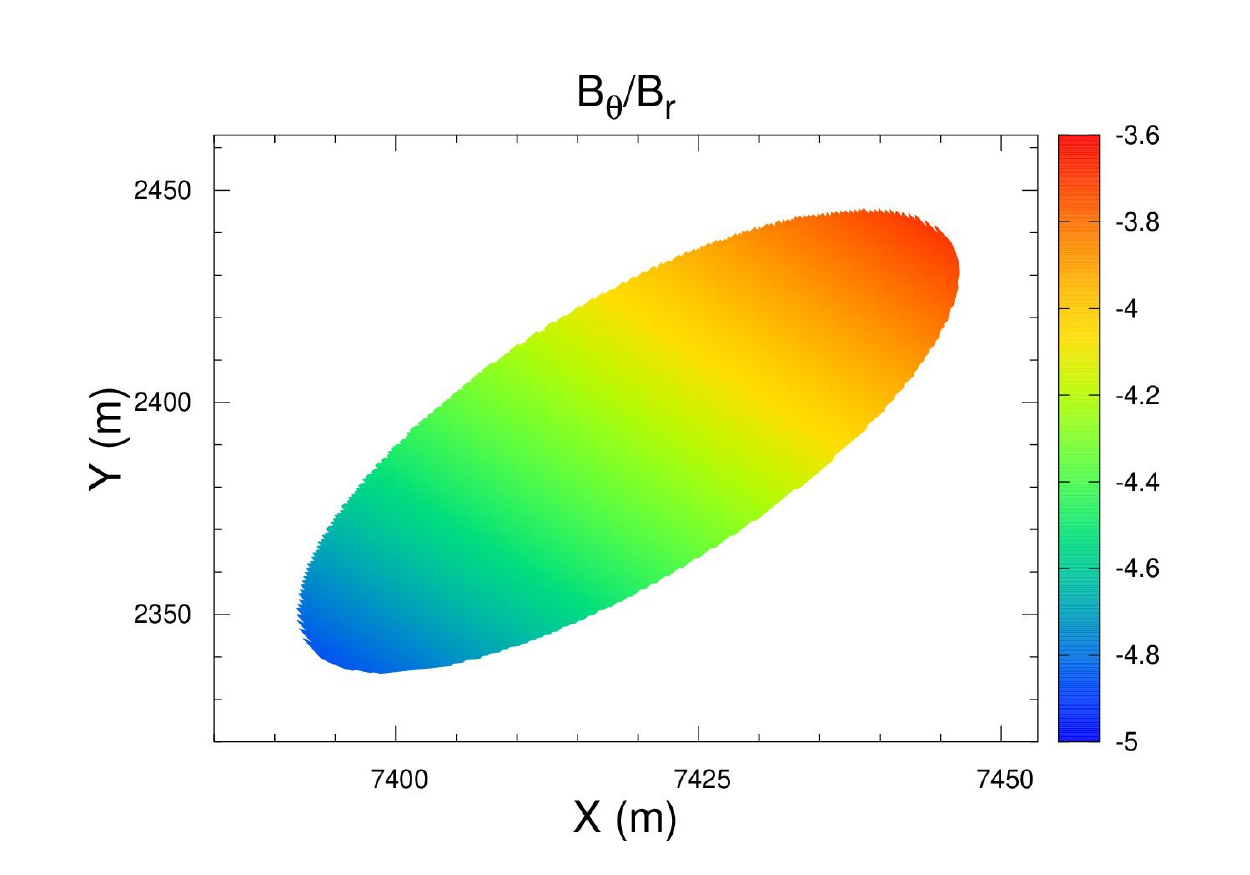}}} &
{\mbox{\includegraphics[scale=0.65,angle=0.]{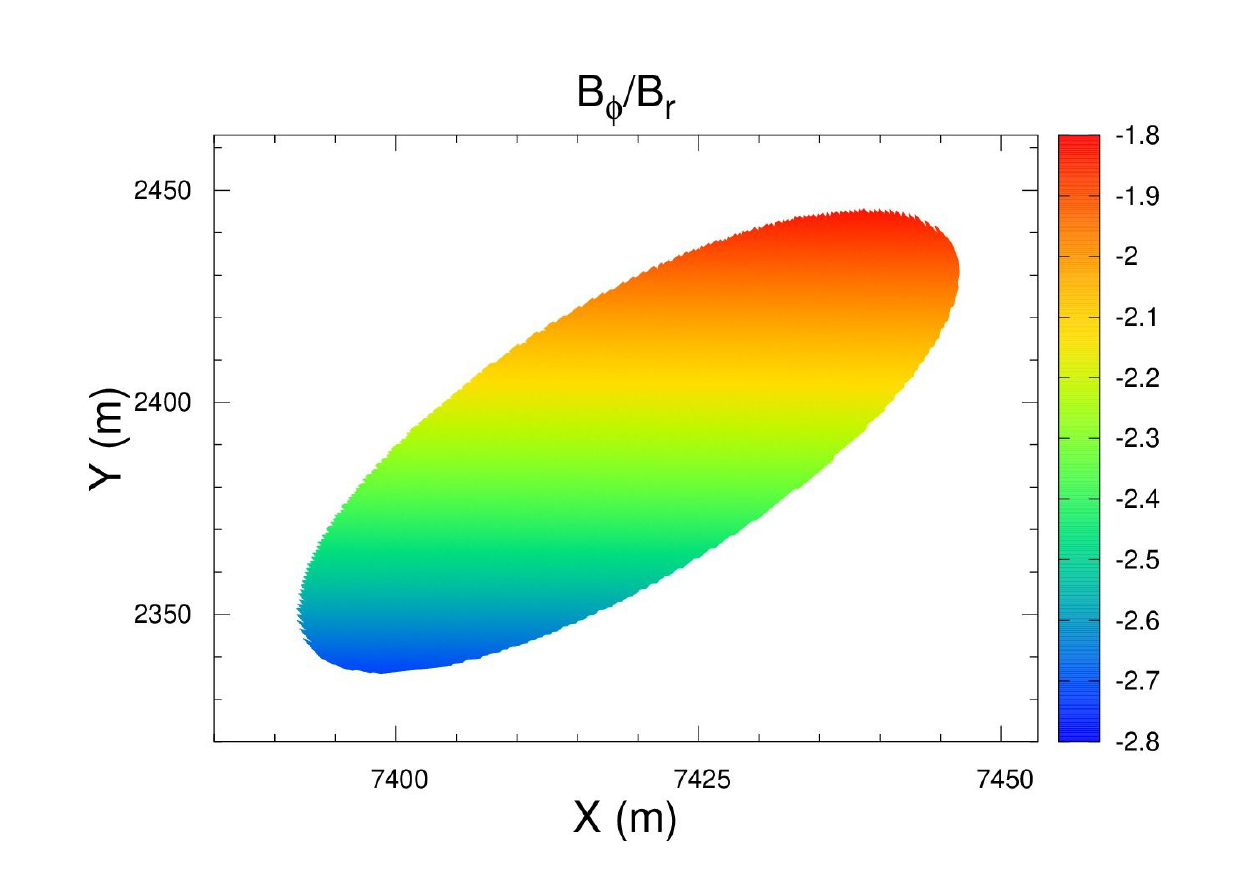}}} \\
{\mbox{\includegraphics[scale=0.65,angle=0.]{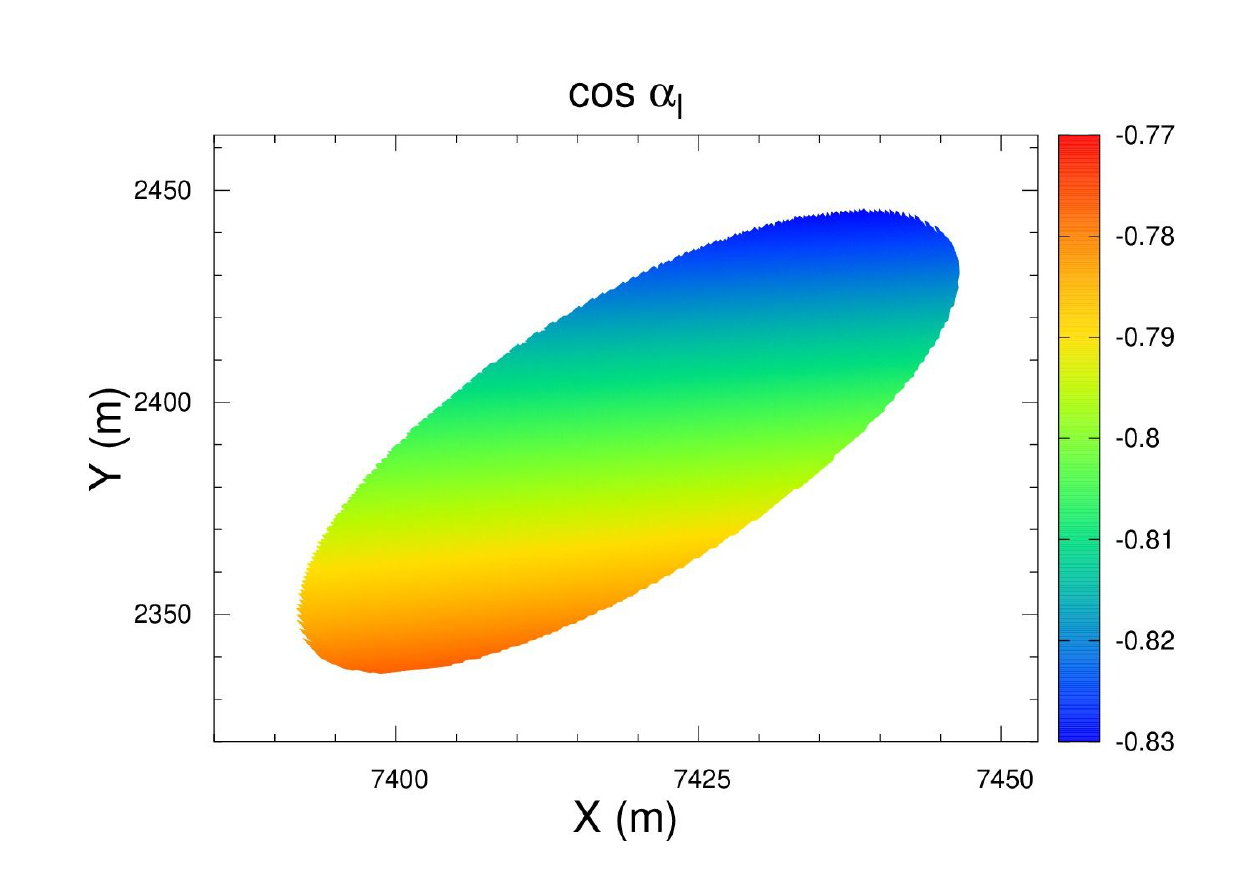}}} &
{\mbox{\includegraphics[scale=0.65,angle=0.]{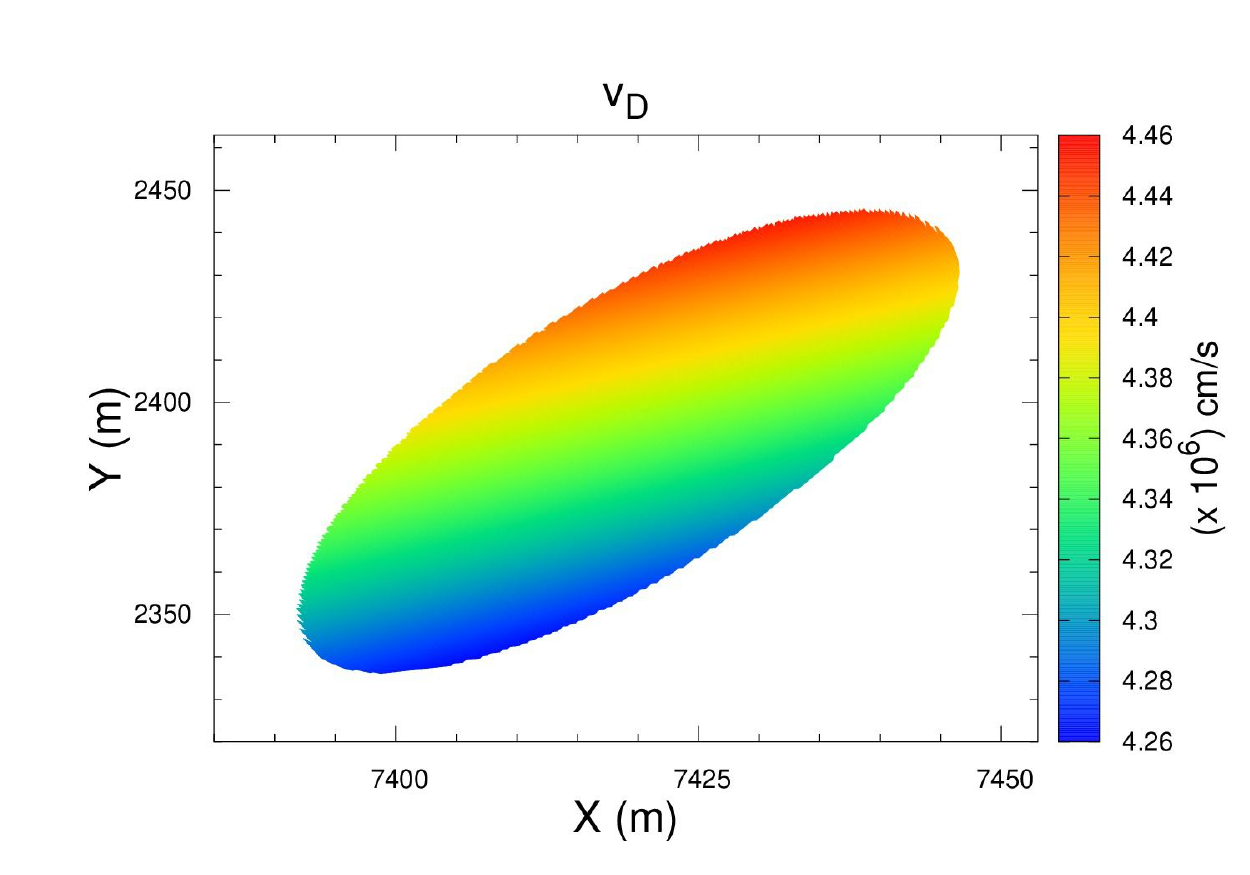}}} \\
\end{tabular}
\caption{The figure shows the estimates of different physical parameters in a 
non-dipolar polar cap consisting of a combination of a star centered dipole 
with inclination angle $\theta_d$ = 45\degr, and a dipole near the surface 
located at $\mathbf{r}_s$ = (0.95$R_S$, 57.08\degr, 20.66\degr) with magnetic
moment $\mathbf{m}_s$ = (0.05$d$, 0\degr, 0\degr). 
The above magnetic field configuration results in low-mixed phase-modulated 
drifting. The different parameters are same as in figure \ref{fig_appcoh}.}
\label{fig_applow}
\end{figure*}

\begin{figure*}
\begin{tabular}{@{}cr@{}}
{\mbox{\includegraphics[scale=0.65,angle=0.]{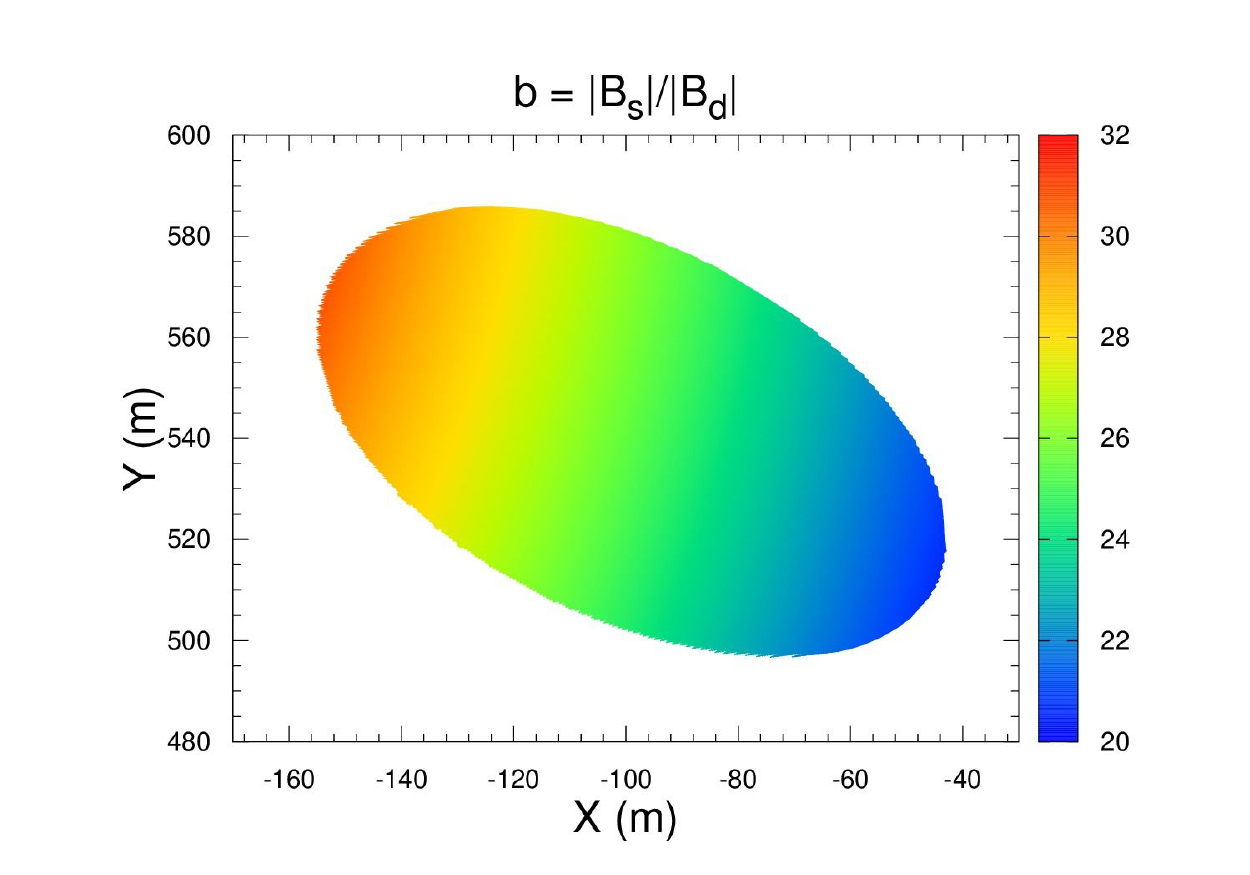}}} &
{\mbox{\includegraphics[scale=0.65,angle=0.]{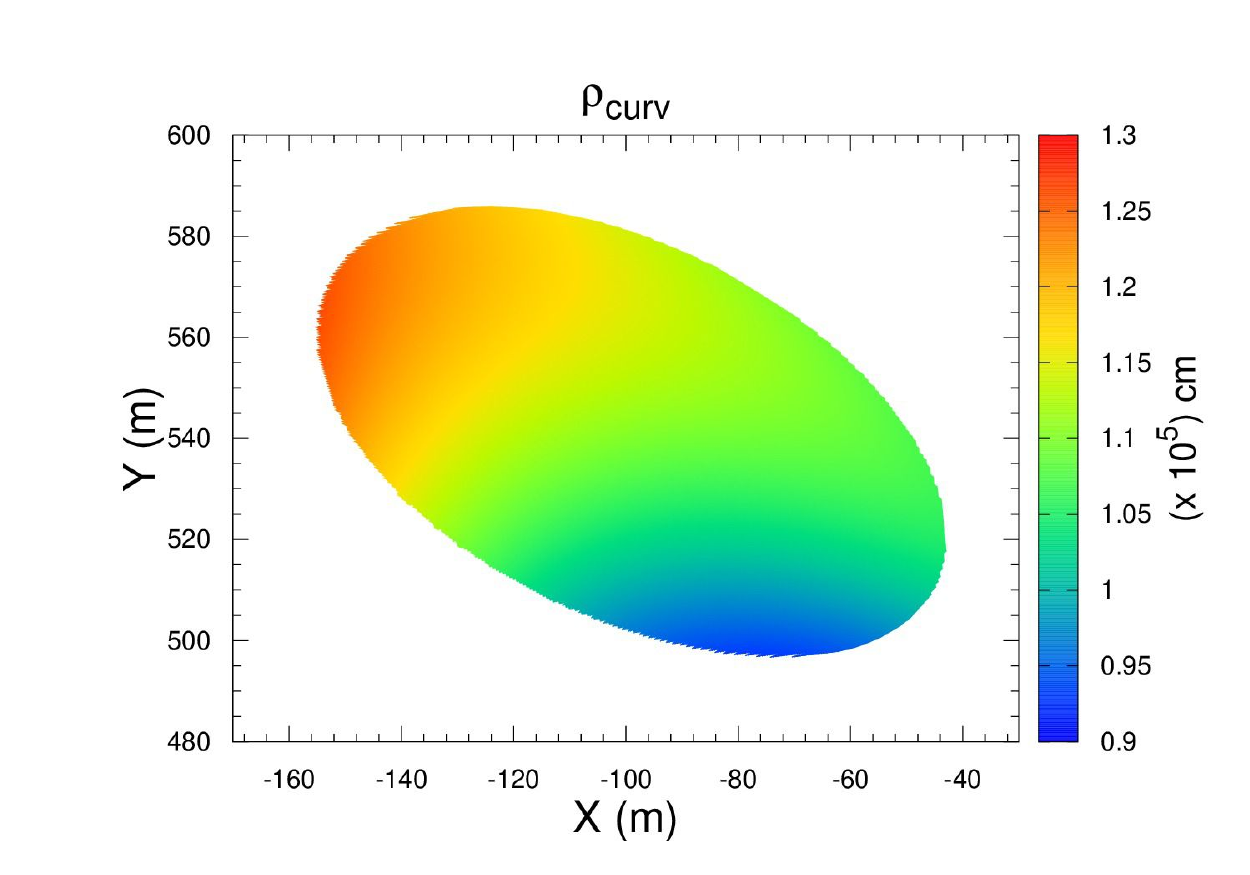}}} \\
{\mbox{\includegraphics[scale=0.65,angle=0.]{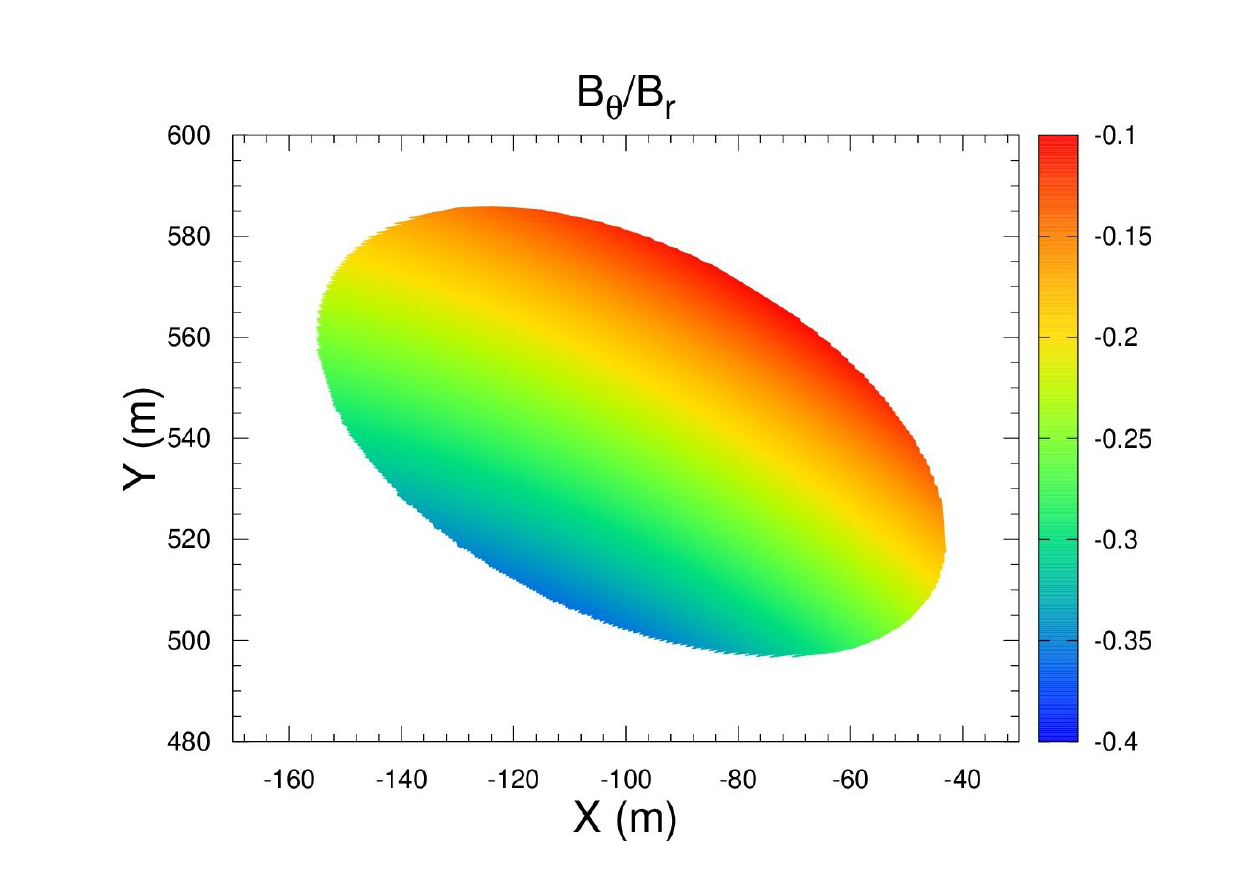}}} &
{\mbox{\includegraphics[scale=0.65,angle=0.]{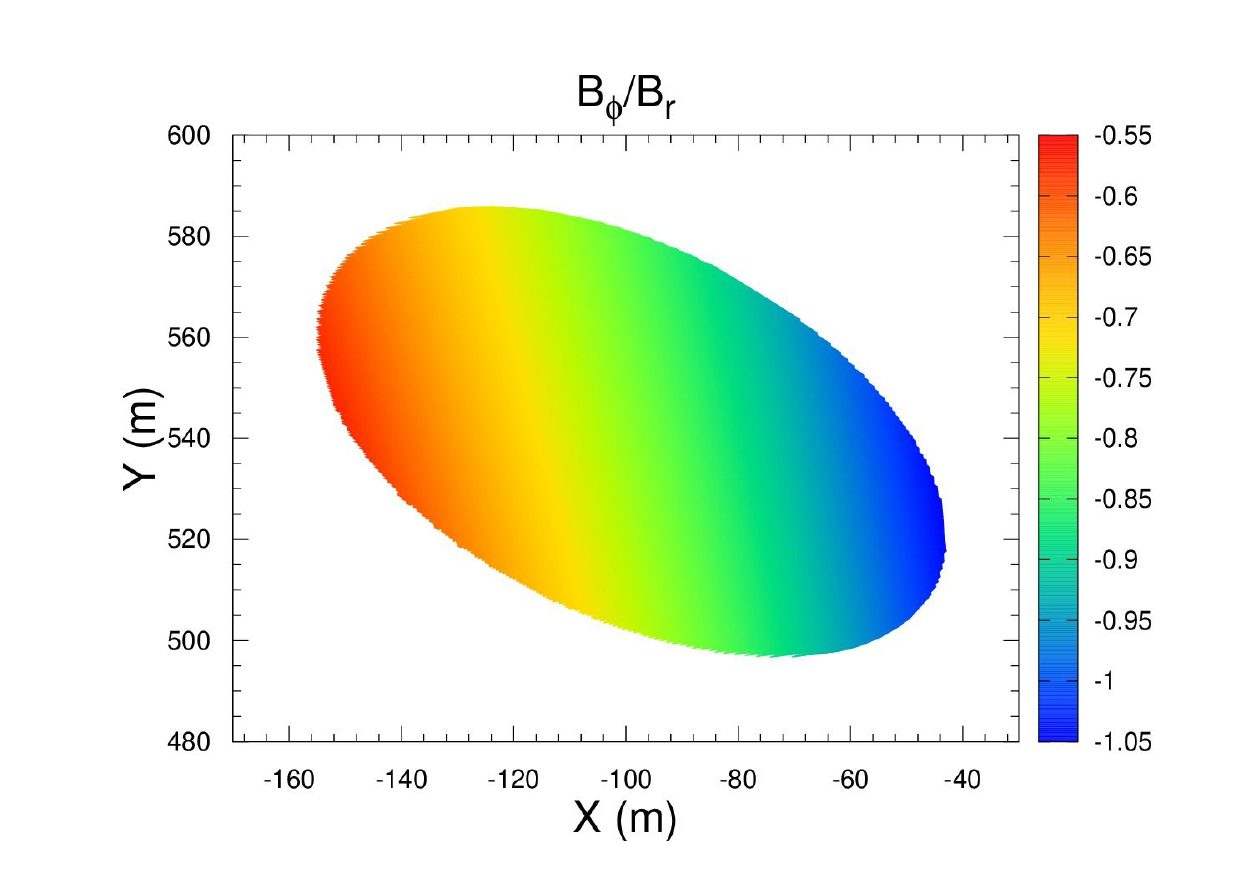}}} \\
{\mbox{\includegraphics[scale=0.65,angle=0.]{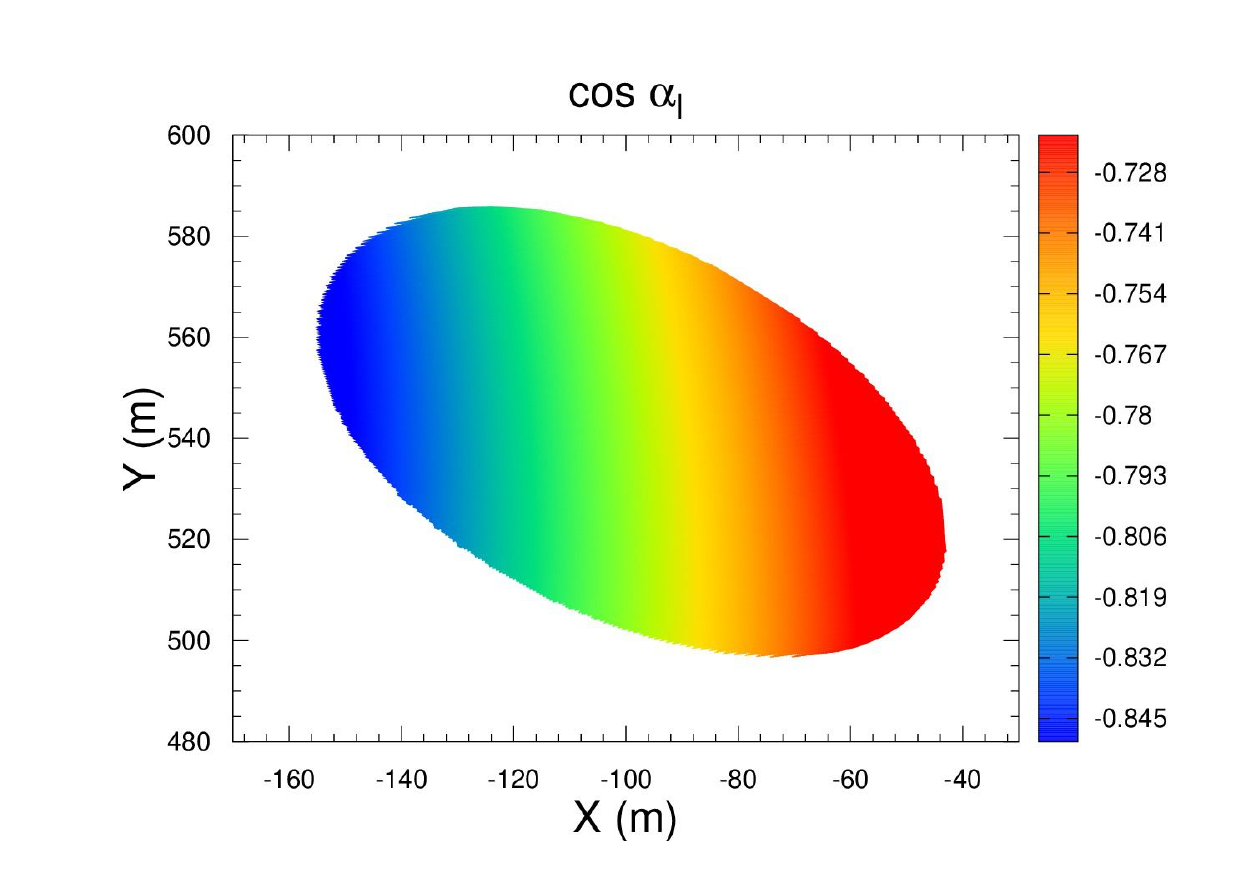}}} &
{\mbox{\includegraphics[scale=0.65,angle=0.]{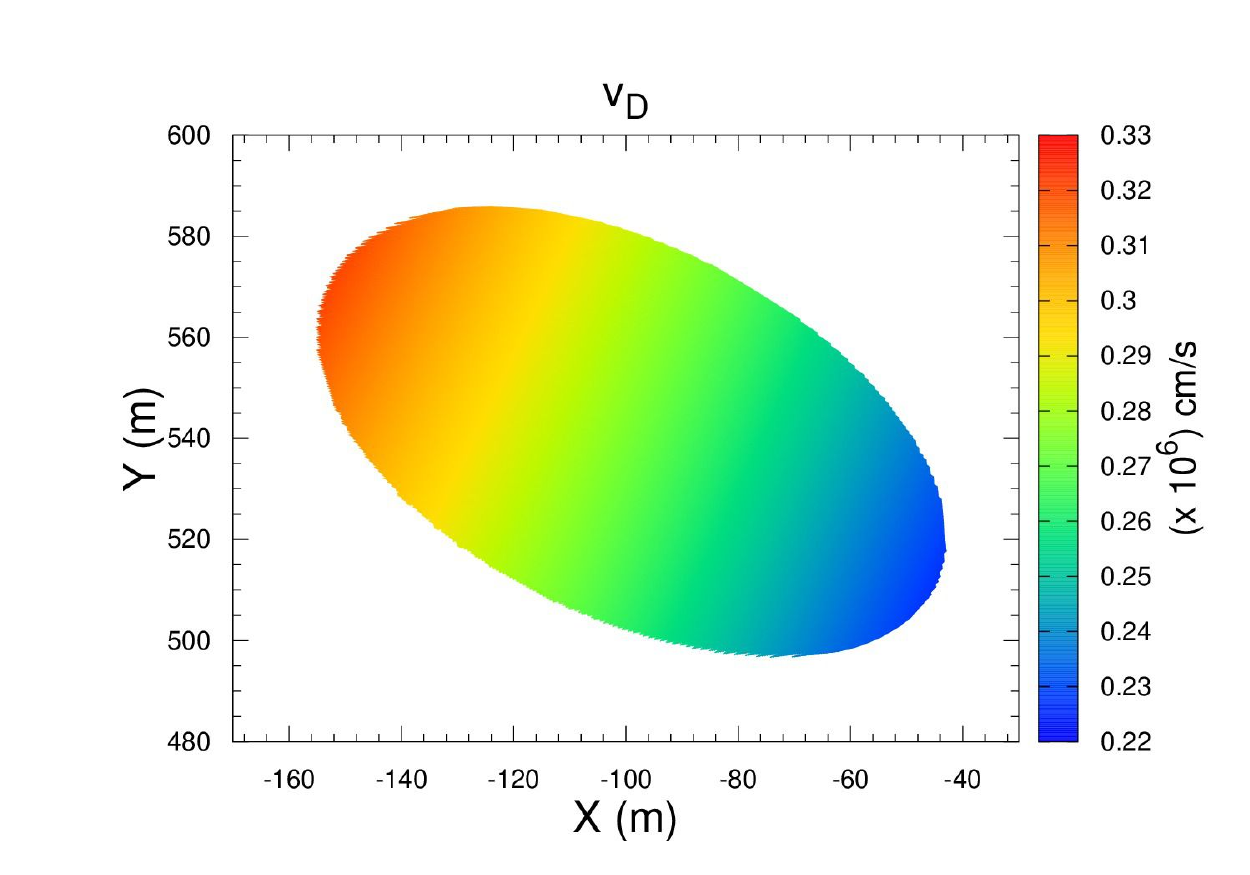}}} \\
\end{tabular}
\caption{The figure shows the estimates of different physical parameters in a 
non-dipolar polar cap consisting of a combination of a star centered dipole 
with inclination angle $\theta_d$ = 5\degr, and a dipole near the surface 
located at $\mathbf{r}_s$ = (0.95$R_S$, 5\degr, 120\degr) with magnetic
moment $\mathbf{m}_s$ = (0.005$d$, 0\degr, 0\degr). The above magnetic field 
configuration results in switching phase-modulated drifting. The different 
parameters are same as in figure \ref{fig_appcoh} and \ref{fig_applow}.}
\label{fig_appswi}
\end{figure*}

The estimates of the different physical parameters presented in the appendix 
\ref{app:magcon} and \ref{app:radcurv}, and is applied to the non-dipolar polar 
cap configurations used in section \ref{sec:driftsiml} to explain the different
drifting behaviour. These include, the ratio between the non-dipolar surface 
field with the equivalent dipolar field defined as $b=B_s/B_d$, the radius of 
curvature, $\rho_c$, shown in appendix \ref{app:radcurv}, the ratio between the
angular components of the local magnetic field with the radial component, 
$B_{\theta}/B_r$ and $B_{\phi}/B_r$, shown in eq.(A1) to eq.(A4), the angle 
$\cos{\alpha_l}$ between the local magnetic field and the z-axis, and the drift
velocity due to corotation, $\mathbf{v}^c_D$, using eq.(\ref{eq_corotvel}). In 
addition, we have also estimated these physical parameters for a purely dipolar
polar cap configuration for comparisons, with the exception of estimating $B_d$
instead of $b$, which is identically 1 by definition in this case. In figure 
\ref{fig_appdip} we present variations of the different quantities across the 
polar cap for a star centered dipole with inclination angle $\theta_d=$15\degr.
Figure \ref{fig_appcoh} shows the respective parameters for a non-dipolar polar
cap defined by a star centered dipole with $\theta_d=$15\degr~and a near 
surface counterpart located at $\mathbf{r}_s$ = (0.95$R_S$, 18.86\degr, 
10.99\degr) with magnetic moment $\mathbf{m}_s$ = (0.001$d$, 0\degr, 0\degr). 
This polar cap configuration is responsible for the coherent phase-modulated 
drifting described in section \ref{sec:cohdrift}. In figure \ref{fig_applow} a 
different non-dipolar polar cap is shown consisting of a star centered dipole 
with $\theta_d=$45\degr~and a near surface counterpart located at 
$\mathbf{r}_s$ = (0.95$R_S$, 57.08\degr, 20.66\degr) with magnetic moment 
$\mathbf{m}_s$ = (0.05$d$, 0\degr, 0\degr). This magnetic configuration is used
to simulate the low-mixed phase modulated drifting in section 
\ref{sec:lowmixdrift}. Finally, in figure \ref{fig_appswi} we show the 
variations of the physical parameters for the non-dipolar polar cap 
corresponding to the switching phase-modulated drifting explored in section 
\ref{sec:swiphsdrift}. The magnetic field configuration consists of a star 
centered dipole with $\theta_d=$5\degr~and a near surface dipole located at 
$\mathbf{r}_s$ = (0.95$R_S$, 5\degr, 120\degr) with magnetic moment 
$\mathbf{m}_s$ = (0.005$d$, 0\degr, 0\degr). All quantities are calculated on
the surface of the neutron star with $R_S$ = 10 km and the variations are 
expressed using a suitable colour scale in the X-Y plane, since the fractional 
extent of the polar cap is least along the z-axis.

Clear differences emerge between the purely dipolar and non-dipolar polar caps 
from these estimates. In case of the dipolar polar cap the magnetic field show
very little variations across the polar cap with less than hundredth of a 
percentage difference from the edge to the center. The configuration has an 
extremum at the field center which is circularly symmetric. In contrast the 
non-dipolar polar caps show much larger variations in $b\sim50-100$\%, which 
changes monotonically from one end of the polar cap to the other. The radius of
curvature is also vastly different between the dipolar polar cap and other 
non-dipolar cases. In the purely dipolar case it shows large variations from 
the edge of the polar cap ($\sim10^8$ cm) to the central regions and is 
circularly symmetric around the magnetic axis. In non-dipolar polar caps the 
radius of curvature varies monotonically from one end to the other with maximum
change around 20-30\%. In all cases the estimated $\rho_c\sim10^5-10^6$ cm, 
which is the preferred curvature for efficient pair production in IAR and 
consequently the sparking process to be sustained (see RS75). The estimates of 
the magnetic field strength and $\rho_c$ across the non-dipolar polar cap 
further highlights that the sparking process can commence at any location with 
identical properties in a realistic polar cap, governed by local thermal 
conditions of the PSG. The angular components of the magnetic field are much 
more prominent in case of non-dipolar polar caps, where $B_{\theta}/B_r$ and 
$B_{\phi}/B_r$ can be several times the radial component. In contrast the 
angular components of the magnetic field in the dipolar polar cap is less than 
1\% of the radial component. A number of studies \citep{2013arXiv1304.4203S,
2020MNRAS.492.2468M} have evaluated the charge motion within the IAR based on 
analytical estimates, with the assumption that the angular components of the 
magnetic field is negligible in the IAR. The large contribution of angular 
components of the magnetic field in non-dipolar polar caps indicate that these 
analytical estimates of particle motion are inadequate and more detailed 
simulations are necessary for studying the rigorous evolution of any sparking 
process. Finally, the estimates of the drift speeds and $\cos{\alpha_l}$ show 
that although they have large diversity for different magnetic field 
configurations, they vary around 10\% across any particular polar cap. Thus 
even in our non-rigorous estimate of sparks lagging behind corotation speed 
with a fraction of $v_D$ specified by $\eta$, the drifting speed across the 
polar cap show maximum 10\% variations. 

\section*{Acknowledgments}
We thank the referee for detailed comments which improved the paper. DM 
acknowledges support and funding from the `Indo-French Centre for the Promotion
of Advanced Research - CEFIPRA' grant IFC/F5904-B/2018.

\bibliographystyle{mn2e}
\bibliography{Driftsiml}

\clearpage
%%%\newpage
%%\pagbreak[four]
%\setcounter{page}{1}
%%\fancyhf{}

\appendix
\section{The surface magnetic field} \label{app:magcon}
The basic estimates of the surface magnetic field configurations from a 
superposition of the star-centered global dipole and the crust-anchored dipole 
moment has been presented in \citet{2002A&A...388..235G}. The primary 
calculations were limited to the 2-dimensional representations. In this 
appendix we provide the more detailed 3-dimensional calculations. We also 
provide estimates for a general configuration of more than one crust-anchored 
dipole moments. The star centered global dipole with dipole moment $\mathbf{d}$
and magnetic inclination angle $\theta_d$, has the magnetic field at a point 
$\mathbf{r} = (r, \theta, \phi)$ given as :
\begin{equation} \label{eq_global}
\begin{split}
 B_r^d & = \frac{2d}{r^3}(\sin{\theta_d}\sin{\theta}\cos{\phi} + \cos{\theta_d}\cos{\theta}) \\
 B_{\theta}^d & = -\frac{d}{r^3}(\sin{\theta_d}\cos{\theta}\cos{\phi} - \cos{\theta_d}\sin{\theta}) \\
 B_{\phi}^d & = \frac{d}{r^3}\sin{\theta_d}\sin{\phi}
\end{split}
\end{equation}
In a more generalised situation the surface fields are non-dipolar in nature 
and can be approximated by one or more crust-anchored dipoles. Assuming the 
possibility of a series of $N$ dipoles with dipole moment $\mathbf{m}_i = 
(m^i, \theta_m^i, \phi_m^i)$ at location $\mathbf{r}_i = (r^i, \theta_r^i, 
\phi_r^i)$ where $i = 1, 2, ..., N$; the components of the radius-vector and 
the local dipole moment for the $i^{th}$ dipole can be expressed as :
\begin{equation} \label{eq_crustrad}
\begin{split}
r_r^i & = r^i(\sin{\theta_r^i}\sin{\theta}\cos{(\phi-\phi_r^i)} + \cos{\theta_r^i}\cos{\theta}),\\
r_{\theta}^i & = r^i(\sin{\theta_r^i}\cos{\theta}\cos{(\phi-\phi_r^i)} - \cos{\theta_r^i}\sin{\theta}), \\
r_{\phi}^i & = -r^i\sin{\theta_r^i}\sin{(\phi-\phi_r^i)}.
\end{split}
\end{equation}
\begin{equation} \label{eq_crustmom}
\begin{split}
m_r^i & = m^i(\sin{\theta_m^i}\sin{\theta}\cos{(\phi-\phi_m^i)} + \cos{\theta_m^i}\cos{\theta}), \\
m_{\theta}^i & = m^i(\sin{\theta_m^i}\cos{\theta}\cos{(\phi-\phi_m^i)} - \cos{\theta_m^i}\sin{\theta}), \\
m_{\phi}^i & = -m^i\sin{\theta_m^i}\sin{(\phi-\phi_m^i)}.
\end{split}
\end{equation}
The above expressions are used to define
\begin{equation*}
\begin{split}
D_i & = (r^i)^2 + r^2 - 2r^ir(\sin{\theta_r^i}\sin{\theta}\cos{(\phi-\phi_r^i)} + \cos{\theta_r^i}\cos{\theta}), \\ 
T_i & = m_r^ir - (m_r^ir_r^i + m_{\theta}^ir_{\theta}^i + m_{\phi}^ir_{\phi}^i).
\end{split}
\end{equation*}
The magnetic field $\mathbf{B}_i = (B_r^i, B_{\theta}^i, B_{\phi}^i)$ for the 
$i^{th}$ dipole can be characterised as follows:
\begin{equation} \label{eq_crustmag}
\begin{split}
B_r^i & = -\frac{1}{D_i^{2.5}}(3T_ir_r^i - 3T_ir + D_im_r^i), \\
B_{\theta}^i & = -\frac{1}{D_i^{2.5}}(3T_ir_{\theta}^i + D_im_{\theta}^i), \\
B_{\phi}^i & = -\frac{1}{D_i^{2.5}}(3T_ir_{\phi}^i + D_im_{\phi}^i).
\end{split}
\end{equation}
The magnetic field can be used to determine the field line of force which are 
obtained in spherical coordinate system by solving the differential equations :
\begin{equation}\label{eq_fieldline}
\begin{split}
\frac{\mathrm{d}\theta}{\mathrm{d}r} & = \frac{B_{\theta}^d + \sum\limits^N_{i=1}B_{\theta}^i}{r\left(B_r^d + \sum\limits^N_{i=1}B_r^i\right)} ~~\equiv ~~\Theta_1, \\
\frac{\mathrm{d}\phi}{\mathrm{d}r} & = \frac{B_{\phi}^d + \sum\limits^N_{i=1}B_{\phi}^i}{r\left(B_r^d + \sum\limits^N_{i=1}B_r^i\right)\sin{\theta}} ~~\equiv ~~\Phi_1.
\end{split}
\end{equation}

%\section{Estimating Physical parameters}

\section{Curvature in Magnetic field} \label{app:radcurv}
Finally, we present a detailed calculation of the curvature of the magnetic 
field lines which is important for the pair production process in the IAR (see 
RS75). Following the specifications of \citet{2002A&A...388..235G} the 
curvature $\Re$ (= 1/$\rho_c$, $\rho_c$ being the radius of curvature) is given
as :

\begin{equation}\label{eq_curv1}
\Re = \left(\frac{\mathrm{d}s}{\mathrm{d}r}\right)^{-3}  \left\vert \left(\frac{\mathrm{d}^2\mathbf{r}}{\mathrm{d}r^2}\frac{\mathrm{d}s}{\mathrm{d}r} - \frac{\mathrm{d}\mathbf{r}}{\mathrm{d}r}\frac{\mathrm{d}^2 s}{\mathrm{d}r^2}\right) \right\vert
\end{equation}
which gives $\Re = (S_1)^{-3} (J_1^2 + J_2^2 + J_3^2)^{1/2}$. The expressions 
are specified as $J_1 = X_2 S_1 - X_1 S_2$; $J_2 = Y_2 S_1 - Y_1 S_2$ and $J_3 
= Z_2 S_1 - Z_1 S_2$, where
\begin{equation}\label{eq_curv2}
\begin{split}
X_1 & = \sin{\theta}\cos{\phi} + r\Theta_1\cos{\theta}\cos{\phi} - r\Phi_1\sin{\theta}\sin{\phi},\\
Y_1 & = \sin{\theta}\sin{\phi} + r\Theta_1\cos{\theta}\sin{\phi} + r\Phi_1\sin{\theta}\cos{\phi}, \\
Z_1 & = \cos{\theta} - r\Theta_1\sin{\theta}, \\
X_2 & = (2\Theta_1 + r\Theta_2)\cos{\theta}\cos{\phi} - (2\Phi_1 + r\Phi_2)\sin{\theta}\sin{\phi} \\ 
    & ~~ - r (\Theta_1^2 + \Phi_1^2)\sin{\theta}\cos{\phi} - 2r\Theta_1\Phi_1\cos{\theta}\sin{\phi},\\
Y_2 & = (2\Theta_1 + r\Theta_2)\cos{\theta}\sin{\phi} + (2\Phi_1 + r\Phi_2)\sin{\theta}\cos{\phi} \\
    & ~~ - r(\Theta_1^2 + \Phi_1^2)\sin{\theta}\sin{\phi} + 2r\Theta_1\Phi_1\cos{\theta}\cos{\phi}, \\
Z_2 & = -2\Theta_1\sin{\theta} - r\Theta_2\sin{\theta}-r\Theta_1^2\cos{\theta}.
\end{split}
\end{equation}
Here $\Theta_1 = \mathrm{d}\theta/\mathrm{d}r$ and $\Phi_1 = 
\mathrm{d}\phi/\mathrm{d}r$ are specified in Eq.(\ref{eq_fieldline}). 
Additionally, $S_1 = \mathrm{d}s/\mathrm{d}r$ and $S_2 = \mathrm{d}^2s/\mathrm{d}r^2$ are given as :
\begin{equation}\label{eq_curv3}
\begin{split}
S_1 & = (1 + r^2\Theta_1^2 + r^2\Phi_1^2\sin^2{\theta})^{1/2}, \\
S_2 & = S_1^{-1} (r\Theta_1^2 + r^2\Theta_1\Theta_2 + r\Phi_1^2\sin^2{\theta}\\
    & ~~~~~~~~~~ + r^2\Phi_1\Phi_2\sin^2{\theta} + r^2\Theta_1\Phi_1^2\sin{\theta}\cos{\theta}). 
\end{split}
\end{equation}
Finally, we have estimated $\Theta_2 = \mathrm{d}\Theta_1/\mathrm{d}r$ and 
$\Phi_2 = \mathrm{d}\Phi_1/\mathrm{d}r$ as follows :
\begin{equation}\label{eq_curv4}
\begin{split}
\Theta_2 = & - \frac{\Theta_1}{r} + \frac{1}{r\left(B_r^d + \sum\limits^N_{i=1}B_r^i \right)}\left(\frac{\mathrm{d}B_{\theta}^d}{\mathrm{d}r} + \sum\limits^N_{i=1}\frac{\mathrm{d}B_{\theta}^i}{\mathrm{d}r}\right)\\
           & - \frac{B_{\theta}^d + \sum\limits^N_{i=1}B_{\theta}^i}{r\left(B_r^d + \sum\limits^N_{i=1}B_r^i \right)^2}\left(\frac{\mathrm{d}B_r^d}{\mathrm{d}r} + \sum\limits^N_{i=1}\frac{\mathrm{d}B_r^i}{\mathrm{d}r}\right),\\
           &    \\
\Phi_2 = & - \frac{\Phi_1}{r} + \frac{1}{r\left(B_r^d + \sum\limits^N_{i=1}B_r^i \right)}\left(\frac{\mathrm{d}B_{\phi}^d}{\mathrm{d}r} + \sum\limits^N_{i=1}\frac{\mathrm{d}B_{\phi}^i}{\mathrm{d}r}\right) \\
           & - \frac{B_{\phi}^d + \sum\limits^N_{i=1}B_{\phi}^i}{r\left(B_r^d + \sum\limits^N_{i=1}B_r^i \right)^2}\left(\frac{\mathrm{d}B_r^d}{\mathrm{d}r} + \sum\limits^N_{i=1}\frac{\mathrm{d}B_r^i}{\mathrm{d}r}\right).\\
\end{split}
\end{equation}
where,
\begin{equation}\label{eq_curv5}
\begin{split}
\frac{\mathrm{d}B_r^d}{\mathrm{d}r} & = - \frac{3B_r^d}{r} - 2B_{\theta}^d\Theta_1 - 2B_{\phi}^d\cos{\theta}~\Phi_1, \\
\frac{\mathrm{d}B_{\theta}^d}{\mathrm{d}r} & = ~~~\frac{3B_{\theta}^d}{r} + \frac{1}{2}B_r^d\Theta_1 + B_{\phi}^d\cos{\theta}~\Phi_1, \\
\frac{\mathrm{d}B_{\phi}^d}{\mathrm{d}r} & = - \frac{3B_{\phi}^d}{r} + \frac{d}{r^3}\sin{\theta_d}\cos{\phi}~\Phi_1. \\
\end{split}
\end{equation}
and
\begin{equation}\label{eq_curv6}
\begin{split}
\frac{\mathrm{d}B_r^i}{\mathrm{d}r} = & - \frac{2.5}{D_i}~B_r^i~\frac{\mathrm{d}D_i}{\mathrm{d}r} - \frac{1}{D_i^{2.5}}\bigg(3\frac{\mathrm{d}T_i}{\mathrm{d}r}r_r^i + 3T_i\frac{\mathrm{d}r^r_i}{\mathrm{d}r} \\
      & - 3\frac{\mathrm{d}T_i}{\mathrm{d}r}r - 3T_i + \frac{\mathrm{d}D_i}{\mathrm{d}r}m^i_r + D_i\frac{\mathrm{d}m_r^i}{\mathrm{d}r}\bigg), \\
\frac{\mathrm{d}B_{\theta}^i}{\mathrm{d}r} = & - \frac{2.5}{D_i}~B_{\theta}^i~\frac{\mathrm{d}D_i}{\mathrm{d}r} - \frac{1}{D_i^{2.5}}\bigg(3\frac{\mathrm{d}T_i}{\mathrm{d}r}r_{\theta}^i + 3T_i\frac{\mathrm{d}r^{\theta}_i}{\mathrm{d}r} \\
      & + \frac{\mathrm{d}D_i}{\mathrm{d}r}m_{\theta}^i + D_i\frac{\mathrm{d}m_{\theta}^i}{\mathrm{d}r}\bigg), \\
\frac{\mathrm{d}B_{\phi}^i}{\mathrm{d}r} = & - \frac{2.5}{D_i}~B_{\phi}^i~\frac{\mathrm{d}D_i}{\mathrm{d}r} - \frac{1}{D_i^{2.5}}\bigg(3\frac{\mathrm{d}T_i}{\mathrm{d}r}r_{\phi}^i + 3T_i\frac{\mathrm{d}r^{\phi}_i}{\mathrm{d}r} \\
      & + \frac{\mathrm{d}D_i}{\mathrm{d}r}m_{\phi}^i + D_i\frac{\mathrm{d}m_{\phi}^i}{\mathrm{d}r}\bigg). \\
      &   \\
\frac{\mathrm{d}D_i}{\mathrm{d}r} & = ~ 2r - 2r_r^i - 2r\frac{\mathrm{d}r_r^i}{\mathrm{d}r}, \\
\frac{\mathrm{d}T_i}{\mathrm{d}r} & = ~ m_r^i + r\frac{\mathrm{d}m_r^i}{\mathrm{d}r} - \bigg(\frac{\mathrm{d}m_r^i}{\mathrm{d}r}r_r^i + m_r^i\frac{\mathrm{d}r_r^i}{\mathrm{d}r} \\
      & ~~~ + \frac{\mathrm{d}m_{\theta}^i}{\mathrm{d}r}r_{\theta}^i + m_{\theta}^i\frac{\mathrm{d}r_{\theta}^i}{\mathrm{d}r} + \frac{\mathrm{d}m_{\phi}^i}{\mathrm{d}r}r_{\phi}^i + m_{\phi}^i\frac{\mathrm{d}r_{\phi}^i}{\mathrm{d}r} \bigg). \\ 
      &   \\
\frac{\mathrm{d}r_r^i}{\mathrm{d}r} = & ~~~ r_{\theta}^i\Theta_1 + r_{\phi}^i\sin{\theta}~\Phi_1, \\
\frac{\mathrm{d}r_{\theta}^i}{\mathrm{d}r} = & - r_r^i\Theta_1 + r_{\phi}^i\cos{\theta}~\Phi_1, \\
\frac{\mathrm{d}r_{\phi}^i}{\mathrm{d}r} = & - r^i\sin{\theta^i_r}\cos{(\phi-\phi_r^i)}~\Phi_1. \\
      &   \\
\frac{\mathrm{d}m_r^i}{\mathrm{d}r} = & ~~~ m_{\theta}^i\Theta_1 + m_{\phi}^i\sin{\theta}~\Phi_1, \\
\frac{\mathrm{d}m_{\theta}^i}{\mathrm{d}r} = & - m_r^i\Theta_1 + m_{\phi}^i\cos{\theta}~\Phi_1, \\
\frac{\mathrm{d}m_{\phi}^i}{\mathrm{d}r} = & - m^i\sin{\theta_m^i}\cos{(\phi-\phi_m^i)}~\Phi_1. \\
\end{split}
\end{equation}

\end{document}